\documentclass[12pt,preprint2]{aastex}
\usepackage{lscape}
\newcounter{ddd}

\newcounter{aaa}

\newcommand{\map}{\hbox{{\sc mappings i}c}}
\newcommand{\nhz}{\hbox{$n_{\rm H}^o$}}
\newcommand{\upz}{\hbox{$U_o$}}
\newcommand{\phih}{\hbox{$\varphi_{\rm H}$}}
\newcommand{\phihet}{\hbox{$\varphi_{\rm HeII}$}}
\newcommand{\hb}{H$\beta$}

\newcommand{\ha}{\hbox{H$\alpha$}}

\newcommand{\heii}{\hbox{He\,{\sc ii}}}

\newcommand{\heiiuw}{\hbox{He\,{\sc ii}\,$\lambda $1640}}

\newcommand{\neiii}{\hbox{[Ne\,{\sc iii}]}}

\newcommand{\neiv}{\hbox{[Ne\,{\sc iv}]}}
\newcommand{\nev}{\hbox{[Ne\,{\sc v}]}}

\newcommand{\fevii}{\hbox{[Fe\,{\sc vii}]}}
\newcommand{\fex}{\hbox{Fe\,{\sc x}]}}

\newcommand{\siiii}{\hbox{Si\,{\sc iii}]}}

\newcommand{\sii}{\hbox{[S\,{\sc ii}]}}

\newcommand{\nii}{\hbox{[N\,{\sc ii}]}}

\newcommand{\cii}{\hbox{C\,{\sc ii}]}}

\newcommand{\ciii}{\hbox{C\,{\sc iii}]}}

\newcommand{\civ}{\hbox{C\,{\sc iv}}}
\newcommand{\civw}{\hbox{C\,{\sc iv}$\,\lambda\lambda $1549}}

\newcommand{\mgii}{\hbox{Mg\,{\sc ii}}}
\newcommand{\niii}{\hbox{N\,{\sc iii}]}}

\newcommand{\niv}{\hbox{N\,{\sc iv}]}}

\newcommand{\nv}{\hbox{N\,{\sc v}}}

\newcommand{\oiii}{\hbox{[O\,{\sc iii}]}}

\newcommand{\oiiiu}{\hbox{O\,{\sc iii}]}}

\newcommand{\oii}{\hbox{[O\,{\sc ii}]}}

\newcommand{\oi}{\hbox{[O\,{\sc i}]}}

\newcommand{\ovi}{\hbox{O\,{\sc vi}}}
\newcommand{\oviw}{\hbox{O\,{\sc vi}$\,\lambda $1035}}

\newcommand{\sed}{{\sc sed}}

\newcommand{\seds}{{\sc sed}s}

\newcommand{\fnu}{\hbox{$F_{\nu}$}}
\newcommand{\nufnu}{\hbox{$\nu F_{\nu}$}}
\newcommand{\fla}{\hbox{$ F_{\lambda}$}}

\newcommand{\ebv}{\hbox{E(B$-$V)}}
\newcommand{\ebvo}{\hbox{E(B-V)$^{I}$}}
\newcommand{\ebvt}{\hbox{E(B-V)$^{II}$}}
\newcommand{\ebvg}{\hbox{E(B-V)$^{GA}$}}

\newcommand{\mic}{\hbox{$\mu{\rm m}$}}
\newcommand{\al}{\hbox{$\alpha$}}

\newcommand{\ad}{\hbox{$\delta$}}

\newcommand{\si}{\hbox{$\sigma$}}
\newcommand{\zq}{\hbox{$z_q$}}

\newcommand{\cms}{\hbox{${\rm cm^{-2}}$}}

\newcommand{\cou}{\hbox{counts\,s$^{-1}$\,keV$^{-1}$}}
\newcommand{\cmc}{\hbox{${\rm cm^{-3}}$}}
\newcommand{\kms}{\hbox{${\rm km\,s^{-1}}$}}

\newcommand{\ufnu}{\hbox{${\rm erg\, cm^{-2}\, s^{-1}}\,{\rm Hz}^{-1}$}}

\newcommand{\Nhv}{\hbox{$10^{20}$\,cm$^{-2}$}}
\newcommand{\umm}{\hbox{${\rm erg}^{-1}\,{\rm cm}^{2}\,{\rm s}\,{\rm \AA}$}}

\newcommand{\HI}{\hbox{H\,{\sc i}}}

\newcommand{\FEKa}{\hbox{Fe\,{K$\alpha$}}}

\newcommand{\Lya}{\hbox{Ly$\alpha$}}

\newcommand{\FeII}{\hbox{Fe\,{\sc ii}}}

\newcommand{\HE}{\hbox{HE\,2347$-$4342}}
\newcommand{\HS}{\hbox{HS\,1700+6416}}
\newcommand{\HStre}{\hbox{HS\,1307+4617}}
\newcommand{\PGelf}{\hbox{PG\,1148+549}}
\newcommand{\PGten}{\hbox{PG\,1008+1319}}
\newcommand{\Pks}{\hbox{Pks\,0232$-$04}}
\newcommand{\tontr}{\hbox{Ton\,34}}

\newcommand{\gx}{\hbox{$\Gamma_{\rm X}$}}
\newcommand{\gla}{\hbox{$\gamma_{\rm \lambda}$}}

\newcommand{\dc}{\hbox{$\delta_{\rm C}$}}

\newcommand{\dmaxo}{\hbox{$\delta_{max}^{I}$}}
\newcommand{\dmaxt}{\hbox{$\delta_{max}^{II}$}}
\newcommand{\dmaxd}{\hbox{$\delta_{max}^{dia}$}}

\newcommand{\bnuvf}{\hbox{$\beta_{\rm NUV}^{dfe}$}}
\newcommand{\bnuvt}{\hbox{$\beta_{\rm NUV}^{II}$}}
\newcommand{\bnuvd}{\hbox{$\beta_{\rm NUV}^{dia}$}}
\newcommand{\bnuvo}{\hbox{$\beta_{\rm NUV}^{I}$}}
\newcommand{\bnuv}{\hbox{$\beta_{\rm NUV}$}}
\newcommand{\bfuvf}{\hbox{$\beta_{\rm FUV}^{dfe}$}}
\newcommand{\bfuvd}{\hbox{$\beta_{\rm FUV}^{dia}$}}
\newcommand{\bfuvo}{\hbox{$\beta_{\rm FUV}^{I}$}}

\newcommand{\bfuv}{\hbox{$\beta_{\rm FUV}$}}

\newcommand{\bx}{\hbox{$\beta_{\rm X}$}}

\newcommand{\bnu}{\hbox{$\beta_{\nu}$}}

\newcommand{\anu}{\hbox{$\alpha_{\nu}$}}
\newcommand{\ax}{\hbox{$\alpha_{\rm X}$}}

\newcommand{\aox}{\hbox{$\alpha_{\rm OX}$}}

\newcommand{\anuv}{\hbox{$\alpha_{NUV}$}}

\newcommand{\Cut}{\hbox{$C_{\rm roll}$}}

\newcommand{\labrk}{\hbox{$\lambda_{brk}^{dfe}$}}
\newcommand{\labrkav}{\hbox{$\bar \lambda_{brk}^{dfe}$}}
\newcommand{\laro}{\hbox{$\lambda_{roll}$}}
\newcommand{\Enexf}{\hbox{$E_{join}^{dfe}$}}
\newcommand{\Enexo}{\hbox{$E_{join}^{I}$}}

\newcommand{\Enexd}{\hbox{$E_{join}^{dia}$}}

\newcommand{\Enexfa}{\hbox{$\bar E^{dfe}_{join}$}}

\newcommand{\Ngal}{\hbox{$N_{H}^{GA}$}}
\newcommand{\Nhx}{\hbox{$N_{H}^{X}$}}

\newcommand{\None}{\hbox{$N_{H}^{I}$}}
\newcommand{\Ntwo}{\hbox{$N_{H}^{II}$}}
\newcommand{\Ndia}{\hbox{$N_{H}^{dia}$}}

\newcommand{\NN}{\hbox{$N_{20}$}}
\newcommand{\NNa}{\hbox{$\bar N_{20}$}}

\slugcomment{accepted for publication in ApJ}

\slugcomment{ \today, acepted}
\shorttitle{Quasar \seds} \shortauthors{Haro-Corzo,  et\,al.}

\begin{document}

%\title{Characterizing the Individual Shape \\for Quasars in the Ionizing Continua UV-X}
\title{Energy distribution  of individual quasars from far-UV to X-rays: I. Intrinsic UV hardness and dust opacities}

\author{Sinhu\'e A. R. Haro-Corzo\altaffilmark{1,2}, Luc Binette\altaffilmark{1}, Yair Krongold\altaffilmark{1},
Erika Benitez\altaffilmark{1},  Andrew Humphrey\altaffilmark{1},
Fabrizio Nicastro\altaffilmark{1,3} \and\ Mario
Rodr\'{\i}guez-Mart\'{\i}nez\altaffilmark{1}
%+
}

%\affil{Instituto de Astronom\'ia, Universidad Nacional Aut\'onoma de M\'exico\\
%Apartado Postal 70264,04510 M\'exico, D.F., M\'exico D.F.}

\altaffiltext{1}{Instituto de Astronom\'ia, Universidad Nacional
Aut\'onoma de M\'exico, Ap. 70264, Cp. 04510,
Ciudad de M\'exico, Estados Unidos Mexicanos.}

\altaffiltext{2}{Instituto de Ciencias Nucleares, Universidad Nacional
Aut\'onoma de M\'exico, Ap. 70543, Cp. 04510, Ciudad de M\'exico,
Estados Unidos Mexicanos. Present email: haro@nucleares.unam.mx }

\altaffiltext{3} {Harvard-Smithsonian Center for Astrophysics, 60
Garden St., Cambridge, MA 02138, USA.}

\begin{abstract}
Using {\em Chandra} and HST archival data, we have studied the
individual Spectral Energy Distribution (\sed) of 11 quasars.
%with
 redshifts $0.3<z<1.8$.
All UV spectra show a spectral break around
1100\,\AA. 5 X-ray spectra show the presence of a ``soft excess''
and 7 spectra showed an intrinsic absorption. We found that for most
quasars a simple extrapolation of the far-UV powerlaw into the X-ray
domain generally lies below the X-ray data and that the big blue
bump and the soft X-ray excess do not share a common physical
origin. We explore the issue of whether the observed \sed\ might be
dust absorbed in the far and near-UV. We fit the UV break, assuming
a powerlaw that is absorbed by cubic nanodiamond dust grains. We
then explore the possibility of a universal \sed\ (with a unique
spectral index) by including further absorption from SMC-like
extinction. Using this approach, satisfactory fits to the spectra
can be obtained. The hydrogen column densities required by either
nanodiamonds or amorphous dust models are all consistent, except for
one object, with the columns deduced by our X-ray analysis, provided
that the C depletion is $\sim0.6$. Because dust absorption implies a
flux recovery in the extreme UV ($<700\,$\AA), our modeling opens
the possibility that the intrinsic quasar \sed\ is much harder and
more luminous in the extreme UV than inferred from the near-UV data,
as required by photoionization models of the broad emission line
region. We conclude that the intrinsic UV \sed\ must undergo a sharp
turn-over before the X-ray domain.
\end{abstract}


\section{Introduction}

The spectra of quasars and Seyfert galaxies show strong emission
lines superimposed onto a bright continuum.
%%($F_{\nu}\propto \nu^{+\alpha}$ where $F_{\nu}$ is the energy flux
%% and  $\alpha$ is the spectral index).
The continuum contains a significant feature in the
optical-ultraviolet region, known as ``the Big Blue Bump'' (BBB).
The emission lines are generally believed to result from
photoionization, in which the far-UV part of the BBB is reprocessed
into line emission.  The shape of the (ionizing) continuum in the
far-UV domain is therefore crucial in the modeling of the observed
intensities of the emission lines. The continuum of Active Galactic
Nuclei (AGN) is furthermore believed to be responsible for the
ionization of gas components observed in absorption, either within
the AGN environment, the so-called ``warm absorbers'' (WA)
\citep[e.g.][]{mathur94}, or within  intergalactic space (e.g. \Lya\
forest and Lyman limit systems). Due to the huge photoelectric
opacity of the Galaxy, the Spectral Energy Distribution (\sed) of
the ionizing continuum, between the Lyman limit and the soft X-rays
(EUV-X), is poorly known. Fortunately, owing to the redshift effect
and the transparency of the quasar environment, it has been possible
to infer the \sed\ of quasars down to $\sim 300$\,\AA. From the
X-ray side, the quasar \sed\ becomes visible again. The physical
relationship between the soft X-rays and the BBB is still a matter
of debate \citep[][]{W04}.

Using the same HST-FOS database as \citet[][ TZ02]{TZ02}, we present
a detailed study of eleven quasars for which a good quality far-UV
spectra exist as well as archived \emph{Chandra} ACIS-S data. Our
objective in this
%%and in previous papers (\citet[][ B05]{a2}, \citet{[][BK07]{binette06})
is to set constraints on the \sed\ behavior in the unknown spectral
domain between the UV and the X-rays.  Any information about this
spectral region, even if indirectly gathered, could improve our
understanding of accretion disk physics as well as of the physical
conditions of the broad emission line region (BELR) and warm
absorbers.

We now briefly review the salient points on what is known about
quasar \sed, in the near and far-UV as well as in the X-ray domain.
We will explore a  novel role for dust in the paradigm that the BBB
is produced by an accretion disk.

\subsection{The far-UV photon crisis} \label{sec:cri}

Satellite and ground-based observations of distant quasars showed
that the BBB peaks in \nufnu\ around $1000$\,\AA\ (rest-frame)
\citep[][hereafter Z97]{OG88,fh91,ZK97}. To characterize the general
trends of the continuum shape, various authors co-added spectra from
samples of quasars at different redshifts. In Fig.\,\ref{fig:seds}
we plot the Sloan Digital Sky Survey (SDSS) composite \sed\ (gray
continuous line) constructed by \citet{vandenberk01}, using 2200 AGN
spectra of redshifts between 0.044 and 4.8. TZ02 showed that the
sharp turnover shortward of $\sim 920$\,\AA\ is mostly due to the
cumulative effect of unresolved intergalactic \Lya\ forest
absorbers, which results in the so-called Lyman valley trough
\citep{moller90}.  To correct this absorption, TZ02 (and previously
Z97) applied a statistical correction to each spectrum before
averaging them. The resulting composite, represented by the red
continuous line in Fig.\,\ref{fig:seds}, consists of 332 HST-FOS
archived spectra of 184 quasars with redshifts between 0.33 and 3.6.
In this case, the resulting \sed\ extends into the far-UV, as one
would expect if this continuum was responsible for the
photoionization of the BELR. A significant break occurs around
1100\,\AA, followed by a steep flux decrease in the far-UV,
equivalent to $\nu^{-1.7}$ ($F_{\nu}\propto \nu^{+\alpha}$).
\citet{KF97} pointed out the difficulties in reproducing the
equivalent widths of the high ionization lines of \heiiuw, \civw\
and \oviw, using a powerlaw as soft as $\nu^{-2}$. Note that the
TZ02 composite with $\nu^{-1.7}$ is somewhat harder than assumed by
the aforementioned authors.

A puzzling aspect is the significant variation in the hardness of
the far-UV continuum among different quasars. Many quasar spectra
reveal an \sed\ significantly steeper than the composite, yet there
is little evidence that the emission lines are very different in
those objects. For instance, the most extreme case is \tontr, which
shows an ultrasteep continuum $\propto \nu^{-5.3}$ but a near-UV
line spectrum not very different from other quasars \citep[][
hereafter BK07]{binette07}. \citet[][ B05]{a2} proposed that
crystalline carbon dust, that is, nanodiamonds\footnote{B05 combined
two flavors of nanodiamonds, meteoritic (i.e. surface-hydrogenated)
and cubic (without surface adsorbates), assuming the small size
regime. The resulting extinction curves A1 and D1 are reproduced in
Fig.\,\ref{fig:exti}.}, might be responsible for the 1100\,\AA\
break, opening the possibility that a rise in flux may take place at
energies higher than 20\,eV. According to this hypothesis, the
absorption trough would indent the continuum only within a
relatively narrow spectral region, between 10 and 15\,eV, followed
by a flux rise at higher energies.  Hence, such absorbed \sed\ would
still produce more photons in the extreme UV beyond 20\,eV than
obtained by extrapolating the TZ02 composite. Examples of \seds\
absorbed by nanodiamond dust are represented by the two dotted lines
in Fig.\,\ref{fig:seds}, corresponding to Models\,I and II, both
defined in \S\,\ref{sec:sedpeind}. The suggestion that AGN may be a
``factory'' of dust \citep{elvis02} or nanodiamonds is not new
\citep{rouan04a,rouan04b}.

\citet{KF97} speculated that there could be a second bump peaking
near 54\,eV that would provide the necessary hard photon luminosity.
In any event, state-of-the art photoionization calculations of the
BELR show a preference for a much harder ionizing \sed\
compared to what is commonly observed in quasars. To
illustrate this, in Fig.\,\ref{fig:seds} we show two distributions
used by \citet[][ KO97]{KB97} in their extensive grid of BELR
models: the short and long-dashed cyan lines correspond to an
exponential cut-off \sed\ of the type $\nu^{-0.5}\exp{-h\nu/kT}$,
with $kT=43$ and 86\,eV, respectively. The Locally Optimally
emitting Cloud model of \citet{baldwin95} and, more recently, of
\citet{CL06} use a similar \sed\footnote{Although \citet{CL06}
find evidence of a very hard \sed\ in RE\,1034+39, they also report
that "the slope appears to become flatter in the FUSE spectrum"
shortward of \ovi, which they qualify as "an unexplained feature"
that is left out of their analysis.}. Also shown in the same figure
is the popular quasar \sed\ of \citet{mf87} (yellow dashed line).
This inconsistency between the \sed\ preferred in photoionization
calculations and the \sed\ actually observed in most quasars is
unfortunate and requires a solution.

An additional aspect to be taken into account is that nearby, less
luminous AGN might possess an intrinsically harder \sed, as reported
by \citet{SK04}, who used FUSE data to construct a composite \sed\
comprising AGN of redshifts $<0.7$.  These authors did not find
evidence of a UV-break, at least for an important fraction of their
objects. The 11 quasars reported in this paper, however, are
consistent with the presence of a break. As in B05, our analysis
will focus on  individual spectra rather than on composites.
%%\sed\ averages, which may in some cases wash away significant spectral features.

\subsection{Constraints provided by the X-ray domain}\label{sec:xco}

Since the X-ray domain undergoes limited absorption as compared to
the Lyman continuum, it provides us with constraints on how the BBB
might behave in the unobservable extreme-UV domain. This unknown
domain is depicted by the shaded area in Fig.\,\ref{fig:seds}. Past
studies \citep[e.g.][ and references therein]{reeves00} revealed
that the typical spectrum of a quasar in the hard X-ray band ($>
2$\,keV) is dominated by a powerlaw-like emission flux of index $\ax
\sim -1$. It has also been found that a gradual upturn occurs below
2\,keV, the so-called X-ray {\it soft excess} \citep{comastri92}.
The latest generation of X-ray satellites (XMM-\emph{Newton} and
\emph{Chandra}), owing to their improved spectral resolution and
sensitivity, confirmed the ubiquity of this excess \citep[e.g.][
hereafter PI05]{GS03,porquet04,pj05}. However, its nature as well as
its relationship with the BBB are still a matter of debate.

\subsubsection{The broken powerlaw X-ray model}\label{sec:bpm}

Considering a sample of 23 low redshift quasars [the \citet{laor97}
subsample of PG quasars] observed with XMM-\emph{Newton},
\citet{bs06} favored a broken powerlaw model to fit both the soft
energy excess and the harder X-ray segment. They found that the two
indices used in their models are well correlated, and they concluded
that the X-ray components in the range 0.3--10\,keV shared a common
origin. These authors also reported that accretion disk parameters
responsible for the UV BBB appeared to be independent of the
0.3\,keV excess.

We averaged the broken powerlaw fits to 13 AGN ($z<0.4$) that
\citet{bs06}  reported as `good fits' (see their Table\,8). Such
average is represented by the two navy blue lines in
Fig\,\ref{fig:seds}. Each line illustrates a predefined
value\footnote{Negative indices imply decreasing flux from the UV to
the X-rays.} of the \aox\ index ($-1.3$ and $-1.5$). The flux from
the TZ02 composite\footnote{Using the observed composite flux rather
than that of a continuum model is preferable since it includes the
contribution from the `small bump' around 2700\,\AA, which is
believed to be due to unresolved multiplet \FeII\ emission
\citep{wills85,yip04}.} at $\sim 2500$\,\AA\ was used to determine
this index. \citet{laor97} and TZ02 suggested that the soft excess
might correspond to an extension of the BBB far-UV powerlaw into the
X-rays. Our Model\,II in Fig.\,\ref{fig:seds}, for instance, would
connect smoothly with the soft excess if extrapolated beyond the
shaded area. Although a tempting proposition, we should beware of
general conclusions derived from composite \seds. A detailed
analysis centered on individual quasar \seds\ leads us to reject the
above premiss in \S\,\ref{sec:sed}.

%As a matter of fact, we find in \S\,\ref{sec:sedpeind} that for each
%of our 11 quasars, none of the far-UV model-fits, assuming
%Model\,II, extrapolates to an X-ray flux compatible with the
%\emph{Chandra} observations.

The \sed\ behavior within the data gap (the shaded area in
Fig.\,\ref{fig:seds}) is not known a\,priori. In order to establish
useful constraints on the different BBB models, we will extrapolate
the UV models across this gap and compare them with the X-ray
observations.

%In all cases, we will assume that the far-UV \sed\ behaves as a
%powerlaw down to the X-rays.

\subsubsection{A thermal model for the X-ray excess}\label{sec:bro}

Using a larger subset of 40 PG (Palomar-Green) quasars, PI05 fit
five different models to the soft X-rays: either one or two
blackbodies (A, E), a multicolor blackbody (B), a bremsstrahlung
emission model (C), and a powerlaw (D). Except for one object, all
these models require the addition of an underlying powerlaw in order
to fit the high energy band. Models A or E were favored for 19
objects, while model\,D was favored for 6 objects only. The soft
excess in some cases shows a top-hat behavior, which favors a
thermal interpretation. The green line in Fig.\,\ref{fig:seds}
represents an average of 13 model fits (blackbody+powerlaw) of
quasars with $z < 0.4$ and of type A and E, using parameters of
Table\,5 from PI05. This average \sed\ has been arbitrarily scaled
to an \aox\ of $-1.45$, and the mean blackbody temperature
describing the excess is $\approx 140$\,eV. PI05 emphasize that the
five functional forms they used must be considered as
phenomenological and not as physically consistent models.

A well resolved rollover (i.e. decreasing flux towards the UV)
of the soft X-ray component, or at least a strong curvature, has
also been reported in
%Mrk\,279 \citep{constantini07},
Ark\,120 \citep{vaughan04}, Mrk\,359 \citep{obrien01} and NGC\,4052
\citep{collinge01}. In a similar fashion to PI05, we adopt in this
Paper a single blackbody to fit the soft X-ray excess.

\subsection{The physical origin of the BBB}\label{sec:mec}

Traditionally, the BBB has been interpreted as a manifestation of
accretion disk emission \citep{shields98,malkan83,edelson86} around
a supermassive blackhole (BH). The presence of a BH in the nucleus
of all grand design galaxies has since been confirmed \citep[e.g. ][
and references therein]{ferrarese00}. A persisting difficulty,
however, is that in the region of the far-UV break, the BBB is not
fitted well by `bare' accretion disk models.  This is illustrated by
a comparison of the composite \sed\ from Z97 with state-of-the-art
disk models such as those of \cite{ha00} (c.f. their Fig.\,22),
which include H and He opacities and non-LTE transfer. A solution
proposed by Z97 and \citet[][ SBG05]{SB05} is the comptonization of
the disk emission by a hot corona above the disk. This corona would
smear the intrinsic disk features such as the Lyman break and result
in a \sed\ with a powerlaw tail in the far-UV. Hence accretion disks
do not necessarily imply a sharp thermal cut-off, as predicted by
bare disk emission models.

For a given \sed\ optical flux, the efficiency in generating {\it
ionizing} photons is higher for hotter accretion disks, and these
imply a fairly hard continuum in the near-UV, with indices $\anuv\
\ga -0.2$. The TZ02 and the \citet{SK04} composites, on the other
hand, reveal a much softer continuum with a mean \anuv\ index of
$-0.67$ and $-0.57$, respectively. In this Paper, we explore whether
dust might be at the origin of this apparent \sed\ softness in the
near-UV. Following the results of \citet{hs04}, who finds evidence
of extinction in AGN similar to that encountered in the Small
Magellanic Cloud (SMC), we will be considering such extinction in
\S\,\ref{sec:sedpeind} to deredden the observed \seds\ until
they match a target spectral index.

As for the X-rays, a popular physical scenario for the {\it hard}
X-ray ($> 2\,$keV) component is that of a hot corona placed above
the accretion disk, which comptonizes the UV to soft X-ray thermal
emission from the disk and up-scatters it into the hard X-ray band
\citep{haardt91}. On the other hand there are many competing
physical mechanisms that have been proposed to explain the {\it soft
excess}. They are reviewed in PI05, \citet{bs06} and
\citet{porquet04}.

%Although we favor a thermal origin in our fits of the excess, our
%conclusions about whether our far-UV models can be extrapolated or
%not into the X-rays do not critically depend on this choice.

\subsection{Objectives of this work}

The issues that will be tackled in this Paper are the following:
using individual spectra rather than composites, we analyze to what
extent the extrapolation of the far-UV results in an X-ray flux
compatible with \emph{Chandra} observations. In the event that the
UV break was the result of crystalline dust absorption, we wish to
know whether the rollover proposed by B05 at 18.5\,eV, followed by a
powerlaw tail in the extreme-UV, is compatible with the observed
X-ray flux. Is this powerlaw tail an acceptable function? Or is a
thermal cut-off more natural? Is SMC-type dust able to account for
the observed wide range in near-UV spectral indices? Are the
columns of material implied by all the dust components compatible
with the intrinsic absorption column inferred from the X-rays? What
is the impact of the dust-absorbed \seds\ studied in this Paper on
the emission spectrum of the Narrow Line Region (NLR)? The framework
that we introduce to explore these issues should be useful for
follow up studies concerned with the far-UV spectral gap in AGN
\seds.

\section{The observational dataset} \label{sec:data}

In this Paper, we study 11 quasars for which high quality spectral
data sets exist in {\it both} the HST-FOS and the \emph{Chandra}
archives. Of these, 9 are radio-loud quasars (RLQ) and 2 radio-quiet
quasars (RQQ). The larger fraction of RLQs is due to small
number statistics of quasars common to both databases. In
Table\,\ref{tab:sam}, we list the object names (Col.\,1) in order of
decreasing X-ray counts. The italic letter shown in Col.\,2 is used
for identification throughout the Paper and in all figures. Other
table entries are the redshift (Col.\,3), the Galactic \HI\ column
density \citep{DL90} (Col.\,4), radio loudness of each source, RQQ
vs RLQ (Col.\,5), and the UV spectral class according to the
nomenclature\footnote{Class\,A spectra show a UV break similar to
that seen in the composite \sed\ of TZ02, while class\,B quasars
show a much steeper flux drop shortward of the break (see
classification scheme in Fig\,2 of B05).\label{foot:cl}} of B05
(Col.\,6). The X-ray journal of observations is also briefly
described and includes the following: the identification number for
each source in the \emph{Chandra} database (Col.\,7), the exposure
time (ksec) of each observation (Col.\,8), the total number of
counts for each spectrum (Col.\,9), the frame-time information
(Col.\,10), and \emph{Chandra}'s observation date (Col.\,11). HST's
observation date is also listed (Col.\,12).

We have postponed the analysis of the 14 quasars   (10 RLQ and 4
RQQ) that showed
evidence of a WA in the UV to a future paper, except for
3C351, for which the detailed analysis is presented in
Appendix\,\ref{sec:rem}. Since it is the only object in common with
the SBG05 sample, its inclusion is intended to facilitate a
comparison of our results with theirs.
%% about the relation between X-rays and far-UV domains.

\subsection{The UV Data} \label{sec:uvda}

The UV spectra for our sample were extracted from the TZ02 database,
which R. C. Telfer was kind enough to provide. It comprises spectra,
mostly HST-FOS\footnote{Note that the TZ02 sample includes 3
HST-STIS and 6 HST-GHRS spectra.}, of 184 quasars, already reduced,
corrected for Galactic dust extinction, de-redshifted and finally
rebinned uniformly. Furthermore, the spectra have been corrected by
TZ02 for the presence of Lyman limit absorbers (down to $\tau >
0.3$), as well as for the \Lya\ absorption valley (caused by the
cumulative absorption from unresolved \Lya\ forest lines, see
\S\,\ref{sec:cri}). We selected spectra that possess the following
characteristics: an underlying continuum of acceptable S/N (see
Col.\,3 in Table\,\ref{tab:uv}, sufficient spectral coverage above
and below the 1100\,\AA\ UV break (see Table\,\ref{tab:uv}), and the
absence of deep absorption lines or similar features from a BALQSO.

The rest-frame UV spectra and their error bars are plotted in $\log
\nu F_{\nu}$ vs. $\log \nu$ in Fig.\ref{fig:uv}. To increase the S/N
and to reduce the smearing from undifferentiated high amplitude
scatter, these spectra have been rebinned to group together 10
points per resolution element.

\subsection{The X-ray Data.} \label{sec:xda}

Our sample includes quasars for which \emph{Chandra}'s observations
were carried out with ACIS-S. The corresponding datasets were freely
available from the \emph{Chandra} X-Ray
Center\footnote{http://chandra.harvard.edu/}.

The files were processed using the CIAO software (version 3.0.2) and
following the
on-line\footnote{http://cxc.harvard.edu/ciao/threads/data.html} data
analysis `threads' provided by the {\it Chandra} X-ray Center in
order to extract source and background spectra.  Redistribution
Matrix Files (RMF) and Auxiliary Response Files (ARF) were generated
by using the CIAO tools {\it mkrmf} and {\it mkarf}, respectively.

We  estimated whether each source was affected by pile-up (at a
level $\ge 10\%$) using the expression $PU=\frac{TC*FT}{ET}$, where
TC is the Total Counts, FT is the frame-time and ET is the
Exposure-Time of each observation (see Table\,\ref{tab:sam}). If $PU
\ge 0.3$, a pile-up correction was carried out. We found that only
quasars $b$, $c$, $g$ and $h$ showed a significant pile-up. For
these, the correction was performed by selecting only counts within
an annulus centered on the source and defined by radii
0.6--3\,arcsec. We took into account that a proper fit requires at
least 1000\,counts. For all other objects, we used a circle of
3\,arcsec radius centered on the source. The selection of this size
aperture guarantees that counts due to extended radio-jets, if
present, are avoided\footnote{We had no means to remove the
contribution from the {\em unresolved}  X-ray jet component close to
the nucleus, which is possibly amplified due to Doppler-beaming
\citep{GS03}. The X-rays measured in RLQs might therefore be
considered upper limits of the nuclear fluxes.} as well as counts
from weaker nearby sources. This procedure was performed by using
either a box or a circle around each non-nuclear X-ray source. For
each source, we generated a background spectrum by considering all
the counts within an annulus defined by the radii of 10 and 20
arcsec, again centered on the object.

%%(Table\,\ref{tab:sam})

\section{Separate model-fitting of the UV and X-ray segments}\label{sec:met}

We use powerlaw segments to characterize the behavior of the energy
distribution in the UV and in the X-rays. We performed spectral
fitting\footnote{The commands used for the different fits are
described in {\tt http://cxc.harvard.edu/ciao/ahelp/xs.html}} of the
11 quasars, using the
Sherpa\footnote{http://cxc.harvard.edu/ciao/ahelp/sherpa.html}
package in CIAO. Each powerlaw segment is expressed by the form
$\fnu \propto \nu^{+\al}$ in flux space \fnu\ (\ufnu) or by the form
$\nufnu \propto \nu^{+\bnu}$ in the customary $\nufnu$
representation. Since most figures are in $\log \nufnu$ vs. $\log
\nu$, we prefer to use the $\bnu$ index throughout most of the text.
Such an index directly reflects the apparent slopes of the powerlaw
segments seen in the figures. We recall that $\anu = \bnu - 1$. A
few authors prefer to use an index defined in \fla\ space ($\fla
\propto \lambda^{+\gla}$), in which case, we have $\gla = -2 -\anu =
-1 -\bnu$. For the X-rays, the photon index \gx\ commonly used in
the literature is given by $\gx = 2 - \bx$ ($\frac{\fnu}{h\nu}
\propto \nu^{-\gx}$, note
 sign convention for \gx).

\subsection{Modeling of the ultraviolet region} \label{sec:uvme}

As in TZ02, we fitted two broken powerlaws, one for the near-UV and
the other for the far-UV, of indices \bnuvf\ and \bfuvf,
respectively\footnote{Note that TZ02 used \anu\ indices, and that
$\bnu =\anu + 1$}.  These preliminary fits assume that no intrinsic
dust is present. We employ the superscript {\it `dfe'} to label
quantities related to the {\it `dust-free'} fits. The wavelength
at which the two broken powerlaws join is defined as \labrk.

Spectral fitting was carried out in two different ways: the first
technique consisted of fitting a broken powerlaw ({\sc bp}) to all
the line-free continuum regions, leaving out the important emission
lines, in a similar fashion as TZ02 and \citet{SK04}. This masking
is equivalent to excluding the following wavelength intervals from
the fit: 750--800, 820--850, 900--1095, 1100--1135, 1150--1265,
1380--1420, 1470--1610, 1830--1950, 2700--2880\,\AA\footnote{Objects
{\it e, g, h, i} and {\it k} required slight modifications of these
blackout intervals, c.f. App.\,\ref{sec:rem}}. The second technique
consisted of fitting the continuum and the emission lines together,
using a line profile consisting of both a narrow and a broad
gaussian component. We found good agreement between the spectral
indices derived using each of the two techniques, and therefore we
consider both reliable. Since we are primarily interested in the
continuum rather than in the lines, we adopted the masking technique
due to its simplicity. Because the continuum break occurs near
1100\,\AA, that is, within the 1100--1135 mask or the \Lya\ mask,
the determination of \labrk\ was complicated and required an initial
eye-ball estimate.  The position of \labrk\ was determined from an
initial fit that did not consider any masking. We then froze \labrk\
and proceeded to determine new values of \bnuvf\ and \bfuvf\ by
least-square-fitting within the unmasked regions only. These values
are presented in Table\,\ref{tab:uv}. The two segments from the best
fitted broken powerlaw are superimposed onto the UV spectra in
Fig.\,\ref{fig:uv} {\it a--k}. The far-UV break near 1100\,\AA\ is
apparent in each quasar spectrum. The mean break position
characterizing our sample is $\labrkav =1165$\,\AA, which is
significantly redward of  the Lyman limit.
%%the energy at which an absorption edge is predicted by some
%%geometrically thin accretion disks \citep[e.g. ][]{kolykhalov89}.
%%%B05 found that nanodiamond extinction can reproduce very well the
%%%position and the steepening of the sharp UV break.

%%Our sample corresponds to only a very small subsample of the values
%%determined by TZ02 (see their Fig.\,14).
The index difference, $\bfuvf - \bnuvf$, gives us a measure of the
steepening across the break.
%%It is listed in Col.\,3 of Table\,\ref{tab:rat}.
The values we determined range from $-1.93$ to $-0.72$, with a {\it
mean} difference value of $-1.39$. The latter  is comparable to the
mean index difference of $-1.22$ and $-0.81$ for the RLQ and RQQ of
the TZ02 sample, respectively.

\subsection{Modeling the X-ray region}\label{sec:xme}

To model the intrinsic X-ray continuum of each quasar, the data were
first rebinned in such a way that the number of counts in each bin
exceeded 20. This had the advantage of preserving the Poisson
statistics when fitting Gaussian profiles and of facilitating the
statistical handling of the background substraction.
%%(otherwise one would have to use the Gehrels statistics whose standard
%%deviation is $1 + (N + 0.75)^{0.5}$, where N is the counts per bin).
The observed continua with their error bars are shown as \cou\ vs.
energy (observer-frame) in the top panels of Figs.\,\ref{fig:x} {\it
a--k}.

The X-ray spectra were fitted in the observer-frame by models that
were folded into the spectral characteristics of the ACIS-S
instrument. We initially fitted a simple powerlaw attenuated by
Galactic absorption of column \Ngal. This initial fit was limited to
the range 2.5--6\,keV (rest-frame), thereby excluding the soft X-ray
region and the \FEKa\ line region around 6.4\,keV. It did {\it not}
consider possible absorption by moderate columns of neutral gas
(i.e. $\le 10^{21}\,$\cms) or by larger columns of ionized gas (i.e.
$\le 10^{23}$\,\cms), both {\it intrinsic} to the quasar \citep{KN03}. The
residuals of this initial fit are presented in the middle subpanels
of Figs.\,\ref{fig:x}\, {\it a--k}.
%%The results of this model are summarized in Col.\,3 of Table\,\ref{tab:x}.

For almost all the quasars, such a simple model could not provide a
satisfactory fit (see Table\,\ref{tab:x}) over the entire energy
range (i.e. 0.3--6\,keV observer-frame). The middle subpanels show
positive as well as negative deviations (with a significance $\ge 2
\sigma$). Shortward of 2.5\,keV, we find either a soft excess or
evidence of absorption gas. Longward of 4\,keV, there is some
evidence of a line emission feature, which might correspond to a
fluorescent transition of neutral iron emission from \FEKa\ at
6.4\,keV. The next step was to repeat the fit, leaving the
spectral index of the powerlaw frozen to the previous value, but now
including a blackbody and/or an intrinsic absorber component in
order to account for the observed residuals (see middle panels). We
left the normalizations of the powerlaw, blackbody and intrinsic
absorption free to vary. We can see in the bottom panels a
noticeable reduction of the residuals at low energies. Note that the
fit to 3C351 (object $c$) requires a warm absorber component, as
shown in App.\,\ref{sec:rem}-$c$.  In this last model and for
the 2 quasars $a$ and $e$ that do not show a clear evidence of soft
excess, we used an {\em f-test} to determine the level of
significance between models with and without a blackbody. The level
of confidence is less than $10^{-3}$, which confirms that including
a blackbody component  for both quasars provides a better fit.

The continuous line in the top subpanels of Fig.\,\ref{fig:x}
represents the final best-fit multi-component model, while the
corresponding residuals are shown in the {\it bottom} subpanels of
Fig.\,\ref{fig:x}. In the particular case of PG\,1115+080, we had to
consider a broken powerlaw that joins at 1.69\,keV (see note $k$ in
App.\,\ref{sec:rem}). Table\,\ref{tab:x} lists all the
parameters of the final model.  Uncertainties at a
2$\sigma$--confidence level regarding the power-law index of all
quasars were derived, considering only the 2.5--6\, keV (rest-frame)
range.

To summarize, the inferred X-ray spectral energy distribution of
each quasar includes the following components: thermal emission from
a blackbody ({\sc zbb}) for objects {\it a, e, f, g, i}, and a
Gaussian Profile ({\sc g}) representing the \FEKa\ line for objects
{\it c, g, i, j}. Evidence was found for neutral gas absorption (of
column \Nhx) intrinsic to the 6 objects {\it a, b, d, e, f, h} and a
WA in the object $c$, while for the remaining 4 objects, we froze
the other parameters at their previous best-fit values and
calculated an upper limit for \Nhx\ based on a confidence level of
2\si.
%These columns appear as upper limits in Table\,\ref{tab:x}.

%%Using ASCA data, Reeves \& Turner (2000) found a correlation between
%%\Nhx\ and redshift, for quasars of redshifts $\ga 0.5$. In
%%Fig.\,\ref{fig:reeves} we plot their values of Reeves \& Turner (2000)
%%and overlay our measurements (filled squares).  Interesting, a similar
%%correlation appears present in our analysis, though it is shifted by
%%0.25\,dex with respect to the least square fit of the Reeves \& Turner data.

\section{Combining the UV and X-ray SEDs in the dust-free case}\label{sec:sed}

In this Section, we assume the environment of quasar to be dust-free
and make use of a broken powerlaw to describe the UV spectrum of each
quasar.

\subsection{Matching the UV and X-ray components}\label{sec:matA}

We used IDL {v.\,5.5} to combine the UV and X-ray segments.  In
Fig.\ref{fig:sed}, we overlay  in \nufnu\ the UV and X-ray model
fits to obtain a single tentative description of each \sed, as a
function of $\nu$ (rest-frame).
%%Since on \nufnu\ there is no effect of the frecuency shift by factor of (1+z)
%%\citep{g97}, we simply plot \nufnu($\nu=\frac{\nu_0}{1+z}$) vs. $\nu_0$.
In order to avoid any clutter due to undifferentiated scatter and
overcrowding, the original data from Figs.\,\ref{fig:x} and \ref{fig:sed} have been left out.  Arrows
near the top of each panel delineates the spectral extent of the original data.

%%For all the UV \seds\ shown in Fig.\,\ref{fig:sed}, we assumed that
%%the environment of the quasar is dust-free, leaving to
%%\S\,\ref{sec:cor} and \S\,\ref{sec:sedpeind} the exploration of dust
%%effects. We now review in detail the information presented in each of the
%%eleven \seds\  {\it a--k} shown.
The near- and far-UV powerlaw  fits are shown as dashed lines, while
the two crossing dotted lines represent a 2$\sigma$ deviation about
the best-fits. We recall that the indices \bnu\ (with \bnu = \anu +
1) are explicitly listed in each panel and represent the actual
slopes\footnote{The conversion from other index definitions such as
\gx\ and \gla\ is obtained as follows: $\bx = 2 -\gx $ and $\bnu =
-\gla - 1$.} in \nufnu\ plots.

As for the X-ray domain, the previous best-fit models are  plotted
as dashed lines. The associated limit variations, assuming a 2\si\
uncertainty for the parameters used, are represented by  dot-dashed
lines. These `fluxed' X-ray models represent the intrinsic \sed\
corrected, not only for Galactic absorption (GA) by column \Ngal,
but also for intrinsic absorption by gas column \Nhx\
(\S\,\ref{sec:xme}).  A vertical dotted-line denotes the {\it lower}
energy boundary of the data.

We determined the energy, \Enexf, of the intercept that occurs when
extrapolating both the far-UV and X-ray segments. In
Fig.\,\ref{fig:sed}, a small circle denotes the position of the
intercept.  The mean intercept energy for the sample is $\Enexfa =
265$\,eV (this average excludes two objects without a defined
intercept, as discussed below). The symbol `@' represents the UV
break position \labrk. We also derived the \aox\ index, that is, the
spectral index of an imaginary powerlaw joining the two flux values
at 2500\,\AA\ and 2\,keV (as denoted by open triangles). This index
is given in Table.\,\ref{tab:sed} and in the inset of
Fig.\,\ref{fig:sed} (along with \labrk\ and \Enexf).  Its value
ranges between $-1.78$ and $-1.11$, with a mean of $-1.46$. The
\aox\ values from samples of QSOs of comparable luminosities tend to
fall in the range $-1.5$ to $-1.6$ \citep{AM87,AW95,GS95,yuan98}.

\subsection{Discussion\,I: the dust-free case}\label{sec:diso}

We now discuss the issue of whether the extrapolation of the far-UV
powerlaw leads to a flux level compatible with the one inferred from
the X-ray model. There are two relevant cases to consider: A-- the
extrapolated far-UV flux model lies {\it below} or {\it at} the
level inferred from the extrapolation of the X-ray model, B-- the
extrapolated far-UV lies {\it above} the level inferred from the
extrapolation of the fluxed X-ray model. In case\,A, at a
phenomenological level, the extrapolation leads to a congruent
result, that is, the extreme-UV flux either vanishes relative to the
soft X-ray component or connects smoothly with it, while in case\,B,
we conclude that a simple extrapolation of the UV powerlaw is
nonsensical.  The far-UV segment in case\,B objects {\it must}
instead undergo an abrupt steepening before reaching the X-ray
domain. With case\,A, the extrapolation of the far-UV segment
does not lead to inconsistencies with the X-ray data, while it does
with case\,B, unless further steepening occurs in the extreme-UV,
possibly in the form of an exponential turn-over.

The statistics are the following: out of 11 quasars, 9 are case\,A
and only 2 case\,B ({\it b, k}). We conclude that the extrapolation
of the far-UV powerlaw leads to a flux level compatible with that
measured in the X-rays, except for two sources. This does not entail
that the continuum in the {\it unobserved} region between the far-UV
and the X-rays necessarily behaves as the particular functional form
considered. The intrinsic \sed\ may still steepen at some unknown
intermediate energy (between 40 and 350\,eV), before the soft X-ray
excess component takes over.
%%300 ang to 35 ang  or 40eV to 340eV
Such a possibility is favored in scenarios in which a thermal
distribution is believed to provide  a more appropriate description
of the BBB.  Finally, absorption by dust can strongly affect the
observed \sed\ in the far-UV. Correcting for such absorption would
lead to a different intrinsic \sed\ and its extrapolation may result
in a different balance between case\,A and  case\,B, as is found in
\S\,\ref{sec:sedpeind}.

Quasar variability is a concern, since the observations in the UV
and X-rays were not simultaneous. The characteristic amplitude of
continuum variability decreases with source luminosity
\citep{pica83,trevese94,hook94,hawkins96,cristiani97}. It is smaller
for quasars than Seyferts. \citet{giveon99} found a B band
\emph{rms} variability of 0.14\,mag, following a 7 year monitoring
campaign of 42 PG quasars of redshift $< 0.4$. Even though such
amplitude is relatively small, there are clear indications that the
amplitude of variations in the far-UV are larger
\citep[e.g.][]{reimers05}.
%In the X-rays, variability is also well
%documented. The most probable amplitude for a quasar is not likely
%to exceed a factor of a few.
As for the X-ray domain,  it has been shown in general terms
that variability anti-correlates with luminosity \citep{nandra97},
and our sample consists of high luminosity quasars. For our sample
in particular, \citet{GS03} reported no short-term high-amplitude
variability for the quasars $b$, $f$, and $i$. As for longer term
variability, these authors reported a decrease of a factor $\simeq
1.5$ over nine years for quasar $b$ and a factor $\simeq 3$ decrease
over twenty years for quasar $f$. It is conceivable that when the
X-ray and UV observations are not concomitant, the distinction
between case\,A and case\,B for a particular quasar may depend on
the temporal limitations of the data. On the other hand, when a
reasonably sized sample is considered, flux variations should have a
neutral impact on the relative proportion of case A vs. B. We note
that the ratio of the extrapolated far-UV model to the observed
X-ray flux, both estimated at the long wavelength limit of the X-ray
data, exceeds $\pm 0.6$\,dex for 7 of the 11 quasars. We
therefore consider it unlikely that our results of a majority of
case\,A's could be reverted owing to AGN variability. The
possibility that the X-ray continuum is significantly contaminated
by an unresolved jet component, however, might have caused real case
B's to be misclassified as case A's. This potential obstacle cannot
be removed without higher spatial resolution and until the issue of
the real nature of the soft excess becomes clearer \citep[see][and
references therein]{GS03}.

%\coms{The two quasars that are not in Case B are not likely so
%because of variability between different observational epochs. In
%effect, a flux variability as high as 50\% produces only a 7 \%
%change in \aox. quasars {\it b} and {\it k} would require a flux
%increase of a factor of 31 and 3, respectively.}{sh}

As an exercise, we compared our results with those of SBG05, who
built comprehensive \seds\ of nearby AGN that extend from the IR up
to the X-rays. Their X-ray data were taken from the ROSAT  `All Sky
Survey' and public PSPC observations.
%%, which are of lower spectral resolution than \emph{Chandra} ACIS-S.
Similarly to us, SBG05 fitted a powerlaw to the far-UV segment. They
represented the ROSAT  data by a simple powerlaw.  Among the 16 AGN,
for which they had X-ray data, we determined from their Fig.\,3 that
6 are case\,A and 10 are case\,B. We do not have a simple
explanation for the higher frequency of case\,B's in the SBG05
sample. One possible explanation is that the soft X-ray excess is
not satisfactorily accounted for by a powerlaw alone. We recall that
we added a blackbody to the underlying powerlaw for our sample, in
order to fit the excess, and that we found a noticeable soft X-ray
bump in\,5 objects (\S\,\ref{sec:met}). Alternatively, since the
SBG05 sample consists of nearby (less luminous) AGN, systematic
differences in the energy distribution could exist
\citep[see][]{SK04} that would result in a larger proportion of
case\,B's in SBG05. Interestingly, we have one object in common with
the SBG05 sample, 3C351 (labeled $c$), which according to their
Fig.\,3 is case\,B, while it turns out to be case\,A in our
analysis. A thermal bump appears not to be necessary to model this
quasar, but we find evidence of a strong WA, which obviously
somewhat complicates a direct comparison with SBG05. The WA model
included in our fit is described in Appendix\,\ref{sec:rem}.

If we consider separately the RQQ or the RLQ composite \seds\ from
TZ02, they both fall in the case\,A category. Therefore, regarding
the dominance of case\,As in the dust-free case, composite \seds\ as
well as individually analyzed spectra lead to similar conclusions.
Case\,A does not entail, however, that the X-ray and far-UV
components share the same physical origin. This was conjectured in
previous works \citep[e.g. ][]{p96,laor97}, in which the soft excess
was considered to be the prolongation of the far-UV component
observed in the composite \sed. The picture that is emerging from
recent works rather suggests the opposite. For instance, using
XMM-\emph{Newton} data, \citet{bs06} and PI05 argue that the soft
X-rays are a distinct component from the BBB (see discussion in
\S\,\ref{sec:xco}). Our analysis supports this interpretation, since
a large fraction of objects (5/11) presents a differentiated thermal
bump in the soft X-rays, and none is found where the soft X-rays
{\it behave} as the prolongation of the far-UV powerlaw.
%%We now revise how these conclusions may be affected when the presence
%%of dust is invoked to explain the break or the relatively soft near-UV
%%index in comparison to that expected from accretion disks.

\section{Combining the UV and X-ray segments in the `break-corrected' case}\label{sec:cor}

\subsection{Competing explanations for the UV break}\label{sec:com}

The composite \sed\  inferred by TZ02 appears  to be too soft
to account for the broad emission lines \citep[see \S\,\ref{sec:cri}
and][]{KF97}. We have initiated an ongoing project to explore
alternative solutions to this problem, assuming that the far-UV
break is more akin to a localized continuum trough, followed by a
marked recovery in the extreme UV, which is the energy region
responsible for the high excitation emission lines. Various
mechanisms that could generate such a trough are summarized by
\citet{binettexian07}. State-of-the-art calculations of the SED for
standard geometrically-thin optically-thick accretion disks do not
reproduce the observed break satisfactorily (\S\,\ref{sec:mec}), but
such models assume a stationary disk with a vertical structure
supported only by gas and radiation pressure. To our knowledge, no
detailed SED calculations assuming a non-stationary disk or with an
accelerating wind have been carried out. Absorption by intergalactic
\HI\ or intergalactic dust have been discarded by \citet{BR03} and
B05, respectively. \citet{eastman83} proposed that \HI\ absorption
by local clouds accelerated up to 0.8$c$ could generate a steepening
of the transmitted continuum.  In order to reproduce modern data,
this model would need to be fine-tuned and extended to the extreme
UV where the flux recovery is expected. More recently, B05 have
proposed that the break could be the result of absorption by
crystalline dust grains {\em local} to the quasars.

More work is needed to falsify some of these competing explanations.
However, whatever the physical origin of the break, once we can
reasonably  reproduce its shape,  we can explore the possibility of
a universal quasar \sed, in which the large variations in observed
spectral indices are the result of absorption by standard dust
models either SMC-like or Galactic. This possibility of a universal
\sed\ is pursued in \S\,\ref{sec:sedpeind}, assuming that the break
is due to crystalline carbon dust, as explained in detail in the
current section. We will assume that a \sed\ turnover takes place,
not at the observed break, but at shorter wavelengths, around
670\,\AA, as in B05.

\subsection{Crystalline dust to model the far-UV break}\label{sec:nd}

B05 showed that absorption by a novel dust component made of
crystalline carbon (i.e. nanodiamonds) could reproduce the position
and detailed shape of the UV break in 50 objects, out of a total
sample of 61 quasars with multi-grating spectra extending
down to at least 900\,\AA\ (rest-frame). An example of such
spectrum is that of \PGelf\ ($z=0.969$) shown in Fig.\,\ref{fig:pg}.
An important prediction of B05 is that a flux rise should take place
near 700\,\AA\  and this rise was observed in \HStre, \PGten\ and
\Pks. BK07 recently presented evidence of a far-UV rise in \tontr,
the quasar with the most extreme break known. The gray dashed-line
in Fig.\,\ref{fig:pg} illustrates the far-UV rise predicted by the
absorption models of B05. We now review specific features of the
crystalline dust model.

\paragraph{Turnover in the EUV-X region.} If the UV-break is the
manifestation of dust absorption, the intrinsic continuum (once
dereddened) cannot rise indefinitely in \nufnu, as implied by a
single index powerlaw. The intrinsic \sed\ must present a turnover
before the X-ray domain.  The \aox\ index provides us with loose
constraints on such a turnover. Typically, \aox\ lies in the range
$-1.4$ to $-1.6$. The radio-loud quasars are X-ray stronger, with an
\aox\ reaching $-1.2$. This index gives us some clues about the drop
in flux between the far-UV and the soft X-rays, but not on where
exactly the actual turnover takes place.
%%assuming the intrinsic \sed\ is harder than $\anuv = -0.5$?

B05 reported evidence of a  rollover  of the continuum in the far-UV
\sed\ of the well studied high redshift quasars \HS\ and \HE\ (both
at $\zq \sim 2.8$). They parameterized the shallow turnover observed
by using the following multiplicative function: $\Cut = (1+
[{\laro}/{\lambda}]^{-f \delta})^{-1/f}$, which produces a
steepening centered on \laro.  \ad\ is the powerlaw index increase
and $f$ a form factor. The cut-off can be progressive or sharp,
depending on $f$. When the \sed\ is multiplied by this function, an
index change (\ad) takes place at wavelength \laro.  The values
inferred from the above two quasars were $f = 2.8$, $\laro=
670$\,\AA\ (18.5\,eV) and an induced steepening of magnitude
$\ad=-1.6$. By incorporating this rollover into their intrinsic
powerlaw \seds, B05 found a significant improvement in the synthetic
simulation of the TZ02 composite. Furthermore, the fit to the far-UV
rise observed in four quasars was also improved (B05, BK07). We will
test whether this functional form, which behaves as a powerlaw in
the unobserved domain, is at least consistent with the X-ray
observations. The rationale behind this form, rather than a
thermal/exponential rollover, is the same as that proposed by Z97 to
model the composite \sed, that is, comptonization of the disk
emission by a hot corona.

\paragraph{A simplified nanodiamond dust model.} B05
could reproduce the UV break of quasars, using two kinds of
crystalline carbon grains: terrestrial cubic and meteoritic. We
hereafter adopt a simplified version of that model, based on a
single albeit modified extinction curve consisting of terrestrial
cubic diamonds only. This new curve D3, shown in
Fig\,\ref{fig:exti}, differs from the D1 curve of B05 in that the
grain size distribution ($\propto a^{-3.5}$) covers a wider size
range of 3--200\,\AA\ (instead of 3--25\,\AA). The main advantage is
that a single extinction curve now suffices to model the break
without the need of meteoritic grains\footnote{Infrared re-emission
by meteoritic nanodiamonds is expected to generate emission bands at
3.43 and 3.53\,\mic\ \citep{VT02,jones04}. Using the spectrum of
3C298 as a test-case, \citet{dediego07} found no evidence of this
emission.}. This is shown in Fig.\,\ref{fig:pg}, where the
extinction by curve D3 provides a satisfactory fit (continuous line)
to the sharp break observed in \PGelf. In the same figure, the
dotted line represents the intrinsic \sed\ adopted for this quasar,
which consists of a powerlaw of index $\bnuv = 0.8$, multiplied by
the function \Cut\ defined above, which produces a shallow rollover
near 670\,\AA. The new fit assuming model D3 is quite similar to the
one obtained by B05 (dashed gray line), who used a combination of
extinction curves A1 and D1. All our dust models were computed
assuming solar metallicity for carbon (C/H$=3.6 \times 10^{-4}$) and
full depletion of carbon onto dust (i.e. \dc = 1.0). This choice is
purely for convenience, as \dc\ is not known a\,priori in AGN.
Diffuse Galactic dust is consistent with a fractional depletion of
$\dc=0.6$ \citep{bookwhittet}. Uncertainties about \dc\ will affect
comparisons between columns derived from dust absorption and those
inferred from the X-rays, as discussion in \S\,\ref{sec:ndres} and
\S\,\ref{sec:req}. The dust-to-gas ratio arising from the C dust is
$0.0031\dc$ \citep[the solar neighborhood value assuming standard
ISM extinction is $\simeq 0.009$,][]{bookwhittet}.

\paragraph{Matching the break-corrected UV segment with the  X-ray SED.}
For each quasar, the column \Ndia\ was varied, until the far-UV
break could be reproduced as closely as possible. Even though the
curve D3 presents a steep  decline redward of the cross-section peak
(Fig.\,\ref{fig:exti}), the extinction is not totally
negligible in the near-UV. To preserve a good fit, one needs to
slightly increase the index \bnuv\ with respect to the value of the
dust-free case. Towards the optical domain, however, the extinction
by nanodiamonds rapidly becomes negligible.

In Fig.\,\ref{fig:sedpeind}, the thick black line represents the
{\it intrinsic} (dereddened) \sed, separately inferred for each
quasar. The `fluxed' X-ray models are the same as before and are
represented by the green dot-dashed lines. To facilitate
comparisons, the dust-free broken powerlaws of Fig.\,\ref{fig:sed}
are repeated (red dashed lines). In Table\,\ref{tab:sed} we list the
values of the intrinsic \bnuvd\ index and other quantities related
to the break-corrected \sed.  All the inferred quantities share the
same superscript {\it `dia'}.

\subsection{Discussion\,II: the break-corrected case}\label{sec:ndres}

After fitting the UV-break with dust-absorbed \seds, we find
that only 4 out of 11 quasars are case\,A ({\it a, e, f, g}).
Therefore, for a majority of objects, the adopted rollover function
\Cut, with the same $\ad=-1.6$ as in B05, results in an extrapolated
flux that is incompatible with the X-ray data. The possibility
that the X-ray continuum is significantly contaminated by an
unresolved jet component would simply reinforce this conclusion.
Either the rollover must steepen further than assumed above, or a
different functional form for the cut-off should be considered, such
as an exponential. For each object found to be case\,B, we
determined the steepening, \dmaxd, that would revert it to case\,A.
The values are listed in Col.\,8 of Table\,\ref{tab:sed} and
correspond only to upper limits, since further reductions of \ad\
would still result in case\,A's. The mean value of \dmaxd\
determined for the 7 case\,B's is $-2.1$, which is significantly
steeper than assumed in B05.

In summary, the dominance of case\,B objects in the
break-corrected case indicates that the functional form adopted for
the rollover should be revised.  Independently of the adopted
form, however, the flux recovery in the extreme-UV results in an
ionizing photon luminosity that is higher than with the dust-free
broken powerlaw discussed in previous \S\,\ref{sec:sed},  a property
further discussed in \S\,\ref{sec:nlr}.

%%OLDER%%%This would lead to smaller equivalent widths of all the major
%emission lines since more ionizing photons reach unhampered the BELR
%than in the alternative case of the dust-free broken powerlaw. As
%discussed in \S\,\ref{sec:nlr}, this would go some distance in
%resolving the photon deficit problem exposed by \citet{KF97}.

%OLD%%%A point in favor of the nanodiamond absorption model is that the
%dust columns required to explain the break are by and large
%compatible with the absorption columns inferred from the X-rays.

%%The soft X-rays are sensitive to absorption by metals present either
%%in the Galaxy (\Ngal) or intrinsic to the quasars (\Nhx).

It is interesting to compare the  nanodiamond dust columns
\Ndia\ in Table\,\ref{tab:nh} with the gas columns, \Nhx,
\emph{intrinsic} to each quasar and inferred from the X-rays
(\S\,\ref{sec:xme}). In the X-ray domain, it makes relatively little
difference whether the metals lie in the dust or consist of free
atoms, at the spectral resolution provided by \emph{Chandra}. If the
assumed dust screen covers both the X-ray and the UV source, the
crystalline dust column \Ndia\ should be less or equal to the column
of (intrinsic) gas inferred from the X-rays. A comparison of the two
columns  show that the condition $\Nhx \ge \Ndia$ is satisfied for
the 7 objects for which \Nhx\ has been measured. It is also in
agreement with the $2\sigma$ upper limits of three other quasars
(but not with the upper limit from quasar $j$). Both column
estimates rely on opacities computed for the solar metallicity case.
Increasing or decreasing all the metals would not affect the column
ratio $\Ndia/\Nhx$. Reducing carbon depletion below the assumed
value of $\dc=1$, however, implies proportionally larger \Ndia\
columns. We find that a depletion in the range $0.1\la\dc\la0.3$
allows the condition $\Ndia/\Nhx \le 1$ to be realized for the 7
objects for which \Nhx\ has been measured.

%The photoelectric cross sections used are from Morrison...

\section{Combining the UV and X-ray segments assuming a universal
SED} \label{sec:sedpeind}

\subsection{The relevance of  testing dust-reddened SEDs}\label{sec:test}

\paragraph{Photoionization requires a harder UV continuum.}
%%Accretion disks are believed to be at the origin of the BBB.
%%This is the starting point of the exercize being carried out in this section.
In their comprehensive review of accretion disks as the ultimate
source of the AGN optical-UV continuum, \citet{koratkar99} noted
that `bare' accretion disks, which are sufficiently hot to produce a
significant fraction of hard ionizing photons, generally have an
optical/UV distribution that is too blue with respect to what is
observed. These authors pointed out that the problem of disk models,
however, does not lie in being able to fit the ``red'' optical/UV
continuum, but rather in how to explain the extreme-UV emission,
needed to power the high excitation emission lines \citep[see
\S\,\ref{sec:cri}, \S\,\ref{sec:com} and][]{KF97}. To our knowledge,
there are no bare disk models that solve both problems
simultaneously. This is the starting point for an exploration of
whether residual dust extinction might be present in almost all
luminous quasars, making their \sed\ appear softer than it really
is. This would allow the fitting of bluer accretion disk models,
which are more efficient in generating a luminous and hard ionizing
continuum.
\paragraph{SMC-type dust favored for SDSS AGN.}
An interesting result from the extensive studies of AGN from the
SDSS by \citet{hs04} and \citet{r03} is that the optical properties
of the dust  in AGN, as inferred from optical and UV colors, are
more akin to the properties found in the SMC than to the
Galaxy\footnote{The extinction curve of Galactic dust is much
shallower than that of the SMC and is characterized by a prominent
2175\,\AA\ absorption bump, absent from the SMC curve.}.
The recent simulations of \citet{w05}  supports these findings.
We adopted the conclusions of \citet{hs04} and calculated  a
SMC-type extinction curve that is based on  amorphous carbon (AC) dust grains.
%%(their parameter $r_s$ that was set at 3 times the Milky Way value).
The model considers the grains to be spherical, with a size
distribution $\propto a^{-3.5}$ within the range of $50 \le a \le
1400$\,\AA. The complex refraction indices $n+ik$ are from
\citet{rm91} for the AC type. We assumed solar metallicity for
carbon and, to be definite, full depletion of carbon onto dust (i.e.
$\dc = 1.0$, see \S\,\ref{sec:nd}). The dust-to-gas ratio arising
from AC dust is $\simeq 0.0034\dc$. We fine-tuned the size range
until our curve closely reproduced the {\it shape} of the SMC
extinction of \citet{P92}, but not its normalization, since the
grain composition and metallicity are different. The resulting curve
is shown as a continuous line in Fig.\,\ref{fig:exti}. For
comparison purposes, we also show a silicate (MgFeSiO$_{4}$) grain
model (long dashed-line) normalized to {\it solar} Si
metallicity\footnote{The dust model of \citet{P92} employed silicate
grains but required 2--3 times higher Si abundance than available in
the SMC interstellar medium. This one reason why we consider more
attractive the amorphous carbon grain composition. We chose to
normalize our dust models to solar C abundances (C/H$=3.6\times
10^{-4}$) in order to be consistent with the metal content assumed
for the opacities \citep{mm83} used in the X-ray model fitting
(\S\,\ref{sec:xme}).}, but with otherwise the same grain size limits
as the Pei model (i.e. 50 -- 2500\,\AA).
%%Note the similarity of the (scaled) silicate extinction curve with that of amorphous C.
The two curves are similar in shape.

\subsection{The hypothesis of a hard energy distribution in the UV}\label{sec:hypo}

Assuming SMC-type extinction and a universal \sed\ that is harder in
the UV than what is directly observed, we proceed to investigate the
amount of AC dust required to provide an acceptable fit of each
observed \sed. We will keep the above nanodiamond dust component, in
order to maintain a satisfactory fit to the UV break.
%\footnote{Because nanodiamond dust only affects the far-UV, the technique used by
%\citet{hs04} was unsuitable to isolate its contribution from that of SMC-type dust.}
%%We also test whether the inferred dust columns are consistent with the
%%absorption columns determined in the X-rays.
%%SBG05 showed that SMC-like extinction could not account for the sharpness of the
%%far-UV break. For that reason, it is necessary to keep the crystalline dust component
%%and  add to it the extinction due to amorphous carbon.
%%We also repeat the test carried out in \S\,\ref{sec:matA} about whether the assumed intrinsic
%%far-UV \sed\ (including the rollover) extrapolates to a flux level consistent with the X-rays.
More specifically, the working hypotheses behind the proposed test are the following:
\begin{list}{\roman{ddd}\,--}{\usecounter{ddd}}
\item{\it A pseudo universal index value for \bnuv.} In their study of radio-quiet
and radio-loud AGN, \citet{gg04} found what appears to be a low end
cut-off in the distribution of near-UV indices. The hardest indices
found corresponded to $\anuv \simeq -0.45$. This value is quite
similar to the optical-to-UV index of the SDSS composite
\citep{vandenberk01}. We define our Model\,I as a powerlaw with such
an index ($\bnuvo=0.55$), multiplied by the function \Cut, for
consistency with previous analysis. Five quasars with a measured
index already harder than $\bnuvo =0.55$ could not be included in
this test and were left out.  The \sed\ from Model\,I is plotted in
Fig.\,\ref{fig:distra} (continuous line). The dotted line represents
the same \sed\ absorbed by a crystalline dust screen with $\Ndia=0.8
\times 10^{20}\,$\cms.
%%the observed \sed\ is already harder than the adopted value.
%%For each quasar, we varied the column density of amorphous carbon
%%included until the reddened flux fit the observed (softer) spectrum.
We also defined a  Model\,II based on an even harder \sed.    B05
encountered a non-negligible fraction of quasars with a much
harder\footnote{B05 found a mean index for Class\,A quasars (see
footnote\,\ref{foot:cl}) of $\anuv=-0.44$ with a dispersion of
0.21.} continuum than the limit proposed by \citet{gg04}, some even
harder than $\bnuv=0.8$. Using the SDSS AGN sample,
\citet{vandenberk01} derived four composite \seds\ of increasing
hardness in the optical-UV. The hardest \sed\ has $\bnuv=0.75$. We
define Model\,II as a powerlaw with a similarly hard index of
$\bnuvt = 0.8$, multiplied by function \Cut. It is plotted in
Fig.\,\ref{fig:distrb} (continuous line) along with the
popular AGN \sed\ published by \citet{mf87}. The dotted line
represents Model\,II after absorption by a crystalline dust
screen of thickness $\Ndia=0.8 \times 10^{20}\,$\cms.
%%This model will obviously require more dust than Model\,I to
%%fit objects with a soft UV continuum.
\item{\it Determination of the AC dust column.} Initially, we  adopt
for each quasar the same column \Ndia\ of crystalline dust required to
reproduce the break (\S\,\ref{sec:cor}) and then determine the
additional column of SMC-type dust needed for the fixed \sed\ Model I
or II (with either $\bnuv\ = 0.55$ or $0.8$, respectively) to fit each
observed spectrum satisfactorily. For two objects ($e$ and $g$) and
only in the case of Model\,II, did we find that the fit to the far-UV
break was affected when SMC-like extinction was included. For both, a
small reduction in crystalline dust column, \Ndia, restored the
quality of the fit blueward of the break. For all the 9 other quasars,
the column \Ndia\ was frozen to the values inferred from the
break-corrected case of \S\,\ref{sec:cor}.

\end{list}

\subsection{Discussion\,III: SMC-like extinction in quasars}\label{sec:disiii}

\subsubsection{Results from dust absorbed  \seds}\label{sec:res}

An interesting result is that, for both \seds\,I or II, a
combination of AC and crystalline dust provides a fit to the UV
spectra as satisfactory  as that from a softer \sed\ with
nanodiamond dust only. This is illustrated by the three quasars {\it
c, f} and {\it g} in Fig.\,\ref{fig:mod2}. The dashed lines
represent amorphous carbon + crystalline dust extinction, assuming
the `universal' Model\,II \sed, while the dotted line represents
absorption by crystalline dust only, assuming the \sed\ index
\bnuvd\ that best fit each individual object (Col.\,6 of
Table\,\ref{tab:sed}). The thin long dashed line represents the
intrinsic Model\,II \sed\ (before extinction) for quasar $c$
(3C351). Neither dust model (with and without AC dust) results in a
perfect fit of the spectra. However, taking into account that the
true continuum level beneath the strong emission lines cannot be
uniquely defined, we consider that both models, with and without AC
dust, are equally successful. The kink around 1770\,\AA\ in the AC
extinction curve (Fig.\,\ref{fig:exti}) results in a noticeable
shoulder shortward of 1750\,\AA\ in the transmitted flux of 3C454.3.
The data are not inconsistent with such a feature. We conclude that
accretion disk \seds\ significantly bluer than observed are an
enticing possibility.
%at least for the quasars in our sample.

Since  two \seds\ of quite different hardness (I, II)  can be fitted
by simply varying the amount of AC dust, we conclude that such a
procedure does not lead to a unique description of the intrinsic
\sed. In Cols.\,6 and 9 of Table\,\ref{tab:nh}, we list the
total\footnote{These columns correspond to the sum of the
nanodiamond and AC dust columns.} absorption columns \None\ and
\Ntwo\ required for a proper fit of each quasar spectrum, assuming
either the \sed\ Model I or II, respectively. The corresponding
color excesses \ebv\ due to intrinsic dust are shown in Cols.\,7 and
10. We note that our \ebv\ values  are relatively small,
reaching at most 0.085 (object $g$), and that they are generally
smaller than the Galactic values (Col.\,3).  Nevertheless, such
small color excesses strongly affect the far-UV SED, as previously
pointed out by \citet{tbg94}, who used an heuristic
extinction curve that resembles the SMC extinction curve to
determine upper limits on the reddening of 7 quasars. The Cols.\,8
and 11 list the column ratios of the absorbing material with respect
to the one determined in the X-ray domain. In
Fig.\,\ref{fig:sedpeind}, the thick cyan and yellow lines represent
the {\it intrinsic} (dereddened) \sed\ corresponding to Models\,I
and II, respectively.
%The `fluxed' X-ray models are the same as before and
%are represented by the green dot-dashed lines.

SBG05 also explored the possibility of dust extinction by SMC-type
dust. They found that the required extinction cannot remove the UV
break without introducing a large scale curvature to the dereddened
optical-to-UV region.  They report that only  2 of their 17 AGN
showed unambiguous evidence of dust extinction. In our study, the
break is specifically accounted for by crystalline dust absorption.
When additional AC extinction is added, it does introduce some
curvature, but not to the extent that it rules out AC extinction
altogether. Presumably, the amount of AC dust required for our test
is less than the amount probed by SBG05.

%Another possibility might be that our quasars are on average less
%absorbed than the more nearby AGN studied by SBG05.

As for the quasar sample studied by TZ02, the near-UV domain between
1300 and 2000\,\AA\ is characterized by a wide range of spectral
indices among quasars.  The dispersion is as large as 0.57 about the
mean value of $\bnuv= 0.31$. It would be interesting to explore
whether variations in the extinction by amorphous carbon  might not
account for most of these variations in \sed\ softness in the TZ02
sample.

In their study of red and reddened quasars from the SDSS,
\citet{vandenberk01} derived one composite \sed\ for significantly
reddened AGN and four presumably unreddened composites that differ
by the hardness of their continuum. These composites are
characterized by optical-UV indices (\bnu) of 0.75, 0.59, 0.46 and
0.24.  The above authors did not rule out that dust extinction may
have an important role in explaining the observed variations in
\bnu\ among their four composites. However, they favor the idea that
the differences are produced by changes in the intrinsic continuum,
based on evidence of emission-line trends with color.

\subsubsection{Requirements on the rollover shape implied  by Models\,I and II}\label{sec:req}

We find that the extrapolation of the Model\,I \sed\ results in
case\,A in only 3 objects ({\it a, e, f}) out of the 6 quasars for
which we could carry out the  test.
%%For 5 objects Model\,I is ill-defined ($b, d, i, j$ and $k$.
%%The four case\,B objects that overpredict the X-ray flux are $c, d, g$ and $h$
%%(in the dust-free case, the two case\,B were $b$ and $k$).
In the case of Model\,II, however, all 11 quasars turned out to be
case\,B. The conclusion therefore is that, while amorphous carbon
extinction favors the existence of an intrinsically harder \sed, it
also points to the need of a steep \sed\ decline in the extreme-UV,
much steeper than the one proposed by B05. Alternatively, we may
consider an exponential turnover as a more appropriate solution.
Since no sharp bend is observed in the spectra of the two quasars
\HS\ \citep{fechner06b} and \HE\ \citep{zheng04} down to $\simeq
250\,$\AA, such a sharp turnover can only take place shortward of
this value. Since bare accretion disks predict a significant opacity
above the ionizing threshold of He$^+$ \citep[e.g. ][]{ha00}, we may
reasonably expect the proposed thermal turnover to occur above the
He$^+$ edge ($<228\,$\AA).

The ratio of the total dust column (crystalline + amorphous) to the
column inferred from the X-rays is given in Col.\,8 and 11 of
Table\,\ref{tab:nh}, for Model\,I and II, respectively.  In the case
of Model\,II, this ratio is smaller than unity for the 7
objects with measured absorption columns \Nhx. For the
remaining 4 quasars, we have only upper limits for \Nhx.
Note that each of them lies among the 5 objects with the
smallest accumulated photon counts in the X-rays. Only for the
object $j$ does the {\it lower limit} column ratio exceed unity (by
a factor of about two). The inferred \Ntwo/\Nhx\ ratio can
also be viewed as the minimum value of \dc\ required to equalize
both columns. We conclude that, except for the 4 objects with upper
\Nhx\ limits, the absorption columns due to dust are overall smaller
than those inferred from the X-rays, provided the fractional
depletion of C is not far from unity. This is consistent with but
does not prove necessarily the concept that the same dust+gas
component covers both continuum emission domains. Based on a
study of BAL quasars, \citet{gallagher07} have recently argued that
the gas in the line-of-sight of the (inner) X-ray source is not the
same than the one along the line-of-sight to the UV emitting
region.

%%correcting N_H^II for the presence of nanodiamonds does not alter the lack of correlation
Since the \aox\ index is sensitive to extinction of the UV-flux at
2500\,\AA, we might expect that it correlates with the dust column.
In Fig.\,\ref{fig:sin}, we plot the column \Ntwo\ versus \aox. There
is no evidence of a correlation. Possibly, the intrinsic
(dereddened) \aox\ varies significantly from object to object, which
would tend to mask such a correlation. Assuming an \aox\ intrinsic
value of $-1.6$, the continuous line illustrates the expected
behavior of \aox\ with increasing dust column. The slope is quite
steep, which indicates that relatively modest variations in \aox\
should be expected in any case.

\section{Comparison of photoionization models for the extended NLR}\label{sec:nlr}

In order to quantify the differences between Models\,I and II and
the TZ02 composite, we computed photoionization calculations for the
{\it extended} NLR (hereafter E-NLR). We favor the low-density
regime ($< 10^{3}$\,\cmc), because the physics of the emission
processes and radiation transfer are simpler and because it is
unclear to us where to position the BELR with respect to the
crystalline dust screen.
%As discussed below, these models can be compared with the E-NLR of radio-galaxies,
%which we assume belong to the same parent population as radio-loud quasars.
%as is appropriate for the resolved {\it extended} NLR (hereafter E-NLR) observed in
%radio-galaxies and Seyfert galaxies.
The context and parameters of the models are described below.

\subsection{The E-NLR of radio-galaxies as a comparison reference}\label{sec:comp}

The emission lines of the E-NLR, observed in projection on the sky
or on the parent galaxy, are well suited for comparison with simple
slab models. The reasons are twofold: \emph{i}- in the case of the
permitted lines, there is no need to perform the difficult task of
separating the NLR from the BELR component, \emph{ii}- the gas
densities of the emission regions seen in projection (outside the
nucleus) are definitely closer to the low-density regime than the
nuclear NLR. The latter is supported by the detailed analysis of
\citet{ferguson97}, who compared a wide range of line ratios from
Seyfert\,I galaxies with those predicted by Locally Optimally
emitting Cloud (LOC) models of the NLR. Their models are quite
successful in reproducing the high-excitation line strengths, the
\oiii\ temperature and the correlation between line widths and
critical density observed in the (unresolved) NLR. As for the E-NLR,
the LOC scenario predicts \sii\ densities consistent with the
low-density regime beyond $\sim 200\,$pc and is able to reproduce
the radial behavior of the \oii/\oiii\ ratio in the E-NLR of
NGC\,4151. These results indicate that \emph{spatially
resolved} emission-line components from the nuclear regions should
be operating in the low-density regime.

The orientation-dependent unification scenario proposed by
\citet{b89}, in which radio-galaxies and radio-loud quasars belong
to the same parent population, justifies our proposed comparison of
E-NLR calculations with the observations of the extended emission
gas from radio-galaxies. This unification scenario has recently
gained further support, following the mid and far-infrared
comparative studies of FR2's and QSRs by \citet{haas04,haas05}, who
used samples that encompass a similar redshift and luminosity range
for both classes.

We assume photoionization by the nuclear UV source as the main
excitation mechanism of the E-NLR. A potential problem is that shock
excitation might be contributing significantly to the emission lines
\citep{dopita96,binette85}.  How to determine the specific
contribution from shocks is still an open question. \citet{laor98}
expressed strong reservations about the overall efficiency of fast
shocks in AGN. \citet{best00} proposed a scheme to separate the
shock-excited case from the photoionized case, which is based on UV
line ratios. In their proposed line-ratio diagram, the photoionized
E-NLR of radio-galaxies occupy a region characterized by the ratios
\ciii/\cii/ (1909\AA/2326\AA) $> 3$ and \neiii/\nev\
(3869\AA/3426\AA) $< 2.4$. In order to provide an observational
dataset for the comparison of our calculations, we chose the
spectrum from the z$=$2.36 narrow-line radio-galaxy 4C$-$00.54,
which satisfies the first criterium ($\ciii/\cii/=8.1$), so as to
minimize the contribution from shocks. The line ratios are listed in
Col.\,3 of Table\,\ref{tab:rat} and the bibliography appears in
footnote of the Table. We emphasize that no attempt has been made
to vary any of the model parameters in order to fit these ratios.
Our main concern, above all, is to compare different \seds\ with one
another.

\subsection{Which SED to use: absorbed or unabsorbed?}\label{sec:abnd}

Since all quasars of redshifts $\ga 1$ found in the TZ02 sample show
a far-UV break, and since the nanodiamond dust columns required to
fit the break according to B05 cover a narrow range\footnote{The
mean dust column derived by B05 for the dominant Class\,A quasars
was $\NNa=1.05$, with a dispersion of 0.29.}, we infer that the
covering factor of the dust is quite uniform and near unity, that
is, all line-of-sights are absorbed by similar amounts of dust.
Hence, the \sed\ to which the NLR of quasars is exposed must display
a UV-break as well. For this reason, we will consider energy
distributions from Models\,I and II that have already been absorbed
by nanodiamonds (rather than the dereddened form). Regarding the AC
dust postulated in \S\,\ref{sec:sedpeind}, since the observed UV
spectral index varies considerably from quasar to quasar (TZ02), we
infer that this hypothetical dust component is not uniform and
presumably located further away from the ionizing source than the
nanodiamond dust. In this case, it is reasonable to expect that the
NLR emission will be dominated by gas photoionized by radiation
emerging from the line-of-sights that are least absorbed by the AC
dust. For this reason, our E-NLR calculations will not consider
\seds\ absorbed by AC dust. The adopted energy distributions
corresponding to Models\,I and II are plotted as dotted lines in
Figs\,\ref{fig:distra} and \ref{fig:distrb}, respectively, while the
\sed\ that approximates the TZ02 composite is represented by the
thick short-dashed line in Fig.\,\ref{fig:distra}. Both the TZ02 and
the Model\,I \seds\ include an X-ray component consisting of a
powerlaw of index $\bx=0$, assuming an \aox\ of $-1.5$.

\subsection{Density stratified slabs}

The results of the modeling of the NLR in Seyfert\,I galaxies by
\citet{ferguson97} suggest that the partially ionized emission zones
contain significant amounts of internal dust, while the fully
ionized regions contain very little. When the ionization
parameter\footnote{The ionization parameter at the face of the slab
is defined as $\upz = \phih/c \nhz$, where $\phih/c$ is the density
of ionizing photons impinging on the slab and \nhz\ the total H
density at the face of the slab.} at the face of the slab, \upz, is
high, radiation pressure becomes significant, and even small amounts
of dust will lead to an internal density stratification. We
therefore allowed our slab model to contain small amounts of
amorphous carbon dust, at a level corresponding to 2\% only
(i.e. $\dc=0.02$),
%\footnote{Hence, the cross-section of AC dust of
%Fig.\,\ref{fig:exti} imported in \map\ was multiplied by 0.02.}
so as to remain consistent with the \citet{ferguson97} results.
\citet{dopita02,groves04a,groves04b} showed that for high \upz\
values, the ensuing density stratification result in an emission
line spectrum that is relatively insensitive to the particular
adopted value. In this case, \upz\ ceases to be a determinant
parameter. This property of stratified models is particularly useful
in the context where we are foremost concerned with the comparison
of different \seds. To be definite, we chose $\upz=0.2$ and a slab
face density of $\nhz=100$\,\cmc. Finally, we adopt the set of solar
abundances from \citet{ag89}

\subsection{Comparison of the three SEDs}

To compute line ratios, we have used the code \map\
\citep{binette85, ferruit97}  and assumed a slab geometry
illuminated on one side. For each ionizing \sed\ that we considered,
we calculated the local equilibrium ionization state of the gas and
integrated the ionization structure inward until less than 1\% of
the gas becomes ionized. We integrated the volume force exerted by
radiation pressure within the slab, assuming hydrostatic equilibrium
as in \citet{dopita02} and \citet{binette97}. Cols\,5 and 6 of
Table\,\ref{tab:rat} list the calculated line ratios with respect to
\hb\ from \seds\ corresponding to Model\,I and II, absorbed by
nanodiamond dust, while Col.\,4 represents calculations with the
TZ02 composite.

Table\,\ref{tab:rat} shows that the high excitation lines are of
similar strengths in the case of Model\,I and of the TZ02 composite.
This is due to the similar hardness of both \seds, as reflected by
the quantity $\phih/\phihet$, which is the ratio of ionizing photons
above 1\,ryd to those above 4\,ryd (see values at bottom of table).
The \sed\ from Model\,II is significantly harder, however, which is
reflected in the calculated ratios.

%Overall, the observations show stronger high excitation lines and a
%hotter \oiii\ temperature than predicted. This is not new and has
%previously been studied by \citet{binette96}. The solution may lie
%in having matter-bounded clouds, as proposed by
%\citet{binette96,wilson97}. The Locally Optimally Emitting cloud
%model of \citet{ferguson97} is another alternative.

For an equal continuum flux  near 1100\,\AA\ (the onset of the
break), the ionizing photon flux, \phih, for Model\,I and II, is
23\% and 52\% higher, respectively, than that from the TZ02
composite. These gains are quite modest, however. If, on the other
hand, the observed softer index of the composite were due to AC dust
absorption (\S\,\ref{sec:sedpeind}), the total gain in \phih\ would
be much higher, reaching values of 62\% and 270\%, respectively
(including previous factors). Increasing the turnover energy beyond
54\,eV, as suggested in \S\,\ref{sec:req}, would result in an
intrinsically harder \sed, which would increase the luminosity of
the high excitation lines such as \heiiuw, \oviw\ and \civw, and
further contribute to resolve the softness problem reported by
\citet{KF97}. By how much the turnover energy should increase would
depend on the precise shape adopted for the turnover, a question
that lies beyond the scope of the current study.

\section{Conclusions}\label{sec:con}

Using archived data from \emph{Chandra} and HST-FOS, we have derived
the UV to X-ray \seds\ of 11 quasars. The aim of our analysis has
been to constrain the behavior of the \sed\ within the domain that
is not directly observable, in the extreme UV. We explored the
possibility that crystalline dust and amorphous carbon dust may
account for the surprising softness of quasar \seds. More
specifically, we have reached the following conclusions:
\begin{list}{\arabic{aaa}-}{\usecounter{aaa}}

\item When we fit the observed UV spectra with a broken
powerlaw directly (dust-free case), we do not find a single object
for which the soft X-rays is the prolongation of the far-UV powerlaw
with the same index. In 9 out of 11 quasars, the extrapolated far-UV
powerlaw lies at a flux level below or near the one actually
observed in the soft X-rays. We find evidence of soft X-ray excess
in 5 objects\footnote{These are Pks\,1127$-$14, B2\,0827+24,
Pks\,1354+19, 3C454.3 and Pks\,1136$-$13.}.

\item We simplify the nanodiamond dust absorption model proposed by B05 to reproduce the far-UV
break. Instead of using two kinds of crystalline carbon grains
(terrestrial cubic and meteoritic), we  obtain a similar fit by
using a dust model consisting of cubic diamonds only, but where the
grain sizes cover a larger range (3--200\,\AA). If the fractional
depletion \dc\ within the gas is $\ga 0.1$, the dust columns
required by {\it nanodiamond dust} models are not excessive, as they
fall below the cold gas columns inferred from the X-ray data.

\item Assuming that the far-UV break is due to nanodiamond dust absorption,
we explored the possibility of a shallow far-UV turnover at 18.5\,eV
to connect the far-UV with the X-rays, as proposed originally by
B05. We find that for most objects the steepening must be more
pronounced than proposed by B05, that is, an index change as steep
as $<-2$. Given that two well studied quasars, \HS\ and \HE, do not
show any sharp break down to 250\,\AA, we also consider that the
proposed rollover should take place at a significantly higher energy
($>50$\,eV) than proposed by B05.

\item We find that adding SMC-type extinction to the extinction
from nanodiamond dust leads to a bluer \sed\ and to a fit of the BBB
of quality comparable to the one achieved with nanodiamond dust
alone. This opens the possibility that the {\it intrinsic} quasar
\sed\ is much harder and luminous in the near and far-UV than
previously considered.

%\item We also find a qualified agreement between the absorption columns
%inferred from the X-rays and from AC dust absorption. Even in the
%case of Model\,II, which assumes an intrinsic powerlaw as hard as
%\bnuvt=0.8, we find that the dust column that is required to fit the
%UV spectrum is generally inferior to the column of cold gas
%determined from the X-rays.

\end{list}

\acknowledgements SARHC was supported by DGEP and CONACyT throughout
his PhD project. This work was also funded by the CONACyT grant
J-49594 and J-50296 and the UNAM PAPIIT grant IN118601. We thank
Chris Done for many constructive comments. We acknowledge the
technical support of Alfredo D\'{\i}az Azuara for configuring the Linux
workstations Deneb, Bishop and Oceania. We thank Diethild Starkmeth
who helped us with proofreading.  This research has made use of the
NASA/IPAC Extragalactic Database (NED) and the SIMBAD database,
which are operated by JPL at CALTECH (under contract with NASA) and
by CDS (in Strasbourg, France), respectively. SARHC thanks the staff
of the Chandra X-Ray Center HelpDesk System, operated by the
Smithsonian Astrophysical Observatory and the Center for
Astrophysics.

\appendix

%\section{c.- 3C351: treatment of the  warm absorber}\label{sec:apA}

\section{Remarks on individual objects}\label{sec:rem}

Additional information and special considerations about each quasar
are described as follow:

{\bf a.- PKS1127-14}. The X-ray observation of this quasar does not
show pile-up. \citet{sb02} detected for the first time in the X-ray
image of this object the presence of jet emission. For the main source
they found a photon index \gx=$1.19\pm0.02$ in the spectral region between 2 and 10\,keV,
consistent with the best-fit value reported in this work within a level of
2\si\ confidence.

{\bf b.- PKS0405-123}. \citep{SF93,BJ93} found the presence of a
\Lya\ absorption line in the UV spectrum of this object. They
associated this absorber with the  rich cluster of galaxies where
the source is located. After the Lyman limit ($\nu >
3.310^{15}$\,Hz) there is an unusual excess with respect to a simple
powerlaw model. \citet{LK95} found that Comptonization of the accretion
disk cannot account for this excess. In our analysis of this source
we created an additional mask between 717$-$882\,\AA\ to avoid the
segment covered by this emission (see \ref{sec:uvme}). We  note,
however, that this does not produce a considerable change in the FUV
index.\\ The X-ray of this source shows jet emission  (removed from
our analysis). The source presented a level of pile-up of 16.7\,\%.
We corrected the spectrum with the procedure described in
\S\,\ref{sec:xda}.

\begin{figure}
\epsscale{1.0} \plotone{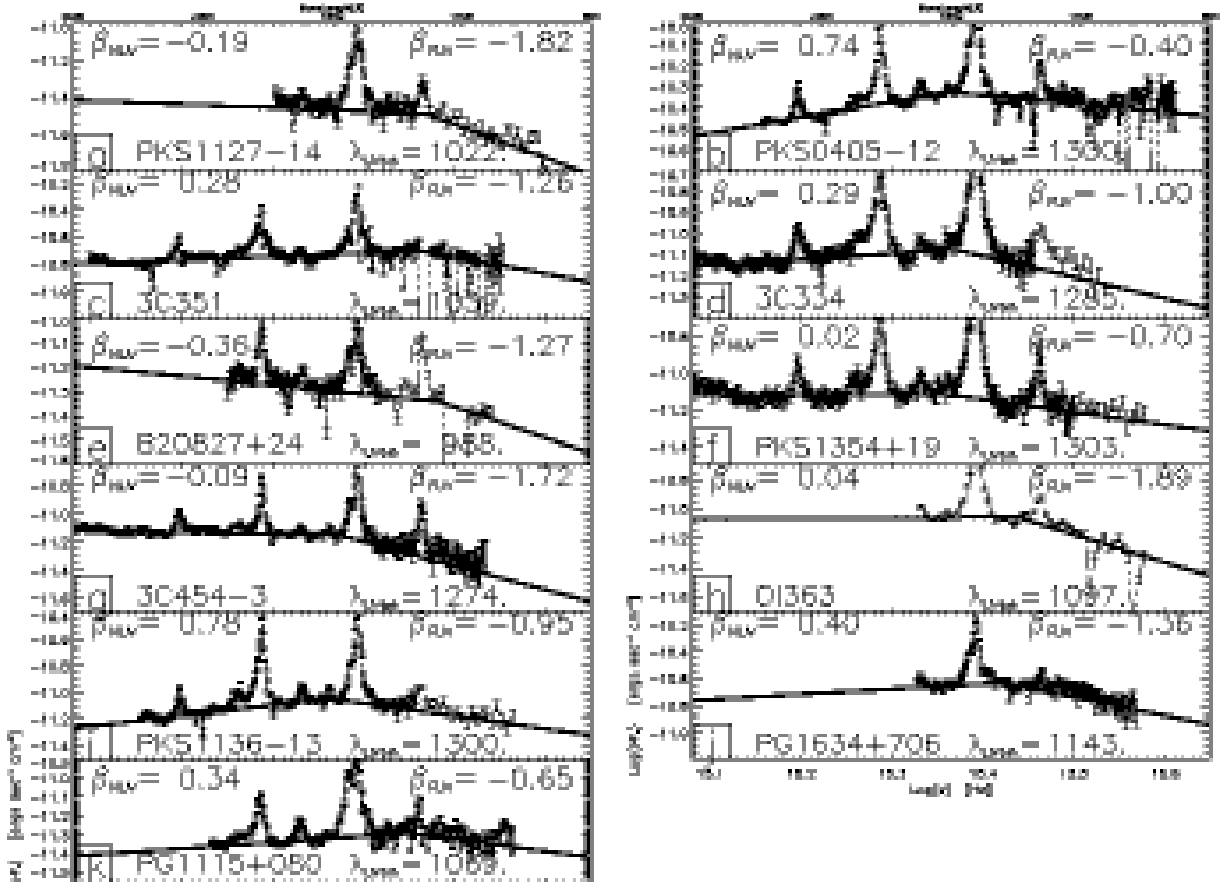} \caption{Rest-frame ultraviolet
HST-FOS spectra of quasars $a$--$k$ in \nufnu\ as a function of
frequency (bottom axis) or wavelength (top axis). Error bars are
superimposed. The data have been rebinned to group 10 points per
resolution element together. Separate powerlaw fits to the near-UV
and far-UV regions are overlaid to the spectra (continuous lines).
The two spectral indices \bnuvd\ and \bfuvd\ and the position of the
UV break (\AA) are shown (see also Table \ref{tab:uv}). }
\label{fig:uv}
\end{figure}

{\bf c.- 3C 351 }. This quasar
%(also called SBS 1704+608)
shows the presence of \civ\ and \Lya\ absorption lines (see
Fig.\,\ref{fig:uv} $c$) with 1700\,\kms\ and 2200\,\kms\ blueshift,
respectively \citep{bb93}. \citet{mathur943c351} and \citet{n99}
find, analyzing ROSAT data, the presence of an associated warm
absorber in the X-ray band. This object provided the first
suggestion for a common UV and X-ray absorber.
%, they found evidence of WA and flux variation (factor of
%1.7) between $\sim 2$ years observations.
Using a {\em Chandra} observation, \citet{hb02} fit the spectrum of
this object with two power laws, attenuated by Galactic and
intrinsic cold absorption. \citet{hb02} rejects a model consisting
of ionized absorption that included only bound-free transitions (the
model {\em absori} in XSPEC). One of the power laws in their model is
associated with the nuclear quasar emission, while the other is
assumed to be related to the jet emission.
%The \Nh\ value reported by this authors is
%consistent of absorbing material)
%reported previously. They assumed use two-power-law model. The fit was
%carry out between 0.4 to 7\,keV (observer-frame).
\\We analyzed the X-ray spectrum of this object using the approach
described in \S\,\ref{sec:xda}. The X-ray observation has $\sim
11$\% of pile-up. Three X-ray prominent hot spots at 25 arcsec
north-east from the main source were also detected. Those were
excluded from our analysis to avoid contamination while getting the
X--ray spectrum.
\\The spectrum of this object, showed the presence
of two data points with deviations larger than 2\si\ near 2.5 keV.
We did not included these points from our initial fit between
2.5--6\,keV as they would change the measured value of \bx.
Therefore, for this object, we fit only the range 2.51--6\,keV. In this
initial fit (that consists of a single powerlaw attenuated by
galactic absorption) we found a spectral index \bx=0.6, consistent
(within 2\si) with the index reported by \citet{hb02}. The
extrapolation of this fit to the whole spectral range (0.3 to 6 keV)
showed that the spectrum is not well described with a single
powerlaw. Strong negative residuals between $0.5<$E $<2$ keV are
present, indicating the possible presence of ionized absorption, as
previously suggested. The residuals further revealed the presence of
an emission line consistent with the \FEKa\ line. We thus fitted the
spectrum with a powerlaw  plus a gaussian (to account for the \FEKa\
line) attenuated by both the WA and Galactic absorption. The WA was
modeled using the code PHASE \citep{KN03} that includes absorption
by both bound-bound and bound-free transitions. A single warm
absorber could not account for the negative residuals, thus the warm
absorber was modeled with two absorption components. This is
consistent with the findings for Seyfert galaxies
\citep[e.g.][]{KN05-1,KN05-2}. The best fit parameters for our fit
are reported in Table\,\ref{tab:x}. These model (d.o.f./$\chi^2 =
224/214$) showed to be better than a model consisting in two power
laws with intrinsic cold absorption (d.o.f./$\chi^2 = 227/255$).
\\
\citet{hb02} reported the possible presence of an emission feature
near 3.2 keV. We also notice the presence of this feature, and like
\citet{hb02} could not constrain its properties with a gaussian
line. As our main interest is to constrain the continuum properties,
no further effort was made in explaining this feature [see
discussion by \citet{hb02}].
\\The above model for this object was carried out in the range between
0.32 to 6\,keV. The region between 0.24 and 0.32\,keV showed several
data points above of the power-law level that could not be fit by
any model with a reasonable physical meaning. Furthermore, most of
these points are located in the spectral region below 0.3\,keV,
where the spectral calibration for {\em Chandra} is not accurate. 
%and we cannot say if they are emission of the quasar. 
Therefore these points were excluded from our analysis.

{\bf d.- 3C334}. This quasar
%(also called 1618+177)
does not present pileup or jet emission in the X-ray domain.

{\bf e.- B2\,0827+24}.  The predefined mask in the UV domain
(\S\,\ref{sec:uvme}) was modified for this quasar. The spectral region
between \Lya\ and \oviw, i.e., between 900--980\,\AA\ was included
to obtain a reliable FUV fit. It was also necessary to ignore the
following regions: 926.61--974.4, 896.5 897.5, 1090 to 1110, 1140.5
to 1142, 1310 to 1316 and 1440 to 1446\,\AA\ due to the presence of
absorption lines. Without these modifications to the mask, a harder
FUV index (\bfuv$=-1.01$) would be obtained in the fits (we report
an index of \bfuv$=-1.27$ in Table\,\ref{tab:uv}). The X-ray
observations of this object show the presence of a jet (removed from
our analysis), and no pile-up.

{\bf f.- Pks\,1354+19}. Also called 4C 19.44, this source did not
presented  pile-up in the X-ray spectrum. Jet emission was removed
to avoid any source of contamination. \citet{GS03} analyzed the same
observation and found variability by a decreasing factor of 3 during
20 years (elapsed between the Chandra and the ROSAT observations). 
They reported an \aox$= -1.38$ and
photon indices of +1.39 and +1.88 for the hard and the soft
spectrum, respectively. We found values consistent with those within
a 2\si\ level of confidence.

{\bf g.- 3C454.3}
%this quasar is a OVV however it does not have variability in spectral
%shape \citep{KM04}.
An additional mask was introduced in the UV spectrum of this quasar
%(also called 2251+158)
between 877--885\,\AA, to remove the absorption features present at
a level exceeding 2\si, and arising from the intergalactic medium.
The X-ray spectrum presented 24\% of pile-up, in spite the object
was observed in use of the sub-array 1/4. The object further shows
X-ray jet emission.

{\bf h.- OI\,363}
%this quasar is called 0738+31.
An additional mask was  introduced in the range between
801--824\,\AA\ of the UV spectrum to remove absorption features that
were present at a level exceeding 2\si, and arising from the
intergalactic medium. In the X-ray region, \citet{ss03} reported  no
pileup. However, with the pileup definition of this work, we found
16.3\% of pileup on the same observation. The source presented
evidence of jet emission.

{\bf i.- Pks\,1136$-$13} The UV spectrum presented emission lines by
\heii\ and \oiiiu\ in the range between 1640--1664 \AA. These were
removed from the analysis in order to obtain a reliable fit of the
continuum. In the X-ray band, the source does not show pile-up, but
jet emission was found.

{\bf j.- PG\,1634+706}
%This source has UV variability, furthermore it presents important
%emission of $Ly\alpha$ and CIV (ref).
The X-ray observation does not present pile-up or jet emission.
\citet{PR04, pj05} analyzed an XMM spectrum of this source between
0.2 and 10\,keV and found a photon index $\sim 2.19$ consistent with
the one reported by us within a level 2\si\ of confidence.
\citet{PR04} further reported the presence of a soft excess, not
required by the {\em Chandra} data analyzed here.

\begin{figure}
\epsscale{.80} \plotone{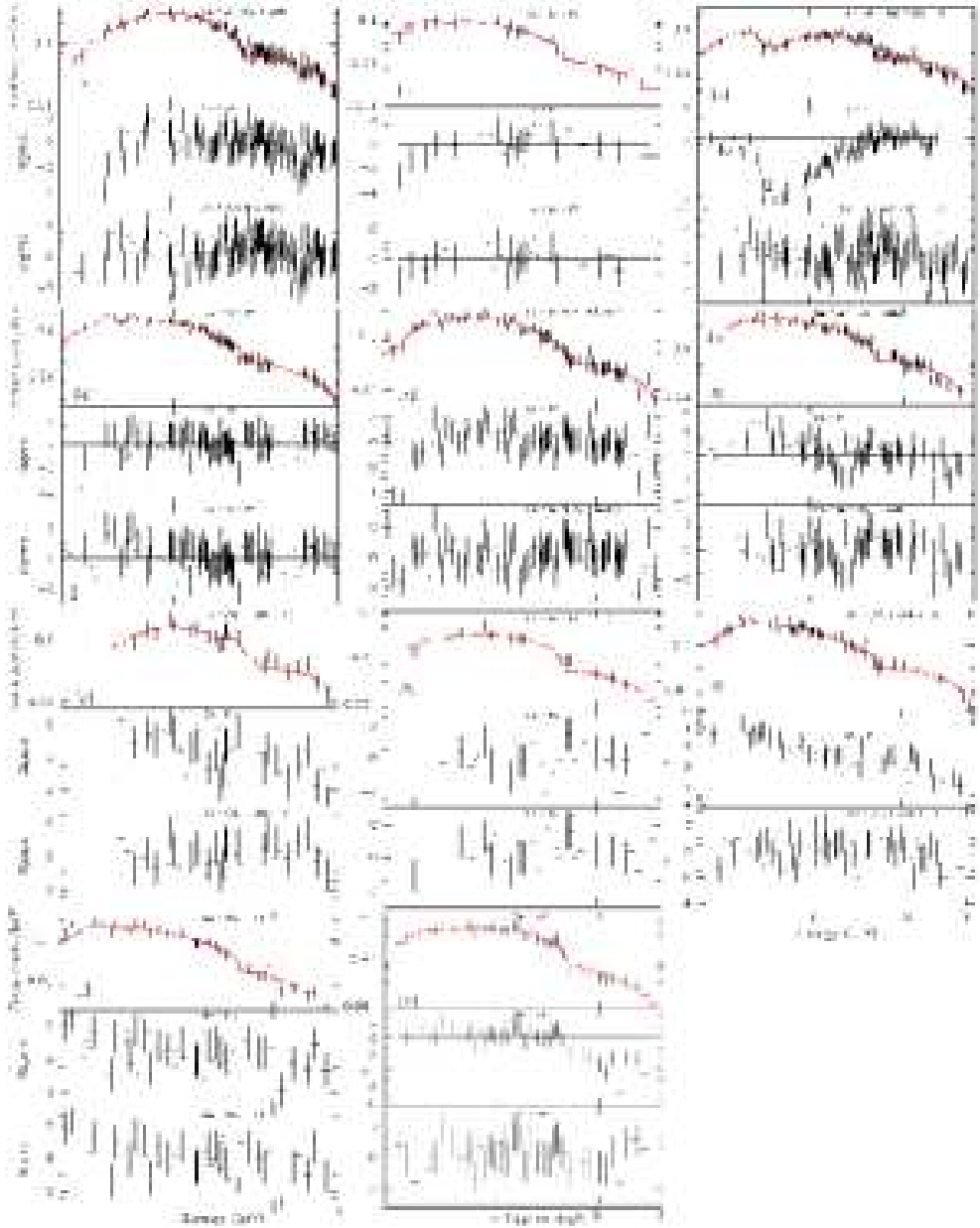} \caption{Observer-frame X-ray
spectra of the previous 11 quasars. The data are in counts/sec/keV.
Top subpanel: final best-fit model superimposed to the observed
count rate, middle subpanel: residuals from the initial model to the
2.5--6\,keV region consisting in a powerlaw attenuated by Galactic
absorption; bottom subpanel: residuals from the multi-component
best-fit model. This Figure is  available in color in the electronic
version of this paper.} \label{fig:x}
\end{figure}

{\bf k.- PG\,1115+080} The UV spectrum of this source presents broad
absorption lines. We then modify the UV mask to ignore these
features in the spectral regions  886 to 892, 1097 to 1105, 1138 to
114 and 1147 to 1149\,\AA. In the case of considering those spectral
ranges a flatter FUV index would be obtained (\bfuvf$=-0.75$,
compared to the one reported in Table \ref{tab:uv} \bfuvf$=-0.65$).
In the X-ray region, the observation did not show pile-up or jet
emission. We note that in the X-ray band this quasar has two close
sources [sources B and C in Fig.\,4 of \citet{young81}] due to
gravitational lensing. These two sources were not present in the
slit aperture of the HST-FOS observation. To be consistent with the
UV data, sources B and C were removed from our analysis.
%We note, however, that this quasar has two close sources
%due to a gravitational lens effect \citep{young81}. These were
%removed from our analysis.
\\This is the only quasar in our sample
that required a broken powerlaw to fit the data. We first fit a
single powerlaw between 2.5 to 6\,keV (rest-frame) absorbed by
Galactic neutral gas, and obtain a spectral index \bx=0.86. However,
the extrapolation of this model above 2 keV resulted in a poor fit
(d.o.f./$\chi^2 = 72/146$). The best continuum model resulted from a
fit with a broken powerlaw attenuated by Galactic absorption
(d.o.f./$\chi^2 = 70/49$). The best fit parameters for this model
are: (1) an index \bx=0.86 in the $0.3<$E$<1.69$\,keV range (the
spectral index was not changed from the previous value); (2) an
index $\bx_{hard}=-0.05^{-0.22}_{+0.25}$ in the $1.69<$E$<6$\,keV
range; and (3) an energy-break at E=1.69 keV.
%exponential cutoff power law reflected from neutral matter \citep{MZ95}
%We found the e-folding energy foldE=8.51 keV, photon flux at 1 keV
%RD.norm=7.18e-5,
%reflection scaling factor RD.relRefl=2.51. We freeze the parameters
%cosine  of inclination angle 0.8, the redshift of the source, both
%abundance  (heavier than He and Fe) with value 1.
\\PI05 analyzed an  XMM-Newton observation of this source. They used
a single powerlaw and obtained a spectral index \bx = 0.15
(\gx=1.85). The difference in the models arises from that we first
performed the analysis in the soft X-ray region between 0.92 and 2.2
keV, and then extrapolated our model to the whole range of
0.3--6\,keV (see \S\,\ref{sec:xme}). This is clear from middle panel
of Fig.\,\ref{fig:x}\,{\it k} where the residuals show a break
around 2\,keV.

%%%%%% Figure 2 here %%%%%%%%
%%%%%%%%%%%%%%%%%%%%%%%%%%%%%%%%%%%%%%%%%%%%%%%%%%%%%%%%%%%%%%%%%%%%%%%%%
%% Note to editor: Please, figure 2 should span TWO columns!!
%%%%%%%%%%%%%%%%%%%%%%%%%%%%%%%%%%%%%%%%%%%%%%%%%%%%%%%%%%%%%%%%%%%%%%%%%

%%%%%%%%%%%%%%%%%%%%%%%%%%%%%%%%%%%%%%%%%%%%%%%%%%%%%%%%%%%%%%%%
%%%%%%%%%%%%%%%%%%%%%%%%%%%%%%%%%%%%%%%%%%%%%%%%%%%%%%%%%%%%%%

\bibliography{sed}

\begin{thebibliography}{103}
\expandafter\ifx\csname natexlab\endcsname\relax\def\natexlab#1{#1}\fi

\bibitem[{{Anders} \& {Grevesse}(1989)}]{ag89}
{Anders}, E., \& {Grevesse}, N. 1989, \gca, 53, 197

\bibitem[{{Anderson} \& {Margon}(1987)}]{AM87}
{Anderson}, S.~F., \& {Margon}, B. 1987, \apj, 314, 111

\bibitem[{{Avni} {et~al.}(1995){Avni}, {Worrall}, \& {Morgan}}]{AW95}
{Avni}, Y., {Worrall}, D.~M., \& {Morgan}, W.~A. 1995, \apj, 454, 673

\bibitem[{{Bahcall} {et~al.}(1993{\natexlab{a}}){Bahcall}, {Bergeron},
  {Boksenberg}, {Hartig}, {Jannuzi}, {Kirhakos}, {Sargent}, {Savage},
  {Schneider}, {Turnshek}, {Weymann}, \& {Wolfe}}]{bb93}
{Bahcall}, J.~N. {et~al.} 1993{\natexlab{a}}, \apjs, 87, 1

\bibitem[{{Bahcall} {et~al.}(1993{\natexlab{b}}){Bahcall}, {Jannuzi},
  {Schneider}, \& {Hartig}}]{BJ93}
{Bahcall}, J.~N., {Jannuzi}, B.~T., {Schneider}, D.~P., \& {Hartig}, G.~F.
  1993{\natexlab{b}}, \apj, 405, 491

\bibitem[{{Baldwin} {et~al.}(1995){Baldwin}, {Ferland}, {Korista}, \&
  {Verner}}]{baldwin95}
{Baldwin}, J., {Ferland}, G., {Korista}, K., \& {Verner}, D. 1995, \apjl, 455,
  L119+, astro-ph/9510080

\bibitem[{{Barthel}(1989)}]{b89}
{Barthel}, P.~D. 1989, \apj, 336, 606

\bibitem[{{Best} {et~al.}(2000){Best}, {R{\"o}ttgering}, \& {Longair}}]{best00}
{Best}, P.~N., {R{\"o}ttgering}, H.~J.~A., \& {Longair}, M.~S. 2000, \mnras,
  311, 23, astro-ph/9908211

\bibitem[{{Binette} {et~al.}(1985){Binette}, {Dopita}, \& {Tuohy}}]{binette85}
{Binette}, L., {Dopita}, M.~A., \& {Tuohy}, I.~R. 1985, \apj, 297, 476

\bibitem[{{Binette} {et~al.}(2007){Binette}, {Haro-Corzo}, {Krongold}, \&
  {Andersen}}]{binettexian07}
{Binette}, L., {Haro-Corzo}, S., {Krongold}, Y., \& {Andersen}, A. 2007, ArXiv
  Astrophysics e-prints, astro-ph/0611011, the Central Engine of Active
  Galactic Nuclei", ed. L. C. Ho and J.-M. Wang (San Francisco: ASP). In press.

\bibitem[{{Binette} \& {Krongold}(2007)}]{binette07}
{Binette}, L., \& {Krongold}, Y. 2007, \apj, submitted

\bibitem[{{Binette} {et~al.}(2005){Binette}, {Magris C.}, {Krongold},
  {Morisset}, {Haro-Corzo}, {de\,Diego}, {Mutschke}, \& {Andersen}}]{a2}
{Binette}, L., {Magris C.}, G., {Krongold}, Y., {Morisset}, C., {Haro-Corzo},
  S., {de\,Diego}, J.~A., {Mutschke}, H., \& {Andersen}, A.~C. 2005, \apj, 631,
  661, astro-ph/0505587, (B05)

\bibitem[{{Binette} {et~al.}(2003){Binette},
  {Rodr{\'{\i}}guez-Mart{\'{\i}}nez}, {Haro-Corzo}, \& {Ballinas}}]{BR03}
{Binette}, L., {Rodr{\'{\i}}guez-Mart{\'{\i}}nez}, M., {Haro-Corzo}, S., \&
  {Ballinas}, I. 2003, \apj, 590, 58

\bibitem[{{Binette} {et~al.}(2006){Binette}, {Wilman}, {Villar-Mart{\'{\i}}n},
  {Fosbury}, {Jarvis}, \& {R{\"o}ttgering}}]{binette06}
{Binette}, L., {Wilman}, R.~J., {Villar-Mart{\'{\i}}n}, M., {Fosbury},
  R.~A.~E., {Jarvis}, M.~J., \& {R{\"o}ttgering}, H.~J.~A. 2006, \aap, 459, 31,
  astro-ph/0607289

\bibitem[{{Binette} {et~al.}(1997){Binette}, {Wilson}, {Raga}, \&
  {Storchi-Bergmann}}]{binette97}
{Binette}, L., {Wilson}, A.~S., {Raga}, A., \& {Storchi-Bergmann}, T. 1997,
  \aap, 327, 909

\bibitem[{{Brocksopp} {et~al.}(2006){Brocksopp}, {Starling}, {Schady}, {Mason},
  {Romero-Colmenero}, \& {Puchnarewicz}}]{bs06}
{Brocksopp}, C., {Starling}, R.~L.~C., {Schady}, P., {Mason}, K.~O.,
  {Romero-Colmenero}, E., \& {Puchnarewicz}, E.~M. 2006, \mnras, 366, 953

\bibitem[{{Casebeer} {et~al.}(2006){Casebeer}, {Leighly}, \& {Baron}}]{CL06}
{Casebeer}, D.~A., {Leighly}, K.~M., \& {Baron}, E. 2006, \apj, 637, 157

\bibitem[{{Collinge} {et~al.}(2001){Collinge}, {Brandt}, {Kaspi}, {Crenshaw},
  {Elvis}, {Kraemer}, {Reynolds}, {Sambruna}, \& {Wills}}]{collinge01}
{Collinge}, M.~J. {et~al.} 2001, \apj, 557, 2, astro-ph/0104125

\bibitem[{{Comastri} {et~al.}(1992){Comastri}, {Setti}, {Zamorani}, {Elvis},
  {Wilkes}, {McDowell}, \& {Giommi}}]{comastri92}
{Comastri}, A., {Setti}, G., {Zamorani}, G., {Elvis}, M., {Wilkes}, B.~J.,
  {McDowell}, J.~C., \& {Giommi}, P. 1992, \apj, 384, 62

\bibitem[{{Cristiani} {et~al.}(1997){Cristiani}, {Trentini}, {La Franca}, \&
  {Andreani}}]{cristiani97}
{Cristiani}, S., {Trentini}, S., {La Franca}, F., \& {Andreani}, P. 1997, \aap,
  321, 123, astro-ph/9610108

\bibitem[{{de\,Diego} {et~al.}(2007){de\,Diego}, {Binette}, {Ogle}, {Andersen},
  {Haro-Corzo}, \& {Wold}}]{dediego07}
{de\,Diego}, J.~A., {Binette}, L., {Ogle}, P., {Andersen}, A.~C., {Haro-Corzo},
  S., \& {Wold}, M. 2007, \aap, submitted

\bibitem[{{Dickey} \& {Lockman}(1990)}]{DL90}
{Dickey}, J.~M., \& {Lockman}, F.~J. 1990, \araa, 28, 215

\bibitem[{{Dopita} {et~al.}(2002){Dopita}, {Groves}, {Sutherland}, {Binette},
  \& {Cecil}}]{dopita02}
{Dopita}, M.~A., {Groves}, B.~A., {Sutherland}, R.~S., {Binette}, L., \&
  {Cecil}, G. 2002, \apj, 572, 753, astro-ph/0203360

\bibitem[{{Dopita} \& {Sutherland}(1996)}]{dopita96}
{Dopita}, M.~A., \& {Sutherland}, R.~S. 1996, \apjs, 102, 161

\bibitem[{{Eastman} {et~al.}(1983){Eastman}, {MacAlpine}, \&
  {Richstone}}]{eastman83}
{Eastman}, R.~G., {MacAlpine}, G.~M., \& {Richstone}, D.~O. 1983, \apj, 275, 53

\bibitem[{{Edelson} \& {Malkan}(1986)}]{edelson86}
{Edelson}, R.~A., \& {Malkan}, M.~A. 1986, \apj, 308, 59

\bibitem[{{Elvis} {et~al.}(2002){Elvis}, {Marengo}, \& {Karovska}}]{elvis02}
{Elvis}, M., {Marengo}, M., \& {Karovska}, M. 2002, \apjl, 567, L107,
  astro-ph/0202002

\bibitem[{{Fechner} {et~al.}(2006){Fechner}, {Reimers}, {Kriss}, {Baade},
  {Blair}, {Giroux}, {Green}, {Moos}, {Morton}, {Scott}, {Shull}, {Simcoe},
  {Songaila}, \& {Zheng}}]{fechner06b}
{Fechner}, C. {et~al.} 2006, \aap, 455, 91, astro-ph/0605150

\bibitem[{{Ferguson} {et~al.}(1997){Ferguson}, {Korista}, {Baldwin}, \&
  {Ferland}}]{ferguson97}
{Ferguson}, J.~W., {Korista}, K.~T., {Baldwin}, J.~A., \& {Ferland}, G.~J.
  1997, \apj, 487, 122, astro-ph/9705083

\bibitem[{{Ferrarese} \& {Merritt}(2000)}]{ferrarese00}
{Ferrarese}, L., \& {Merritt}, D. 2000, \apjl, 539, L9, astro-ph/0006053

\bibitem[{{Ferruit} {et~al.}(1997){Ferruit}, {Binette}, {Sutherland}, \&
  {Pecontal}}]{ferruit97}
{Ferruit}, P., {Binette}, L., {Sutherland}, R.~S., \& {Pecontal}, E. 1997,
  \aap, 322, 73

\bibitem[{{Francis} {et~al.}(1991){Francis}, {Hewett}, {Foltz}, {Chaffee},
  {Weymann}, \& {Morris}}]{fh91}
{Francis}, P.~J., {Hewett}, P.~C., {Foltz}, C.~B., {Chaffee}, F.~H., {Weymann},
  R.~J., \& {Morris}, S.~L. 1991, \apj, 373, 465

\bibitem[{{Gallagher} \& {Everett}(2007)}]{gallagher07}
{Gallagher}, S.~C., \& {Everett}, J.~E. 2007, ArXiv Astrophysics e-prints,
  astro-ph/0701076

\bibitem[{{Gambill} {et~al.}(2003){Gambill}, {Sambruna}, {Chartas}, {Cheung},
  {Maraschi}, {Tavecchio}, {Urry}, \& {Pesce}}]{GS03}
{Gambill}, J.~K., {Sambruna}, R.~M., {Chartas}, G., {Cheung}, C.~C.,
  {Maraschi}, L., {Tavecchio}, F., {Urry}, C.~M., \& {Pesce}, J.~E. 2003, \aap,
  401, 505

\bibitem[{{Gaskell} {et~al.}(2004){Gaskell}, {Goosmann}, {Antonucci}, \&
  {Whysong}}]{gg04}
{Gaskell}, C.~M., {Goosmann}, R.~W., {Antonucci}, R.~R.~J., \& {Whysong}, D.~H.
  2004, \apj, 616, 147

\bibitem[{{Giveon} {et~al.}(1999){Giveon}, {Maoz}, {Kaspi}, {Netzer}, \&
  {Smith}}]{giveon99}
{Giveon}, U., {Maoz}, D., {Kaspi}, S., {Netzer}, H., \& {Smith}, P.~S. 1999,
  \mnras, 306, 637, astro-ph/9902254

\bibitem[{{Green} {et~al.}(1995){Green}, {Schartel}, {Anderson}, {Hewett},
  {Foltz}, {Brinkmann}, {Fink}, {Truemper}, \& {Margon}}]{GS95}
{Green}, P.~J. {et~al.} 1995, \apj, 450, 51

\bibitem[{{Groves} {et~al.}(2004{\natexlab{a}}){Groves}, {Dopita}, \&
  {Sutherland}}]{groves04a}
{Groves}, B.~A., {Dopita}, M.~A., \& {Sutherland}, R.~S. 2004{\natexlab{a}},
  \apjs, 153, 9, astro-ph/0404175

\bibitem[{{Groves} {et~al.}(2004{\natexlab{b}}){Groves}, {Dopita}, \&
  {Sutherland}}]{groves04b}
------. 2004{\natexlab{b}}, \apjs, 153, 75, astro-ph/0404176

\bibitem[{{Haardt} \& {Maraschi}(1991)}]{haardt91}
{Haardt}, F., \& {Maraschi}, L. 1991, \apjl, 380, L51

\bibitem[{{Haas} {et~al.}(2004){Haas}, {M{\"u}ller}, {Bertoldi}, {Chini},
  {Egner}, {Freudling}, {Klaas}, {Krause}, {Lemke}, {Meisenheimer},
  {Siebenmorgen}, \& {van Bemmel}}]{haas04}
{Haas}, M. {et~al.} 2004, \aap, 424, 531, astro-ph/0406111

\bibitem[{{Haas} {et~al.}(2005){Haas}, {Siebenmorgen}, {Schulz}, {Kr{\"u}gel},
  \& {Chini}}]{haas05}
{Haas}, M., {Siebenmorgen}, R., {Schulz}, B., {Kr{\"u}gel}, E., \& {Chini}, R.
  2005, \aap, 442, L39, astro-ph/0509340

\bibitem[{{Hardcastle} {et~al.}(2002){Hardcastle}, {Birkinshaw}, {Cameron},
  {Harris}, {Looney}, \& {Worrall}}]{hb02}
{Hardcastle}, M.~J., {Birkinshaw}, M., {Cameron}, R.~A., {Harris}, D.~E.,
  {Looney}, L.~W., \& {Worrall}, D.~M. 2002, \apj, 581, 948

\bibitem[{{Hawkins}(1996)}]{hawkins96}
{Hawkins}, M.~R.~S. 1996, \mnras, 278, 787

\bibitem[{{Hook} {et~al.}(1994){Hook}, {McMahon}, {Boyle}, \& {Irwin}}]{hook94}
{Hook}, I.~M., {McMahon}, R.~G., {Boyle}, B.~J., \& {Irwin}, M.~J. 1994,
  \mnras, 268, 305

\bibitem[{{Hopkins} {et~al.}(2004){Hopkins}, {Strauss}, {Hall}, {Richards},
  {Cooper}, {Schneider}, {Vanden Berk}, {Jester}, {Brinkmann}, \&
  {Szokoly}}]{hs04}
{Hopkins}, P.~F. {et~al.} 2004, \aj, 128, 1112

\bibitem[{{Hubeny} {et~al.}(2000){Hubeny}, {Agol}, {Blaes}, \& {Krolik}}]{ha00}
{Hubeny}, I., {Agol}, E., {Blaes}, O., \& {Krolik}, J.~H. 2000, \apj, 533, 710

\bibitem[{{Humphrey} {et~al.}(2006){Humphrey}, {Villar-Mart{\'{\i}}n},
  {Fosbury}, {Vernet}, \& {di Serego Alighieri}}]{humphrey06}
{Humphrey}, A., {Villar-Mart{\'{\i}}n}, M., {Fosbury}, R., {Vernet}, J., \& {di
  Serego Alighieri}, S. 2006, \mnras, 369, 1103, astro-ph/0602504

\bibitem[{{Iwamuro} {et~al.}(2003){Iwamuro}, {Motohara}, {Maihara}, {Kimura},
  {Eto}, {Shima}, {Mochida}, {Wada}, {Imai}, \& {Aoki}}]{iwamuro03}
{Iwamuro}, F. {et~al.} 2003, \apj, 598, 178, astro-ph/0308062

\bibitem[{{Jones} {et~al.}(2004){Jones}, {d'Hendecourt}, {Sheu}, {Chang},
  {Cheng}, \& {Hill}}]{jones04}
{Jones}, A.~P., {d'Hendecourt}, L.~B., {Sheu}, S.-Y., {Chang}, H.-C., {Cheng},
  C.-L., \& {Hill}, H.~G.~M. 2004, \aap, 416, 235

\bibitem[{{Koratkar} \& {Blaes}(1999)}]{koratkar99}
{Koratkar}, A., \& {Blaes}, O. 1999, \pasp, 111, 1

\bibitem[{{Korista} {et~al.}(1997{\natexlab{a}}){Korista}, {Baldwin},
  {Ferland}, \& {Verner}}]{KB97}
{Korista}, K., {Baldwin}, J., {Ferland}, G., \& {Verner}, D.
  1997{\natexlab{a}}, \apjs, 108, 401, (KO97)

\bibitem[{{Korista} {et~al.}(1997{\natexlab{b}}){Korista}, {Ferland}, \&
  {Baldwin}}]{KF97}
{Korista}, K., {Ferland}, G., \& {Baldwin}, J. 1997{\natexlab{b}}, \apj, 487,
  555

\bibitem[{{Krongold} {et~al.}(2003){Krongold}, {Nicastro}, {Brickhouse},
  {Elvis}, {Liedahl}, \& {Mathur}}]{KN03}
{Krongold}, Y., {Nicastro}, F., {Brickhouse}, N.~S., {Elvis}, M., {Liedahl},
  D.~A., \& {Mathur}, S. 2003, \apj, 597, 832

\bibitem[{{Krongold} {et~al.}(2005{\natexlab{a}}){Krongold}, {Nicastro},
  {Brickhouse}, {Elvis}, \& {Mathur}}]{KN05-2}
{Krongold}, Y., {Nicastro}, F., {Brickhouse}, N.~S., {Elvis}, M., \& {Mathur},
  S. 2005{\natexlab{a}}, \apj, 622, 842

\bibitem[{{Krongold} {et~al.}(2005{\natexlab{b}}){Krongold}, {Nicastro},
  {Elvis}, {Brickhouse}, {Mathur}, \& {Zezas}}]{KN05-1}
{Krongold}, Y., {Nicastro}, F., {Elvis}, M., {Brickhouse}, N.~S., {Mathur}, S.,
  \& {Zezas}, A. 2005{\natexlab{b}}, \apj, 620, 165

\bibitem[{{Laor}(1998)}]{laor98}
{Laor}, A. 1998, \apjl, 496, L71+, astro-ph/9802164

\bibitem[{{Laor} {et~al.}(1997){Laor}, {Fiore}, {Elvis}, {Wilkes}, \&
  {McDowell}}]{laor97}
{Laor}, A., {Fiore}, F., {Elvis}, M., {Wilkes}, B.~J., \& {McDowell}, J.~C.
  1997, \apj, 477, 93, astro-ph/9609164

\bibitem[{{Lee} {et~al.}(1995){Lee}, {Kriss}, {Davidsen}, \& {Zheng}}]{LK95}
{Lee}, G., {Kriss}, G.~A., {Davidsen}, A.~F., \& {Zheng}, W. 1995, Bulletin of
  the American Astronomical Society, 27, 846

\bibitem[{{Malkan}(1983)}]{malkan83}
{Malkan}, M.~A. 1983, \apj, 268, 582

\bibitem[{{Mathews} \& {Ferland}(1987)}]{mf87}
{Mathews}, W.~G., \& {Ferland}, G.~J. 1987, \apj, 323, 456

\bibitem[{{Mathur} {et~al.}(1994{\natexlab{a}}){Mathur}, {Wilkes}, {Elvis}, \&
  {Fiore}}]{mathur943c351}
{Mathur}, S., {Wilkes}, B., {Elvis}, M., \& {Fiore}, F. 1994{\natexlab{a}},
  \apj, 434, 493

\bibitem[{{Mathur} {et~al.}(1994{\natexlab{b}}){Mathur}, {Wilkes}, {Elvis}, \&
  {Fiore}}]{mathur94}
------. 1994{\natexlab{b}}, Bulletin of the American Astronomical Society, 26,
  885

\bibitem[{{Moller} \& {Jakobsen}(1990)}]{moller90}
{Moller}, P., \& {Jakobsen}, P. 1990, \aap, 228, 299

\bibitem[{{Morrison} \& {McCammon}(1983)}]{mm83}
{Morrison}, R., \& {McCammon}, D. 1983, \apj, 270, 119

\bibitem[{{Nandra} {et~al.}(1997){Nandra}, {George}, {Mushotzky}, {Turner}, \&
  {Yaqoob}}]{nandra97}
{Nandra}, K., {George}, I.~M., {Mushotzky}, R.~F., {Turner}, T.~J., \&
  {Yaqoob}, T. 1997, \apj, 476, 70, astro-ph/9608170

\bibitem[{{Nicastro} {et~al.}(1999){Nicastro}, {Fiore}, {Perola}, \&
  {Elvis}}]{n99}
{Nicastro}, F., {Fiore}, F., {Perola}, G.~C., \& {Elvis}, M. 1999, \apj, 512,
  136

\bibitem[{{Obrien} {et~al.}(1988){Obrien}, {Gondhalekar}, \& {Wilson}}]{OG88}
{Obrien}, P.~T., {Gondhalekar}, P.~M., \& {Wilson}, R. 1988, \mnras, 233, 845

\bibitem[{{O'Brien} {et~al.}(2001){O'Brien}, {Page}, {Reeves}, {Pounds},
  {Turner}, \& {Puchnarewicz}}]{obrien01}
{O'Brien}, P.~T., {Page}, K., {Reeves}, J.~N., {Pounds}, K., {Turner},
  M.~J.~L., \& {Puchnarewicz}, E.~M. 2001, \mnras, 327, L37, astro-ph/0109346

\bibitem[{{Page} {et~al.}(2004){Page}, {Reeves}, {O'Brien}, {Turner}, \&
  {Worrall}}]{PR04}
{Page}, K.~L., {Reeves}, J.~N., {O'Brien}, P.~T., {Turner}, M.~J.~L., \&
  {Worrall}, D.~M. 2004, \mnras, 353, 133

\bibitem[{{Pei}(1992)}]{P92}
{Pei}, Y.~C. 1992, \apj, 395, 130

\bibitem[{{Pica} \& {Smith}(1983)}]{pica83}
{Pica}, A.~J., \& {Smith}, A.~G. 1983, \apj, 272, 11

\bibitem[{{Piconcelli} {et~al.}(2005){Piconcelli}, {Jimenez-Bail{\'o}n},
  {Guainazzi}, {Schartel}, {Rodr{\'{\i}}guez-Pascual}, \&
  {Santos-Lle{\'o}}}]{pj05}
{Piconcelli}, E., {Jimenez-Bail{\'o}n}, E., {Guainazzi}, M., {Schartel}, N.,
  {Rodr{\'{\i}}guez-Pascual}, P.~M., \& {Santos-Lle{\'o}}, M. 2005, \aap, 432,
  15, astro-ph/0411051

\bibitem[{{Porquet} {et~al.}(2004){Porquet}, {Reeves}, {O'Brien}, \&
  {Brinkmann}}]{porquet04}
{Porquet}, D., {Reeves}, J.~N., {O'Brien}, P., \& {Brinkmann}, W. 2004, \aap,
  422, 85, astro-ph/0404385

\bibitem[{{Puchnarewicz} {et~al.}(1996){Puchnarewicz}, {Mason},
  {Romero-Colmenero}, {Carrera}, {Hasinger}, {McMahon}, {Mittaz}, {Page}, \&
  {Carballo}}]{p96}
{Puchnarewicz}, E. {et~al.} 1996, \mnras, 281, 1243

\bibitem[{{Reeves} \& {Turner}(2000)}]{reeves00}
{Reeves}, J.~N., \& {Turner}, M.~J.~L. 2000, \mnras, 316, 234, astro-ph/0003080

\bibitem[{{Reimers} {et~al.}(2005){Reimers}, {Hagen}, {Schramm}, {Kriss}, \&
  {Shull}}]{reimers05}
{Reimers}, D., {Hagen}, H.-J., {Schramm}, J., {Kriss}, G.~A., \& {Shull}, J.~M.
  2005, \aap, 436, 465, astro-ph/0504014

\bibitem[{{Richards} {et~al.}(2003){Richards}, {Hall}, {Vanden Berk},
  {Strauss}, {Schneider}, {Weinstein}, {Reichard}, {York}, {Knapp}, {Fan},
  {Ivezi{\'c}}, {Brinkmann}, {Budav{\'a}ri}, {Csabai}, \& {Nichol}}]{r03}
{Richards}, G.~T. {et~al.} 2003, \aj, 126, 1131

\bibitem[{{Rouan} {et~al.}(2004{\natexlab{a}}){Rouan}, {Gratadour},
  {Cl{\'e}net}, \& {Gendron}}]{rouan04a}
{Rouan}, D., {Gratadour}, D., {Cl{\'e}net}, Y., \& {Gendron}, E.
  2004{\natexlab{a}}, in SF2A-2004: Semaine de l'Astrophysique Francaise,
  479--+

\bibitem[{{Rouan} {et~al.}(2004{\natexlab{b}}){Rouan}, {Lacombe}, {Gendron},
  {Gratadour}, {Cl{\'e}net}, {Lagrange}, {Mouillet}, {Boisson}, {Rousset},
  {Fusco}, {Mugnier}, {S{\'e}chaud}, {Thatte}, {Genzel}, {Gigan}, {Arsenault},
  \& {Kern}}]{rouan04b}
{Rouan}, D. {et~al.} 2004{\natexlab{b}}, \aap, 417, L1, astro-ph/0312094

\bibitem[{{Rouleau} \& {Martin}(1991)}]{rm91}
{Rouleau}, F., \& {Martin}, P.~G. 1991, \apj, 377, 526

\bibitem[{{Scott} {et~al.}(2004){Scott}, {Kriss}, {Brotherton}, {Green},
  {Hutchings}, {Shull}, \& {Zheng}}]{SK04}
{Scott}, J.~E., {Kriss}, G.~A., {Brotherton}, M., {Green}, R.~F., {Hutchings},
  J., {Shull}, J.~M., \& {Zheng}, W. 2004, \apj, 615, 135

\bibitem[{{Shang} {et~al.}(2005){Shang}, {Brotherton}, {Green}, {Kriss},
  {Scott}, {Quijano}, {Blaes}, {Hubeny}, {Hutchings}, {Kaiser}, {Koratkar},
  {Oegerle}, \& {Zheng}}]{SB05}
{Shang}, Z. {et~al.} 2005, \apj, 619, 41, (SBG05)

\bibitem[{{Shields} {et~al.}(1998){Shields}, {Wobus}, \& {Husfeld}}]{shields98}
{Shields}, G.~A., {Wobus}, L., \& {Husfeld}, D. 1998, \apj, 496, 743,
  astro-ph/9711210

\bibitem[{{Siemiginowska} {et~al.}(2002){Siemiginowska}, {Bechtold},
  {Aldcroft}, {Elvis}, {Harris}, \& {Dobrzycki}}]{sb02}
{Siemiginowska}, A., {Bechtold}, J., {Aldcroft}, T.~L., {Elvis}, M., {Harris},
  D.~E., \& {Dobrzycki}, A. 2002, \apj, 570, 543

\bibitem[{{Siemiginowska} {et~al.}(2003){Siemiginowska}, {Stanghellini},
  {Brunetti}, {Fiore}, {Aldcroft}, {Bechtold}, {Elvis}, {Murray}, {Antonelli},
  \& {Colafrancesco}}]{ss03}
{Siemiginowska}, A. {et~al.} 2003, \apj, 595, 643, astro-ph/0306129

\bibitem[{{Spinrad} {et~al.}(1993){Spinrad}, {Filippenko}, {Yee}, {Ellingson},
  {Blades}, {Bahcall}, {Jannuzi}, {Bechtold}, \& {Dobrzycki}}]{SF93}
{Spinrad}, H. {et~al.} 1993, \aj, 106, 1

\bibitem[{{Telfer} {et~al.}(2002){Telfer}, {Zheng}, {Kriss}, \&
  {Davidsen}}]{TZ02}
{Telfer}, R.~C., {Zheng}, W., {Kriss}, G.~A., \& {Davidsen}, A.~F. 2002, \apj,
  565, 773, (TZ02)

\bibitem[{{Trevese} {et~al.}(1994){Trevese}, {Kron}, {Majewski}, {Bershady}, \&
  {Koo}}]{trevese94}
{Trevese}, D., {Kron}, R.~G., {Majewski}, S.~R., {Bershady}, M.~A., \& {Koo},
  D.~C. 1994, \apj, 433, 494, astro-ph/9407003

\bibitem[{{Tripp} {et~al.}(1994){Tripp}, {Bechtold}, \& {Green}}]{tbg94}
{Tripp}, T.~M., {Bechtold}, J., \& {Green}, R.~F. 1994, \apj, 433, 533

\bibitem[{{Van Kerckhoven} {et~al.}(2002){Van Kerckhoven}, {Tielens}, \&
  {Waelkens}}]{VT02}
{Van Kerckhoven}, C., {Tielens}, A.~G.~G.~M., \& {Waelkens}, C. 2002, \aap,
  384, 568

\bibitem[{{Vanden Berk} {et~al.}(2001){Vanden Berk}, {Richards}, {Bauer},
  {Strauss}, {Schneider}, {Heckman}, {York}, {Hall}, {Fan}, {Knapp},
  {Anderson}, {Annis}, {Bahcall}, {Bernardi}, {Briggs}, {Brinkmann}, {Brunner},
  {Burles}, {Carey}, {Castander}, {Connolly}, {Crocker}, {Csabai}, {Doi},
  {Finkbeiner}, {Friedman}, {Frieman}, {Fukugita}, {Gunn}, {Hennessy},
  {Ivezi{\'c}}, {Kent}, {Kunszt}, {Lamb}, {Leger}, {Long}, {Loveday}, {Lupton},
  {Meiksin}, {Merelli}, {Munn}, {Newberg}, {Newcomb}, {Nichol}, {Owen}, {Pier},
  {Pope}, {Rockosi}, {Schlegel}, {Siegmund}, {Smee}, {Snir}, {Stoughton},
  {Stubbs}, {SubbaRao}, {Szalay}, {Szokoly}, {Tremonti}, {Uomoto}, {Waddell},
  {Yanny}, \& {Zheng}}]{vandenberk01}
{Vanden Berk}, D.~E. {et~al.} 2001, \aj, 122, 549

\bibitem[{{Vaughan} {et~al.}(2004){Vaughan}, {Fabian}, {Ballantyne}, {De Rosa},
  {Piro}, \& {Matt}}]{vaughan04}
{Vaughan}, S., {Fabian}, A.~C., {Ballantyne}, D.~R., {De Rosa}, A., {Piro}, L.,
  \& {Matt}, G. 2004, \mnras, 351, 193, astro-ph/0402660

\bibitem[{{Vernet} {et~al.}(2001){Vernet}, {Fosbury}, {Villar-Mart{\'{\i}}n},
  {Cohen}, {Cimatti}, {di Serego Alighieri}, \& {Goodrich}}]{vernet01}
{Vernet}, J., {Fosbury}, R.~A.~E., {Villar-Mart{\'{\i}}n}, M., {Cohen}, M.~H.,
  {Cimatti}, A., {di Serego Alighieri}, S., \& {Goodrich}, R.~W. 2001, \aap,
  366, 7, astro-ph/0010640

\bibitem[{{Whittet}(2003)}]{bookwhittet}
{Whittet}, D. 2003, {Dust in the Galactic Enviroment (2d ed.)} (Institute of
  Physics Publishing, Bristol and Philadelphia. Series in Astronomy and
  Astrophysics)

\bibitem[{{Wilkes}(2004)}]{W04}
{Wilkes}, B. 2004, in ASP Conf. Ser. 311: AGN Physics with the Sloan Digital
  Sky Survey, 37

\bibitem[{{Willott}(2005)}]{w05}
{Willott}, C.~J. 2005, \apjl, 627, L101

\bibitem[{{Wills} {et~al.}(1985){Wills}, {Netzer}, \& {Wills}}]{wills85}
{Wills}, B.~J., {Netzer}, H., \& {Wills}, D. 1985, \apj, 288, 94

\bibitem[{{Yip} {et~al.}(2004){Yip}, {Connolly}, {Vanden Berk}, {Ma},
  {Frieman}, {SubbaRao}, {Szalay}, {Richards}, {Hall}, {Schneider}, {Hopkins},
  {Trump}, \& {Brinkmann}}]{yip04}
{Yip}, C.~W. {et~al.} 2004, \aj, 128, 2603, astro-ph/0408578

\bibitem[{{Young} {et~al.}(1981){Young}, {Deverill}, {Gunn}, {Westphal}, \&
  {Kristian}}]{young81}
{Young}, P., {Deverill}, R.~S., {Gunn}, J.~E., {Westphal}, J.~A., \&
  {Kristian}, J. 1981, \apj, 244, 723

\bibitem[{{Yuan} {et~al.}(1998){Yuan}, {Siebert}, \& {Brinkmann}}]{yuan98}
{Yuan}, W., {Siebert}, J., \& {Brinkmann}, W. 1998, \aap, 334, 498

\bibitem[{{Zheng} {et~al.}(2004){Zheng}, {Kriss}, {Deharveng}, {Dixon}, {Kruk},
  {Shull}, {Giroux}, {Morton}, {Williger}, {Friedman}, \& {Moos}}]{zheng04}
{Zheng}, W. {et~al.} 2004, \apj, 605, 631, astro-ph/0312557

\bibitem[{{Zheng} {et~al.}(1997){Zheng}, {Kriss}, {Telfer}, {Grimes}, \&
  {Davidsen}}]{ZK97}
{Zheng}, W., {Kriss}, G.~A., {Telfer}, R.~C., {Grimes}, J.~P., \& {Davidsen},
  A.~F. 1997, \apj, 475, 469, (Z97)

\end{thebibliography}

\clearpage \onecolumn

%%%%%%%%%%%%%%%%%%%%%%%%%%%%%%%%%%%%%%%%%%%%%%%%%%%%%
%%%%%%%%%%%%%%F I G U R E%%%%%%%%%%%%%%%%%%%%%%%%%%%%%%%%%%
%%%%%%%%%%%%%%%%%%%%%%%%%%%%%%%%%%%%%%%%%%%%%%%%%%%%%%%

\begin{figure}
\epsscale{0.75} \plotone{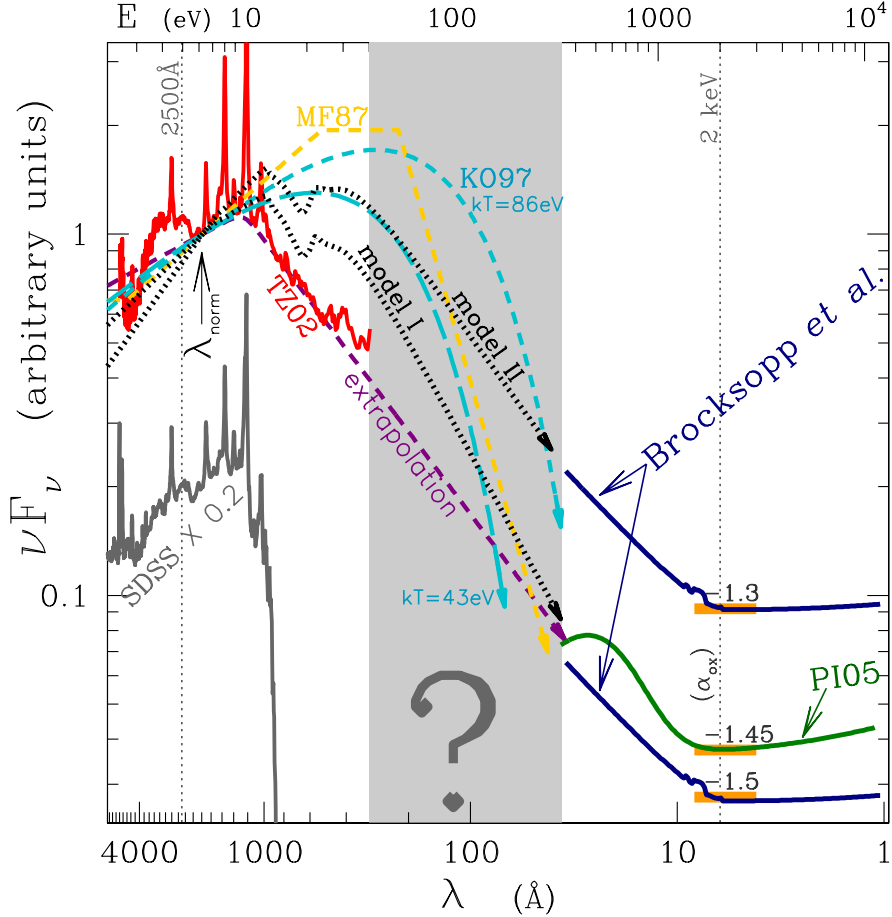} \caption{Comparison of
different \sed\ components in the UV and the X-rays: y-axis \nufnu,
lower axis: wavelength (\AA), top axis energy (eV). All \seds\ have
been normalized to unity at $\lambda_{norm.}=2000\,$\AA. The SDSS
AGN composite (gray solid line) was multiplied by 0.2 to avoid
overcrowding.  The red line represents the composite quasar \sed\
from TZ02 that combines RQQ and RLQ quasars. The dashed purple line
represents a broken powerlaw approximation of the TZ02 composite,
using index values of 0.31 and $-0.76$, for \bnuv\ and \bfuv,
respectively. We also overlay the following theoretical \seds: the
\citet{mf87} \sed\ (yellow dashed line),  Models I and II absorbed
by crystalline dust and introduced in \S\,\ref{sec:sedpeind} (dotted
black lines), the thermal-like \seds\ from KO97 with an exponential
cut-off at $kT=43$ and 86\,eV (short and long-dashed cyan lines,
respectively). To avoid cluttering, UV \seds\ end in a pointed arrow
before the X-rays. The gray shaded area represents the wavelength
region where scarce information exists. We can imagine different
ways by which the extreme-UV connects within this data gap with the
soft X-ray component. For a given X-ray \sed, the index \aox\ can be
used to determine its flux level with respect to the UV domain. The
thick orange markers at 2\,keV delineate X-ray fluxes corresponding
to arbitrary \aox\ values of $-1.3$, $-1.45$ and $-1.5$, defined
with respect to the 2500\,\AA\ flux of the TZ02 composite.
%, which does not leave out the small UV bump \citep{wills85}.
As examples of AGN \seds\ in the X-ray domain, we show two popular
model fits: a broken-powerlaw (navy blue line) and a
blackbody+powerlaw (green line) fit. They correspond to unweighed
averages of fits performed by \citet{bs06} and PI05, respectively
(see \S\,\ref{sec:xco}). The blackbody+powerlaw X-ray model is
favored in a larger number of AGN (PI05), and is adopted throughout
this Paper.
%%The  lines represent the average of 13 broken powerlaw fits of low
%%redshift quasar spectra by \citet{bs06}.
  }
\label{fig:seds}
\end{figure}

\begin{figure}
\epsscale{1.2} \plotone{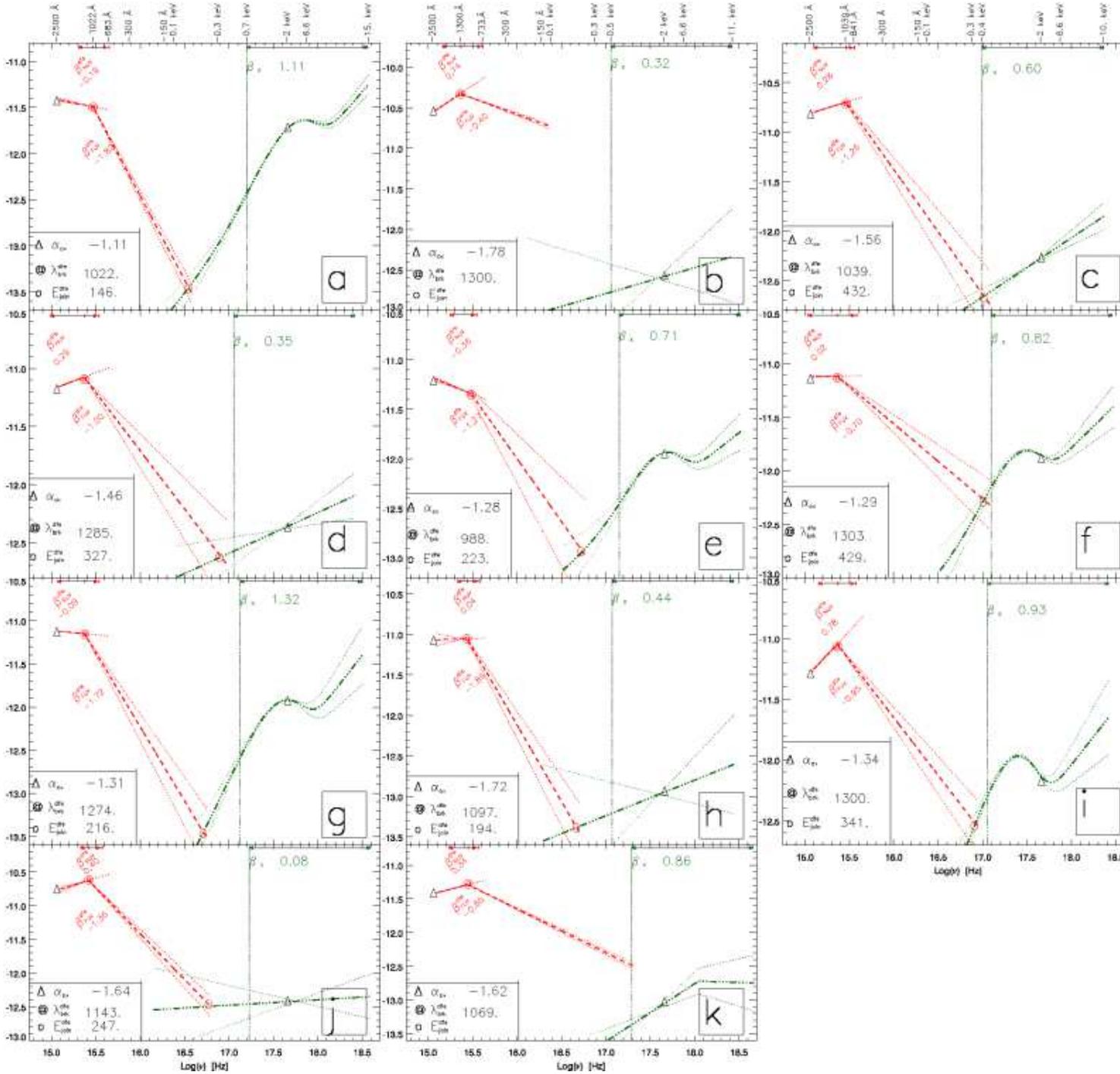} \caption{Overlay  in \nufnu\  vs.
$\nu$ of the UV  and X-ray spectral fits that were separately shown
in Fig.\ref{fig:uv} and Fig.\ref{fig:x}, respectively. Black arrows
at the top of each panel indicate the wavelength coverage of the
original data. Quasar environment is assumed dust-free. Powerlaws of
indices \bnuvf, \bfuvf\ and \bx\ (equivalent to actual slopes in
\nufnu\ plots) are used to describe the near-UV, far-UV and hard
X-ray segments, respectively. A few representative wavelength and
energy markers are shown above the top axis.  For both the UV and
X-ray segments, dashed lines represent the best-fit models.
Converging dotted lines indicate variations about these fits,
assuming a 2$\sigma$ uncertainty for the input parameters, except
for \labrk, which was held constant. The symbol `@' denotes the UV
break position, \labrk, and the circle the energy, \Enexf, where the
intercept of the extrapolated UV and X-ray segments takes place. The
bottom-left inset lists the values of \labrk, \Enexf\ and \aox. Open
triangles mark the position corresponding to 2500\,\AA\ and 2\,keV.
This Figure is  available in color in the electronic version of this
paper.} \label{fig:sed}
\end{figure}

\begin{figure}
\epsscale{.80} \plotone{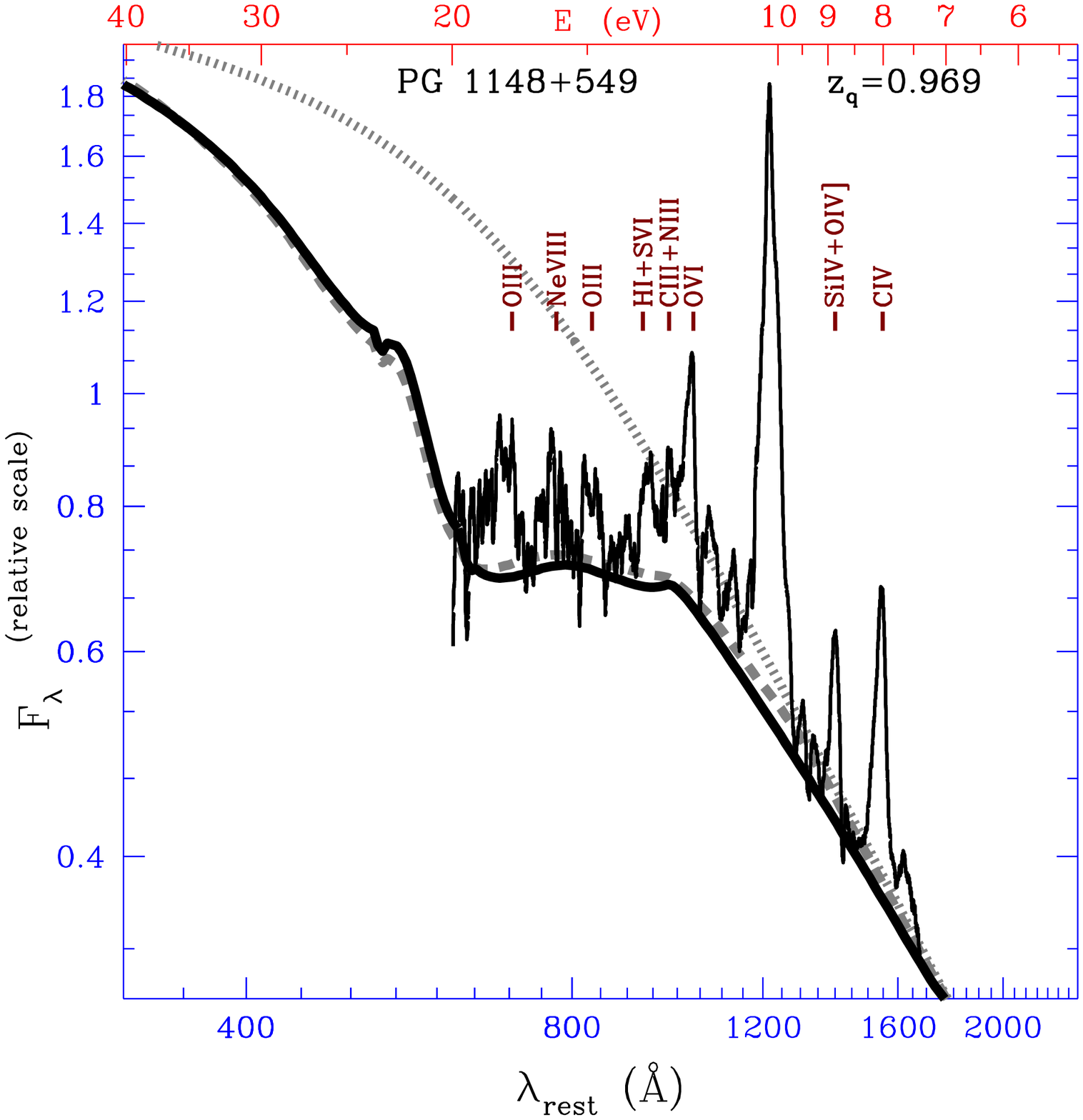} \caption{Rest-frame spectrum
in \fla\ of  \PGelf\  (from B05), multiplied by the scaling factor
$0.45 \times 10^{14}$\,\umm. Pointers indicate the position of
relevant emission lines above the continuum. The dotted line
represents the intrinsic \sed, which consists of a powerlaw with
\bnuvt = 0.8, multiplied by the rollover function \Cut\ as defined
in \S\,\ref{sec:nd}. (Coincidentally, it is equivalent to Model\,II
defined in \S\,\ref{sec:hypo}.) Its flux is unity at the Lyman
limit. The black continuous line represents the above \sed\ {\it
absorbed} by nanodiamond dust, assuming the extinction curve D3
shown in Fig.\,\ref{fig:exti} with \NN=1.0, while the dashed gray
line represents a model with \NN=0.8 that combines the A1 and D1
curves (40\% and 60\%, respectively), as in B05. } \label{fig:pg}
\end{figure}

\begin{figure}
\epsscale{.80} \plotone{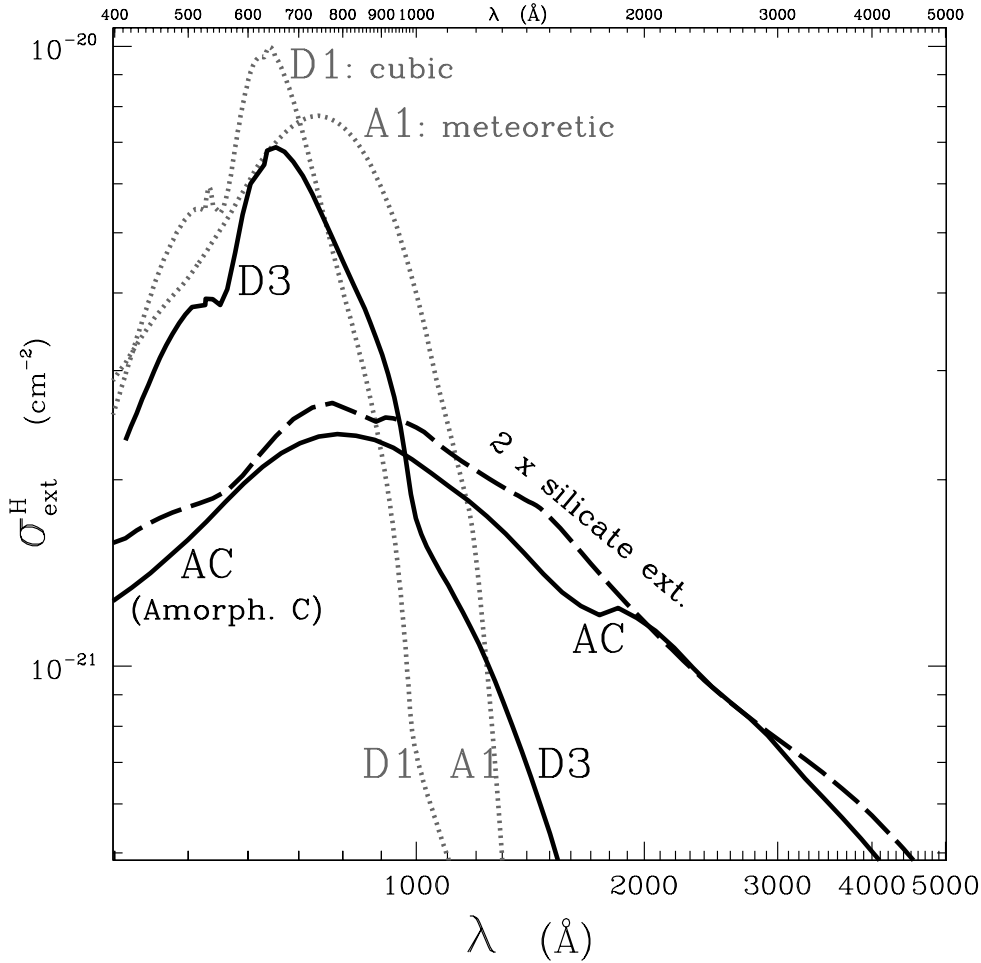} \caption{Cross-sections (normalized
to H) of different dust models as a function of wavelength.
Extinction curves labeled A1 (meteoritic nanodiamond) and D1
(terrestrial cubic diamond) were used by B05 to reproduce the far-UV
break of quasars. Curve D3 (continuous black line) is used
throughout this Paper to fit the quasar break near 1100\,\AA\ (see
\S\,\ref{sec:nd} and Fig.\,\ref{fig:pg}). It consists of terrestrial
nanodiamonds that cover a wider range of 3--200\,\AA\ as compared to
the D1 curve (3--25\,\AA).  The long-dashed line is an extinction
model of the SMC, using silicate grains with the same size range as
in \citet{P92}. The normalization assumes solar metallicity of Si,
and the resulting extinction curve has been scaled by a factor 2 for
comparison purposes. We will adopt a different dust model that
behaves similarly to that of the Pei model, but consists  of
amorphous carbon grains instead (solid line labeled AC).  The
normalization for AC assumes solar abundance of C and full depletion
onto grains ($\dc=1$). The curve AC will be used to deredden the
{\it near-UV} continua of quasars, as described in
\S\,\ref{sec:hypo}.} \label{fig:exti}
\end{figure}

%\footnote{To derive rest-frame luminosity distribution, the
%above scaled distribution should be divided by the quoted scaling
%factor and multiplied by $(1+z) 4\pi D^2_l$, where $D_l$ is the luminosity distance.}

\begin{figure}
\epsscale{1.2}
\plotone{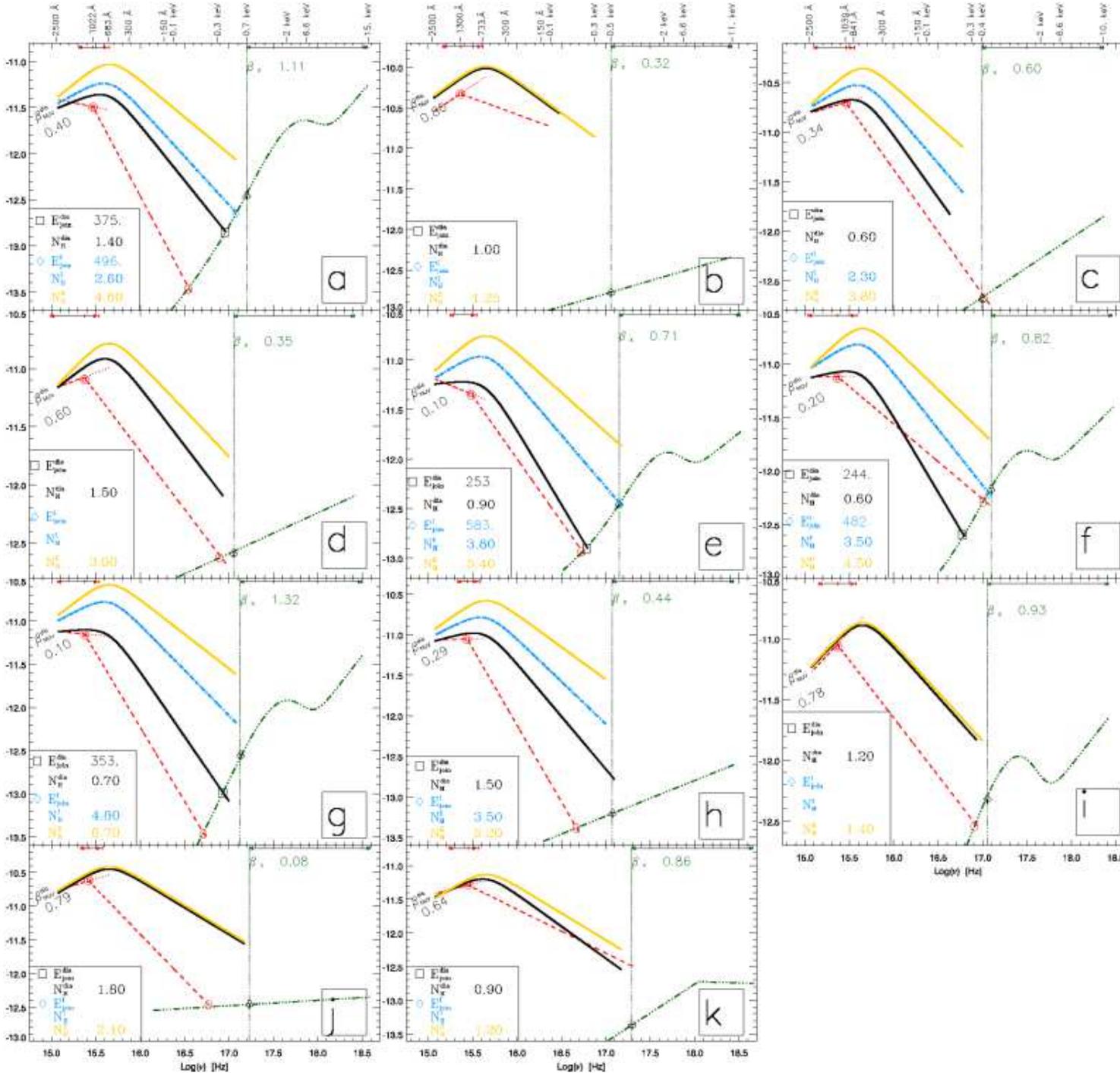}

\caption{Overlay  in \nufnu\ vs. $\nu$ of the X-ray spectral segment
(green dash-dotted line) with different UV models. The thick black
line: overlay of the intrinsic UV \sed\  of index \bnuvd\ needed to
fit the break assuming nanodiamond dust only (\S\,\ref{sec:cor});
the cyan line: overlay of the intrinsic UV \sed\ Model\,I ($\bnuvo =
0.55$) that fits the UV spectra when absorbed by both amorphous
carbon and nanodiamond dust (\S\,\ref{sec:hypo}); the yellow line:
overlay of the intrinsic UV \sed\ Model\,II ($\bnuvt = 0.8$) that
fits the UV spectra when absorbed by both amorphous carbon and
nanodiamond dust (\S\,\ref{sec:hypo}). All UV \seds\ include a
rollover at 670\,\AA\ defined by function \Cut, which steepens the
powerlaw index by amount $-1.6$ (\S\,\ref{sec:nd}). Other symbols
have the same meaning as in Fig.\,\ref{fig:sed}. For comparison
purposes, we also overlay the dust-free broken powerlaw models of
Fig.\,\ref{fig:sed} (red dashed line, \S\,\ref{sec:sed}). The
absorption columns \Ndia, \None, \Ntwo\ and \Nhx\ are shown in the
bottom inset and are listed in Table\,\ref{tab:nh}. The intercept
energies of the UV \seds\ with the extrapolated X-ray segment are
also shown in the inset and in Table\,\ref{tab:sed}.}
\label{fig:sedpeind}
\end{figure}

\begin{figure}
\epsscale{.80} \plotone{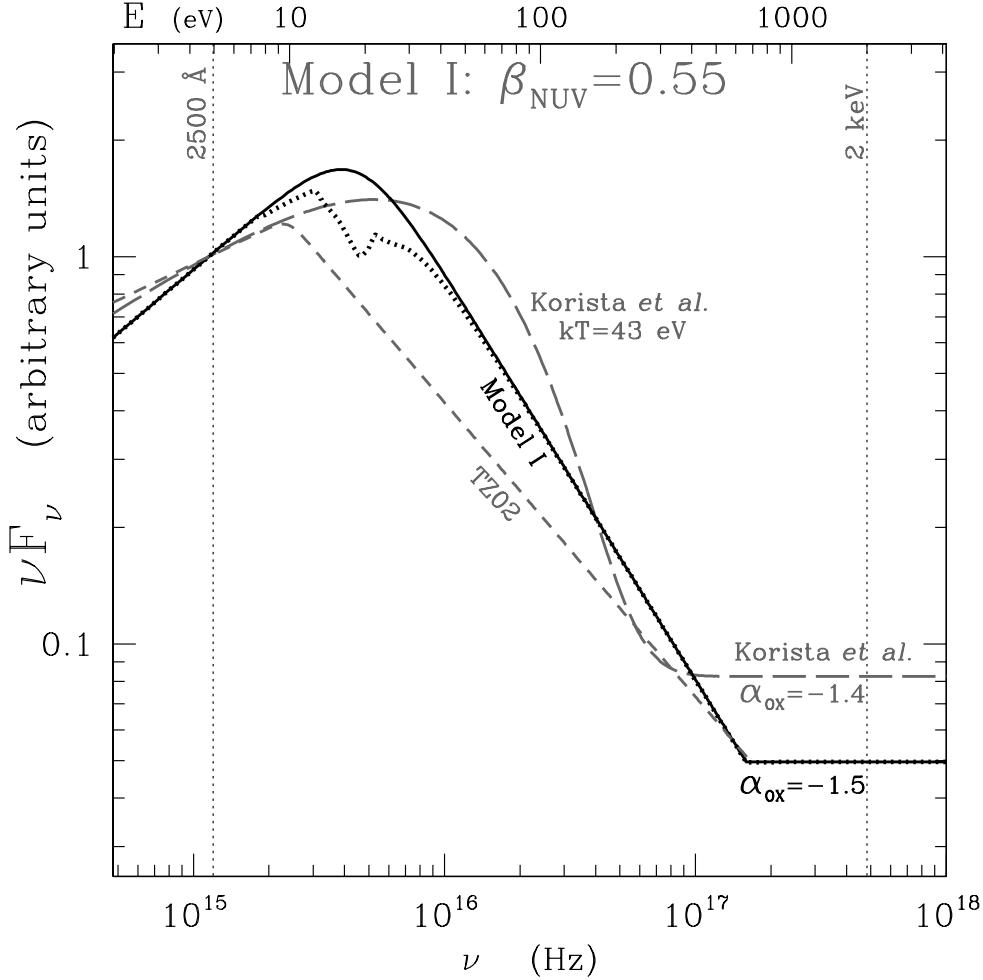} \caption{Comparison of
Model\,I with other spectral energy distributions.  The continuous
line represents the  intrinsic \sed\  from Model\,I, as defined in
\S\,\ref{sec:hypo}. It consists of a powerlaw with $\bnuvo = 0.55$,
multiplied by the turnover function \Cut\ defined in
\S\,\ref{sec:nd}.  This turnover at 18.5\,eV ($4.48 \times
10^{15}$\,Hz) consists of an index steepening (by $\ad = -1.6$),
bringing the far-UV index progressively towards $\bfuvo=-1.05$. The
dotted line represents the same \sed, absorbed by nanodiamond dust
only, assuming \NN=0.8 and the extinction curve D3
(Fig.\,\ref{fig:exti}). The long dashed line represents the softer
of the two \seds\ used by KO97 in their BELR grid of models. It
peaks at 22\,eV. The short dashed line is a broken powerlaw
approximation of the composite \sed\ of TZ02 (with $\bnuv=0.31$ and
$\bfuv=-0.76$). In order to include the X-rays in the
photoionization calculations of \S\,\ref{sec:nlr}, a truncated
powerlaw of index \bnu\ = 0.0 ($\gx=-2$) was appended to Model\,I
and the TZ02 composite, so as to obtain an \aox\ of $-1.5$ in both
cases. All distributions are normalized to unity at 2500\,\AA. }
\label{fig:distra}
\end{figure}

\begin{figure}
\epsscale{.80} \plotone{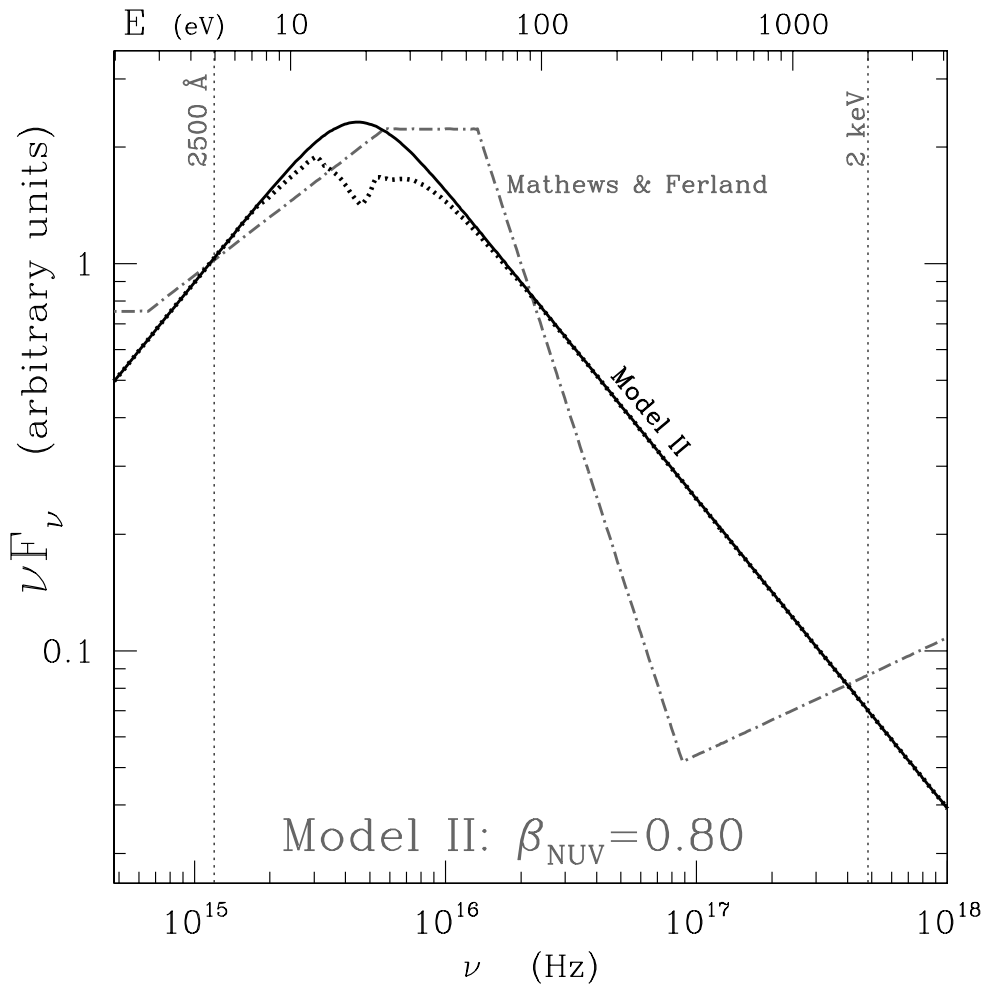} \caption{Comparison of Model\,II
with different spectral energy distributions. The continuous line
represents the   \sed\  Model\,II, as defined in \S\,\ref{sec:hypo}.
It consists of a powerlaw with $\bnuvt = 0.80$, multiplied by the
same turnover function \Cut\ as defined in \S\,\ref{sec:nd}. The
dotted line represents the same \sed\ absorbed by nanodiamond dust
only (extinction curve D3), assuming \NN=0.8. The dot-dashed line
line represents the AGN \sed\ proposed by \citet{mf87}. All
distributions are normalized to unity at 2500\,\AA. }
\label{fig:distrb}
\end{figure}

\begin{figure}
\epsscale{.80} \plotone{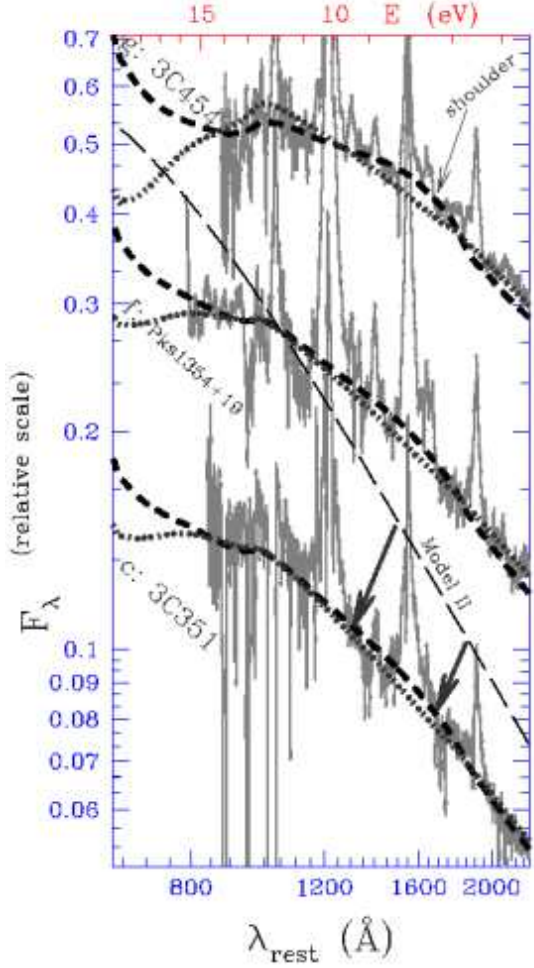} \caption{Spectral energy
distributions in \fla\ as a function of $\lambda$ for the quasars
$c$: 3C351, $f$: Pks\,1354+19 and $g$: 3C454. The dotted lines
correspond to absorption by crystalline carbon grains only, assuming
an intrinsic powerlaw index \bnuvd\ of 0.28, 0.02 and $-0.09$,
respectively. The thick short-dashed lines represent absorption
models that incorporate amorphous as well as crystalline carbon
grains. The \sed\ corresponds to Model\,II for all three objects,
that is a powerlaw of index $\bnuvt=0.8$ in the UV with a rollover
at 670\,\AA. The thin long-dashed line represents such an intrinsic
\sed, normalized to fit the spectrum of 3C351 when absorbed by dust.
} \label{fig:mod2}
\end{figure}

\begin{figure}
\epsscale{.80} \plotone{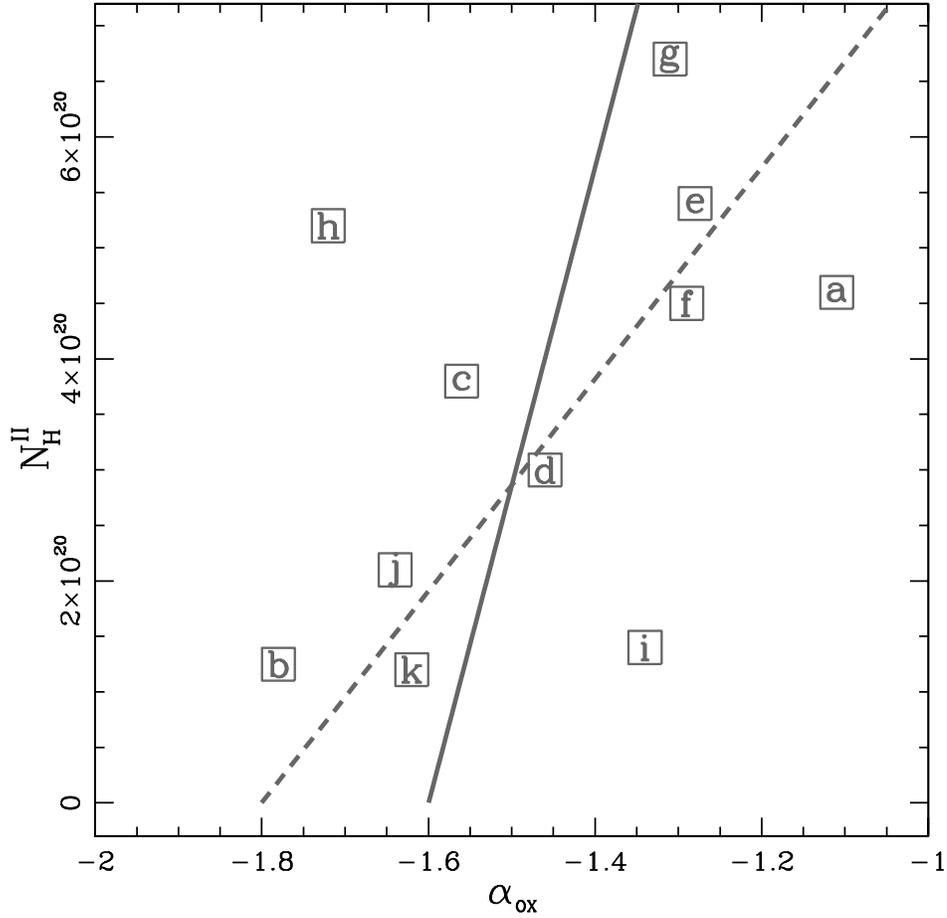} \caption{For each quasar, the
absorption column inferred by fitting Model\,II, \Ntwo, versus
measured \aox. Using the absorption cross-section of AC, the
continuous line gives the expected behavior of \aox\ if all objects
had the same intrinsic value of $-1.6$. The dashed line illustrates
the hypothetical case of a  threefold increase of the dust column,
assuming an intrinsic index of $-1.8$. } \label{fig:sin}
\end{figure}

\clearpage

\begin{deluxetable}{lcllccrrrrrl}
\tabletypesize{\scriptsize}
\rotate
\tablecaption{Observation Log for the 11 quasars \label{tab:sam}}
\tablewidth{0pt}
\tablehead{
\colhead{Common}&\colhead{Fig}  &\colhead{z}
&\colhead{\Ngal\tablenotemark{(a)}} &\colhead{Radio-}
&\colhead{Spectral} &\colhead{\emph{Chandra}}  &\colhead{Exp. Time}    &\colhead{Total}&\colhead{Frame}&\colhead{\emph{Chandra}}&\colhead{HST}\\

\colhead{Name}  &\colhead{Label}&       &\colhead{\Nhv} &\colhead{loudness}         &\colhead{Class\tablenotemark{(b)}}&\colhead{ID}&\colhead{(ksec)}&\colhead{Counts\tablenotemark{c}}&\colhead{Time\tablenotemark{d}}&\colhead{Date}&\colhead{Date}\\
\tableline
\colhead{(1)}&\colhead{(2)}&\colhead{(3)}&\colhead{(4)}&\colhead{(5)}&\colhead{(6)}&\colhead{(7)}&\colhead{(8)}&\colhead{(9)}&\colhead{(10)}&\colhead{(11)}&\colhead{(12)}
}
\startdata
Pks\,1127$-$14  &{\it a}    &1.18   &4.09   &RLQ    &A  &866    &27.3   &16370  &0.4    &2000-05-28 &1993-01-01\\
Pks\,0405$-$123 &{\it b}    &0.57   &3.8    &RLQ    &A  &2131   &8.6    &11994  &0.4    &2001-07-22 &1991-07-01\\
3C351   &{\it c}    &0.3712 &2.49   &RLQ    &A  &2128   &50.92  &9450   &1.7    &2001-08-24 &1991-10-22\\
3C334   &{\it d}    &0.56   &4.28   &RLQ    &A  &2097   &32.4   &8928   &0.85   &2001-08-22 &1991-09-07\\
B2\,0827+24 &{\it e}    &0.939  &3.62   &RLQ    &A  &3047   &18.2   &6914   &0.4    &2002-05-07 &1997-10-28\\
Pks\,1354+19    &{\it f}    &0.720  &2.23   &RLQ    &A  &2140   &9. &6445   &0.4    &2001-01-08 &1992-02-26\\
3C454.3 &{\it g}    &0.850  &6.41   &RLQ    &B  &3127   &4.9    &4716   &0.8    &2002-11-06 &1991-09-11\\
OI\,363 &{\it h}    &0.63   &4.18   &RLQ    &B  &377    &27.6   &4291   &3.2    &2000-10-10 &1997-05-15\\
Pks\,1136$-$13  &{\it i}    &0.560  &3.6    &RLQ    &A  &2138   &8.9    &3720   &0.4    &2000-11-30 &1992-01-30\\
PG\,1634+706    &{\it j}    &1.330  &4.54   &RQQ    &A  &1269   &10.8   &3113   &0.4    &1999-08-21 &1991-11-03\\
PG\,1115+080    &{\it k}    &1.718  &4.01   &RQQ    &A  &363    &26.4   &1814   &3.2    &2000-06-02 &1997-01-22\\

\enddata
\tablenotetext{a}{Webtool {\sc colden}
http://cxc.harvard.edu/toolkit/colden.jsp, which is based on radio
\HI maps of \citet{DL90}.}

\tablenotetext{b}{Spectral class based on the  far-UV break,
following B05.}

\tablenotetext{c}{Extracted from Sherpa's command ``show" using a 3
arcsec circle region around the source.}

\tablenotetext{d}{See table with CDD frame-time (sec) for standard
Subarrays http://cxc.harvard.edu/proposer/POG/.}
\end{deluxetable}

\clearpage

\begin{deluxetable}{lcrrrrllllll}
\tabletypesize{\scriptsize}
\rotate
\tablecaption{Broken powerlaw fit to the observed UV continuum\label{tab:uv}}
\tablewidth{0pt}
\tablehead{
\colhead{Common}&\colhead{Object}&\colhead{S/N\tablenotemark{a}} &\colhead{Spectral}  &\colhead{\labrk}
&\colhead{Near-UV\tablenotemark{b}} &\colhead{Far-UV\tablenotemark{b}}&\colhead{$A$\tablenotemark{b}}\\
\colhead{Name}
&\colhead{Label}&&\colhead{Range}&(\AA)&\colhead{\bnuvf}&\colhead{\bfuvf}
&($10^{-15}$) \\
\tableline
\colhead{(1)}   &\colhead{(2)} &\colhead{(3)}
&\colhead{(4)}              &\colhead{(5)} &\colhead{(6)} &\colhead{(7)}&\colhead{(8)}}

\startdata
Pks\,1127$-$14      &a &8.5 &681-1498     &1022   &$-0.19^{-0.06}_{+0.05}$    &$-1.82\pm0.08$     &3.60\\
Pks0405$-$123   &b&18.5&733-2083  &1300   &$0.74\pm0.01$  &$-0.40\pm0.03$     &38.1\\
3C351   &c&10.3&840-2386  &1039   &$0.28\pm0.01$  &$-1.26\pm0.20$     &20.1\\
3C334   &d&8.9&884-3075  &1285   &$0.29\pm0.01$  &$-1^{-0.24}_{+0.23}$   &6.55\\
B2\,0827+24     &e&9.5&861-1690  &988    &$-0.36\pm0.07$     &$-1.27\pm0.41$     &4.87\\
Pks\,1354+19    &f&9.3&788-2780  &1303   &$0.02\pm0.02$  &$-0.70^{-0.16}_{+0.14}$    &6.03\\
3C454.3     &g&11.9&877-2575  &1274   &$-0.09\pm0.01$     &$-1.72^{-0.27}_{+0.25}$    &5.32\\
OI\,363     &h&13.6&789-1406  &1097   &$0.04\pm0.25$  &$-1.89\pm0.28$     &8.73\\
Pks\,1136$-$13  &i&12.2&815-2100  &1300   &$0.78\pm0.02$  &$-0.95\pm0.14$     &6.94\\
PG\,1634+706    &j&12.5&809-1412  &1143   &$0.40\pm0.13$  &$-1.36\pm0.11$     &21.0\\
PG\,1115+080    &k&15.8&817-1759  &1069   &$0.34\pm0.01$  &$-0.65\pm0.02$     &4.62\\
\enddata
\tablenotetext{a}{Signal to Noise (\fnu/error-bar) at 1300 \AA.}

\tablenotetext{b}{The powerlaw is normalized as follows:  $\fnu =
A\; ({\nu}/{\nu_{ref}})^{+\alpha_{\nu}} = A \;
({\nu}/{\nu_{ref}})^{(\beta_{\nu} -1)} $, where $\nu_{ref} = c/1197
\times 10^{-8}$\,cm. We recall that \bnu\ is defined in the \nufnu\
plane, with $\bnu = \anu + 1$. }
\end{deluxetable}

\clearpage

\begin{deluxetable}{lllrrrrrrrrr}
\tabletypesize{\scriptsize}
\setlength{\tabcolsep}{0.06in}
\rotate
\tablecaption{X-ray spectral fits (observer-frame)\label{tab:x}}
\tablewidth{0pt}
\tablehead{

\colhead{Common}    &\colhead{Fig.}  &\colhead{Best Fit}    &\colhead{PL}   &   &\colhead{Neu. Gas}&\colhead{ d.o.f/ $\chi^2$\tablenotemark{(d)}}  &\colhead{Ionized Gas}  &\colhead{zBB } &\colhead{G}    &\colhead{d.o.f./$\chi^2$/Q}\\
\colhead{Name}  &\colhead{Lab}    &  \colhead{Model Descr.} &\colhead{\bx\tablenotemark{a}} &\colhead{N\tablenotemark{b}}   &\colhead{{\Nhx}\tablenotemark{(c)}} & & \colhead{Log\,$U$ / \Nhx / $z$}    &\colhead{kT/N\tablenotemark{(e)}}  &\colhead{fwhm/pos/A\tablenotemark{(f)}}    &\\
\\
\colhead{(1)}   &\colhead{(2)}  &\colhead{(3)}  &\colhead{(4)}  &\colhead{(5)}  &\colhead{(6)}  &\colhead{(7)}  &\colhead{(8)}  &\colhead{(9)}  &\colhead{(10)}&\colhead{(11)}
}
\startdata
Pks\,1127$-$14  &a  &GA*IA*(PL+zBB) &$1.11\pm0.12$  &4.02   &$27.34^{-7.13}_{+8.8}$& 285/608 &\nodata    &0.63/2.44  &\nodata    &282/272/0.65\\
Pks\,0405$-$123 &b  &GA*IA*PL   &$ 0.32^{-0.66}_{+0.70}$    &1.54   &$4.09^{-7.94}_{+4.23}$&49/33 &\nodata    &\nodata    &\nodata    &48/28/0.99\\
3C351   &c  &GA*IA1*IA2*(PL+G)  &$0.60\pm0.16$  &2.8    &\nodata& 232/900   &$1.38\pm0.03/363.1\pm1.07/0.36$    &\nodata    &0.08/2.66/0.23 &224/214/0.67\\
    &   &   &   &   &  & &$0.90\pm0.21/26.3\pm1.32/0.32$\\
3C334   &d  &GA*IA*PL   &$0.35\pm0.23$  &2.60   &$4.68^{-1.45}_{+1.53}$&177/229  &\nodata   &\nodata    &\nodata    &176/149/0.93\\
B2\,0827+24 &e  &GA*IA*(PL+zBB) &$0.71\pm0.22$  &2.95   &$13.05^{-5.52}_{+7.26}$ &172/219   &\nodata    &0.46/1.13  &\nodata    &169/128/0.99\\
Pks\,1354+19    &f  &GA*IA*(PL+zBB) &$0.82\pm0.23$  &4.95   &$6.44^{-12.66}_{+7.71}$ &166/206   &\nodata    &0.28/1.54  &\nodata    &163/143/0.86\\
3C454.3 &g  &GA*(PL+zBB+G)  &$1.32\pm0.40$  &1.89   &$ \le 43.38$ &67/117  &\nodata    &0.43/1.49  &0.42/2.14/1.20 &62/44/0.96\\
OI\,363 &h  &GA*IA*(PL) &$0.44\pm0.72$  &0.66   &$7.21^{-11.47}_{+6.05}$&61/54    &\nodata    &\nodata   &\nodata    &60/46/0.76\\
Pks\,1136$-$13  &i  &GA*(PL+zBB+G)  &$0.93\pm0.36$  &2.30   &$\le 2.40$&220/396 &\nodata    &0.25/1.35  &0.01/ 3.35/2.57    &115/90/0.71\\
PG\,1634+706    &j  &GA*(PL+G)  &$0.08^{-0.39}_{+0.40}$ &2.6    &$\le 1.16$&93/92 &\nodata    &\nodata    &0.001/2.84/28.92   &90/88/0.54\\
PG\,1115+080    &k  &GA*BP  &$0.86\pm0.36$ \tablenotemark{g}    &1.67
&$\le 14.28$ &71/77   &\nodata    &\nodata    &\nodata&70/49/0.97 \\
\enddata

\tablenotetext{a}{Index of powerlaw fits carried out, considering
the 2.5--6 keV (rest-frame) region only. We recall that $\beta$
indices are defined in $\nufnu $ vs $\nu$, with $\bx = \ax + 1 = 2 -
\gx$.}

\tablenotetext{b}{Powerlaw normalization defined at 1\,keV in
$10^{-4}$ photons $keV^{-1} cm^{-2} s^{-1}$.}

\tablenotetext{c}{Column density of Neutral Gas \Nhx $\times$ \Nhv\ due to intrinsic
absorption and associated $2\sigma$ error.}

\tablenotetext{d}{Statistics during initial fit; Power Law attenuated by Galactic absorption
and extrapolated between 0.3 to 6 keV (Observed Frame).}

\tablenotetext{e}{Black Body Temperature (keV) and normalization in
units of $10^{-5} L_{39}/D^2_{10}$, where $L_{39}$ is the source
luminosity in units of $10^{39}$ ergs $s^{-1}$ and $D_{10}$ is the
distance to the source in units of 10\,kpc.}

\tablenotetext{f}{1-D unnormalized Gaussian function. Listed:
full-width at half-maximum (keV), mean position (keV) and  amplitude
($10^{-4}$).}

\tablenotetext{g}{Broken Power Law. Spectral Index after break (1.69
keV) is \bx $= -0.05^{-0.22}_{+0.25}$ and  energy of reference at
0.5\,keV. }
\end{deluxetable}

\begin{deluxetable}{lcllllllllccccl}
\tabletypesize{\scriptsize}
\rotate
\tablecaption{Parameters for the four different UV \sed\ Models  \label{tab:sed}}
%\tablecolumns{15}
\tablewidth{0pt} \tablehead{ \multicolumn{3}{c}{Quasars} &
\colhead{\phm{}} & \multicolumn{2}{c}{Dust-free Model} &
\colhead{\phm{}} & \multicolumn{3}{c}{Break-corrected Model} &
\colhead{\phm{}} & \multicolumn{2}{c}{AC Dust Model\,I} &
\colhead{\phm{}} & \multicolumn{1}{c}{Model\,II}
\\
\cline{1-3}\cline{5-6} \cline{8-10} \cline{12-13}\cline{15-15}{\vspace{0.1mm}}\\
\colhead{Common}    &   \colhead{Object}  & \colhead{\aox}  &  \phm{}       &   \colhead{\bnuvf}    &   \colhead{\Enexf}    &    \phm{}  &  \colhead{\bnuvd}    &   \colhead{\Enexd}    &   \colhead{\dmaxd}    &   \colhead{\phm{}}    &       \colhead{\Enexo}    &   \colhead{\dmaxo}    &  \colhead{\phm{}}     &   \colhead{\dmaxt}
\\
\colhead{Name}  &   \colhead{Label} &\phm{} & \phm{}  & \phm{}  & \colhead{eV}  &  \phm{}   & \phm{} & \colhead{eV}     & \phm{}    & \phm{} &\phm{}    & \colhead{eV}  &   \phm{}   &\phm{} \\
%\\&    &   &   &   &   &   &   &   &   &   & \\
\tableline\\
\colhead{(1)}   &   \colhead{(2)}   &  \colhead{(3)} & \phm{} &
\colhead{(4)}   &   \colhead{(5)}   &     & \colhead{(6)}   &
\colhead{(7)} &   \colhead{(8)}   & &       \colhead{(9)}   &
\colhead{(10)}  & & \colhead{(11)}   {\vspace{0.1mm}} } \startdata
Pks\,1127$-$14      &   a  &$-1.11$ &   \phm{}      &   -0.19   &   146  &  &   0.4     &   375     &   -1.21   &   \phm{}      &   496     &   -1.39   &   \phm{}      & $-1.78$\\

Pks\,0405$-$123     &   b  &$-1.78$ &   \phm{}  &
0.74    &   \nodata &   &   0.8     &   \nodata
&   -2.73   &   \phm{}  &   \tablenotemark{(a)} & \tablenotemark{(a)}   &   \phm{}  & $-2.86$   \\
3C351   &   c  &$-1.56$ &   \phm{}  &   0.28    &   432     &   &   0.34    &   \nodata     &   -1.87   &   \phm{}  &   \nodata     &   -2.22   &   \phm{}  & $-2.60$       \\
3C334   &   d  &$-1.46$ &   \phm{}  &   0.29    &   327     &  &    0.6     &   \nodata     &   -1.85   &   \phm{}  & \tablenotemark{(a)}   & \tablenotemark{(a)}   &   \phm{}  &   $-2.15$ \\
B2\,0827+24     &   e  &$-1.28$ &   \phm{}  &   -0.36   &   223  &  &   0.1     &   253     &   -0.97   &   \phm{}  &   583     &   -1.60   &   \phm{}  & $-2.00$\\
Pks\,1354+19    &   f  &$-1.29$ &   \phm{}  &   0.02    &   429  &  &   0.2     &   244     &   -1.0    &   \phm{}  &   482 &   -1.54   &   \phm{}  & $-1.90$   \\
3C454.3     &   g  &$-1.31$ &   \phm{}  &   -0.09   &   216     &  &    0.1     &   353     &   -1.10   &   \phm{}  &  \nodata  &   -1.81   &   \phm{}  & $-2.20$   \\
OI\,363     &   h &$-1.72$  &   \phm{}  &   0.04    &   194     &  &    0.29    &   \nodata     &   -1.92   &   \phm{}  &   \nodata     &   -2.32   &   \phm{}  & $-2.72$\\
Pks\,1136$-$13  &   i  &$-1.34$ &   \phm{}  &   0.78    &   341 &   &   0.78    &   \nodata     &   -1.83   &   \phm{}  & \tablenotemark{(a)}   & \tablenotemark{(a)}   &   \phm{}  &   $-1.89$ \\
PG\,1634+706    &   j  &$-1.64$ &   \phm{}  &   0.40    &   247 &   &   0.79    &   \nodata     &   -2.17   &   \phm{}  & \tablenotemark{(a)}   & \tablenotemark{(a)}   &   \phm{}  &   $-2.16$ \\
PG\,1115+080    &   k  &$-1.62$ &   \phm{}  &   0.34    &   \nodata &   &   0.64    &   \nodata     &   -2.07   &   \phm{}  & \tablenotemark{(a)}   & \tablenotemark{(a)}   &   \phm{}  &   $-2.23$\\

\enddata
\tablenotetext{a}{Model\, could not be applied to these quasars,
since their near-UV index is already harder than the target value of
$\bnuvo=0.55$.}
\end{deluxetable}

%%%%%%%%%%%%%%%%%%%%%%%%%%%%%%%%%%%%%%%%%%%%%%%%%%%%%

\begin{deluxetable}{lcllllllllllllll}
\tabletypesize{\scriptsize}
\rotate
\tablecaption{Absorption columns due to dust and metals \label{tab:nh}}
%\tablecolumns{16}
\tablewidth{0pt}
\tablehead{
\multicolumn{1}{c}{Quasars} &  \colhead{\phm{}} &
\multicolumn{2}{c}{Galactic Abs.} & \colhead{\phm{}} &
\multicolumn{1}{c}{X-rays} & \colhead{\phm{}} &
\multicolumn{1}{c}{ND Dust} & \colhead{\phm{}} &
\multicolumn{3}{c}{AC Dust Model\,I} & \colhead{\phm{}} &
\multicolumn{3}{c}{AC Dust Model\,II}
\\
\cline{1-1}\cline{3-4} \cline{6-6} \cline{8-8}\cline{10-12}\cline{14-16}\\

\colhead{Object}    & \colhead{\phm{}}      &   \colhead{\Ngal}     &   \colhead{\ebvg} &   \phm{}  &\colhead{\Nhx} &  \phm{}   &   \colhead{\Ndia} &\phm{}     &   \colhead{\None} &   \colhead{\ebvo} &   \colhead{\None/\Nhx}    &    &  \colhead{\Ntwo} &   \colhead{\ebvt} &   \colhead{\Ntwo/\Nhx}    \\
\colhead{Label} &   \phm{}
&\colhead{\Nhv} & \phm{}    &\phm{} &\colhead{\Nhv} &  \phm{} &
\colhead{\Nhv}      &   \phm{}  &\colhead{\Nhv}     &
\phm{}  &\phm{} &\phm{} &\colhead{\Nhv}     &\phm{}
&\phm{}     \\
\tableline \colhead{(1)}   &\phm{}     &   \colhead{(2)}   &
\colhead{(3)}   &       &\colhead{(4)}  &\phm{} &   \colhead{(5)} &
\phm{}  &   \colhead{(6)}   &   \colhead{(7)}   & \colhead{(8)}   &
\phm{} &   \colhead{(9)}  &   \colhead{(10)} &   \colhead{(11)} }
\startdata
    a   &   \phm{}      &   4.09    &   0.069   &   \phm{}      &27.3      &   \phm{}  &   1.4     &   \phm{}      &   2.6     &   0.017   &   0.095   &  &    4.6     &   0.045   &   0.17    \\
    b   &   \phm{}  &   3.8     &   0.064   &   \phm{}  &4.1   &   \phm{}  &   1.0 &   \phm{}  &   \nodata & \nodata   &   \nodata &  &    1.25    &   3.6$\times10^{-3}$  &   0.31    \\
    c   &   \phm{}  &   2.49    &   0.042   &
\phm{}  &389  &   \phm{} &    0.60    &   \phm{}  & 2.3     & 0.024
&   0.0059  &   &   3.8     &   0.045
&   0.010   \\
    d   &   \phm{}  &   4.28    &   0.073   &   \phm{}  &4.7   &   \phm{} &    1.5     &   \phm{}  &   \nodata &   \nodata &   \nodata     &   &   3.0     &   0.021   &   0.64    \\
    e   &   \phm{}  &   3.62    &   0.061   &   \phm{}  &13.1  &   \phm{} &    0.90    &   \phm{}  &   3.8     &   0.041   &   0.29    &   &   5.4     &   0.065   &   0.41    \\
    f   &   \phm{}  &   2.23    &   0.038   &   \phm{}  &6.4   &   \phm{} &    0.60    &   \phm{}  &   3.5     &   0.041   &   0.54    &  &    4.5     &   0.055   &   0.69    \\
    g   &   \phm{}  &   6.41    &   0.11    &   \phm{}  &$ \le 43.4$   &   \phm{} &    0.70    &   \phm{}  &   4.8     &   0.053   &   $>$0.11     &  &    6.7 &       0.085   &   $>$0.15     \\
    h   &   \phm{}  &   4.18    &   0.071   &   \phm{}  &7.2   &   \phm{} &    1.5     &   \phm{}  &   3.5     &   0.029   &   0.48    & & 5.2     &   0.054   &   0.72    \\
    i   &   \phm{}  &   3.6     &   0.061   &   \phm{}  &$\le 2.4$     &   \phm{} &    1.2     &   \phm{}  &   \nodata & \nodata   &   \nodata &  &    1.4     &   2.3$\times10^{-3}$  &   $>$0.58     \\
    j   &   \phm{}  &   4.54    &   0.077   &   \phm{}  &$\le 1.2$     &   \phm{} &    1.8     &   \phm{}  &   \nodata & \nodata   &   \nodata     &  &    2.1     &   4.3$\times10^{-3}$  &   $>$1.8     \\
    k   &   \phm{}  &   4.01    &   0.068   &   \phm{}  &$\le 38.5$    &   \phm{}  &   0.90    &   \phm{}  &   \nodata &  \nodata  &  \nodata  &  &    1.2     &   4.3$\times10^{-3}$  &   $>$0.031    \\

\enddata
\end{deluxetable}

\clearpage

\begin{table*} \caption[]{Line ratios from
photoionization calculations$^{\mathrm{a}}$ } \label{tab:rat}
%\tabletypesize{\scriptsize}
\begin{tabular}{llcccc}
\noalign{\smallskip}\hline\hline\noalign{\smallskip} Line ID
            & $~\;\lambda$  & E-NLR$^{\mathrm{b}}$ & TZ02 & Model\,I$^{\mathrm{c}}$ & Model\,II$^{\mathrm{c}}$ \\
   (1) & ~(2) & (3) & (4) & (5) & (6)  \\
            \noalign{\smallskip}
            \hline
            \noalign{\smallskip}
%%%%%%
\Lya   & 1216 & 7.1$^{\mathrm{d}}$     & 12.3    & 13.7   & 13.8  \\
\hb    & 4861 & 1.0     & 1.00    & 1.00   & 1.00  \\
\ha    & 6563 & 2.9     & 2.93    & 2.92   & 2.91  \\
\heii  & 1640 & 0.58    & 1.2     & 1.3    & 1.7   \\
\heii  & 4686 & --      & 0.15    & 0.17   & 0.22  \\
\cii   & 2326 & 0.10    & 0.05    & 0.03   & 0.05  \\
\ciii  & 977  & --      & 0.029   & 0.020  & 0.042 \\
\ciii  & 1909 & 0.26    & 0.5     & 0.5    & 0.8   \\
\civ   & 1549 & 0.79    & 1.4     & 1.3    & 2.1   \\
\nii   & 6583 & 0.83    & 0.21    & 0.18   &  0.23  \\
\niii  & 991  & --      & 0.0018  & 0.0012 & 0.0026 \\
\niii  & 1749 & --      & 0.024   & 0.023  &  0.038 \\
\niv   & 1485 & --      & 0.12    & 0.11   &  0.18  \\
\nv    & 1240 & 0.34    & 0.36    & 0.33   &  0.54  \\
\oi    & 6300 & 0.29    & 0.19    & 0.09   &  0.16 \\
\oii   & 3727 & 1.3     & 0.56    & 0.54   &  0.69 \\
\oiii  & 4363 & --      & 0.057   &  0.060 & 0.10  \\
\oiii  & 5007 & 9.8     & 10      &  11    &  14   \\
\oiiiu & 1665 & 0.10    & 0.084   & 0.075  &  0.15 \\
\ovi\  & 1035 & --      & 1.3     & 1.1    & 2.6   \\
\sii   & 6723 & 0.60    & 0.36    & 0.18   &  0.27 \\
%\siv  & 1406 & --      & 0.022   & 0.012  & 0.028 \\
\mgii  & 2800 & --      & 0.29    & 0.22   & 0.32  \\
\siiii & 1887 & --      & 0.0023  & 0.0022 &   0.0051\\
%\siiv & 1397 & --      & 0.022   & 0.015  & 0.043  \\
\neiii & 3869 & --      & 0.57    & 0.66   & 0.87   \\
\neiv  & 2424 & 0.61    & 0.2     & 0.2    &  0.27  \\
\nev   & 3426 & --      & 0.61    & 0.63   & 0.93   \\
\fevii & 6087 & --      & 0.00085 & 0.00089& 0.0011 \\
\fex   & 6374 & --      & 0.0076  & 0.0046 & 0.025  \\
%%%%%%%%%%% average qties
\noalign{\smallskip}
\phihet/\phih\ & --  & -- &  0.088 & 0.093 & 0.127 \\
%$\phih/F_{912}$ & $10^{26}$ & \phm{} & 0.86 &  0.97& 1.14 \\
%$\phih/F_{2500}$ & $10^{25}$ & \phm{} & 1.91 & 3.53 & 6.32\\
$\left<{\rm T_e}\right>$ & (K) & -- & 9450  & 9300 & 10100 \\
$\left<{\rm n_e}\right>$ & (\cmc) & -- &  595 & 565 & 560 \\
\noalign{\smallskip}   \hline\hline
\end{tabular}
   \end{table*}
   %%%  footnotes of table listed below
\clearpage
\begin{list}{}{}
\item[$^{\mathrm{a}}$] Line ratios are expressed relative to \hb.
All models are ionization bounded and assume a `front' density $\nhz
=10^2$\,\cmc, solar metallicities and dust grains made of amorphous
carbon corresponding to a depletion of only $\dc=0.02$ (see
\S\,\ref{sec:abnd}).
\item[$^{\mathrm{b}}$] Line fluxes, relative to \hb, for the $z=2.36$ narrow-line
radio-galaxy 4C$-$00.54, taken from \citet{vernet01, iwamuro03} and
Humphrey et al. (in prep.).  The relatively high-ionization state in
the E-ELR of this source ($\ciii/\cii=8.1$) (see
\S\,\ref{sec:comp}), taken together with the unperturbed kinematics
(FWHM$\sim700$\,\kms\ and large radio size ($\sim200$\,kpc),
suggests that it is not strongly affected by shock-ionization
\citep{humphrey06}. Since H$\beta$ was not detected in the spectrum
of this source, we have assumed that H$\alpha$/H$\beta=2.9$. The
line ratios are corrected for Galactic reddening using $\ebv =
0.026$ \citep{vernet01}.
\item[$^{\mathrm{c}}$] Both \seds\ labeled Model\,I and II have been
absorbed by nanodiamond dust and are plotted as dotted lines in
Figs.\,\ref{fig:distra} and \ref{fig:distrb}, respectively.
\item[$^{\mathrm{d}}$] The \Lya\ is most likely absorbed, either by
internal dust or possibly by an extended but local \HI\ absorber,
which is a common phenomenon in high redshift radio-galaxies
\citep[][ and references therein]{binette06}.
\end{list}

%%%%%%%%%%%%%%%%%%%%%%%%%%%%%%%%%%%%%%%%%%%%%%%%%%
%%%%%%%%%%%%%%%% END %%%%%%%%%%%%%%%%%%%%%%%%%%%%%
\end{document}